\documentclass[aoas]{imsart}

\RequirePackage{amsthm, amsmath, amsfonts, amssymb, bm, mathtools, mathrsfs}
\RequirePackage[authoryear]{natbib}
\usepackage{array, booktabs}
\usepackage{subcaption}
\usepackage[many]{tcolorbox}
\usepackage{multirow} 

\usepackage{siunitx}
\usepackage{tikz}

\usepackage{clipboard}
\newclipboard{output-myclipboard} 

\newcommand{\supplementnumbering}{
  \renewcommand{\thesection}{S\arabic{section}}
  \renewcommand{\theequation}{S\arabic{equation}}
  \renewcommand{\thefigure}{S\arabic{figure}}
  \renewcommand{\thetable}{S\arabic{table}}
  \renewcommand{\thepage}{Page S\arabic{page}}
}

\usepackage{hyperref}

\startlocaldefs
\usetikzlibrary{arrows.meta}
\usetikzlibrary{decorations}
\usetikzlibrary{decorations.pathreplacing, positioning}
\tikzset{
    center coordinate/.style={
            execute at end picture={
                    \path ([rotate around={180:#1}]perpendicular cs: horizontal line through={#1},
                    vertical line through={(current bounding box.east)})
                    ([rotate around={180:#1}]perpendicular cs: horizontal line through={#1},
                    vertical line through={(current bounding box.west)});}}}

\newcommand{\1}{\mathbf{1}}
\renewcommand{\d}{\, \mathrm{d}}
\newcommand{\A}{\mathbf{A}}

\newcommand{\C}{\mathbf{C}}
\newcommand{\D}{\mathbf{D}}

\newcommand{\F}{\mathbf{F}}

\renewcommand{\H}{\mathbf{H}}
\newcommand{\I}{\mathbf{I}}

\newcommand{\K}{\mathbf{K}}
\renewcommand{\L}{\mathbf{L}}

\newcommand{\N}{\mathbf{N}}
\renewcommand{\P}{\mathbf{P}}
\newcommand{\Q}{\mathbf{Q}}
\newcommand{\R}{\mathbf{R}}

\newcommand{\V}{\mathbf{V}}

\newcommand{\X}{\mathbf{X}}
\newcommand{\Y}{\mathbf{Y}}

\renewcommand{\b}{\mathbf{b}}
\newcommand{\cb}{\mathbf{c}}

\renewcommand{\u}{\mathbf{u}}
\renewcommand{\v}{\mathbf{v}}
\newcommand{\w}{\mathbf{w}}
\newcommand{\y}{\mathbf{y}}
\newcommand{\h}{\mathbf{h}}
\newcommand{\x}{\mathbf{x}}
\newcommand{\z}{\mathbf{z}}
\newcommand{\m}{\mathbf{m}}
\newcommand{\0}{\mathbf{0}}

\newcommand{\balpha}{\bm{\alpha}}
\newcommand{\bbeta}{\bm{\beta}}
\newcommand{\bdelta}{\bm{\delta}}
\newcommand{\bgamma}{\bm{\gamma}}
\newcommand{\blambda}{\bm{\lambda}}
\newcommand{\bkappa}{\bm{\kappa}}

\newcommand{\btheta}{\bm{\theta}}

\newcommand{\bphi}{\bm{\phi}}

\newcommand{\bmu}{\bm{\mu}}

\newcommand{\Gau}{\mathrm{Gau}}

\newcommand{\Exp}{\mathbb{E}}

\newcommand{\diag}{\operatorname{diag}}
\newcommand{\vech}{\operatorname{vech}}

\newcommand{\fricunit}{~\textrm{M}~\textrm{Pa}~\textrm{m}^{-1/3}~\textrm{yr}^{1/3}}
\newcommand{\fricunitsq}{~\textrm{M}~\textrm{Pa}^2~\textrm{m}^{-2/3}~\textrm{yr}^{2/3}}
\newcommand{\stiffnessunit}{~\textrm{M}~\textrm{Pa}~\textrm{yr}^{1/3}}

\DeclareMathOperator*{\argmin}{arg\,min}

\DeclarePairedDelimiter\abs{\lvert}{\rvert}%
\DeclarePairedDelimiter\norm{\lVert}{\rVert}%

\DeclarePairedDelimiterX{\divergence}[2]{(}{)}{%
    #1\;\delimsize\|\;#2%
}
\newcommand{\KL}{\textrm{KL} \divergence}

\DeclareTextSymbolDefault{\dh}{T1} 

\makeatletter
\let\oldabs\abs
\def\abs{\@ifstar{\oldabs}{\oldabs*}}
\let\oldnorm\norm
\def\norm{\@ifstar{\oldnorm}{\oldnorm*}}
\makeatother

\endlocaldefs

\begin{document}

\begin{frontmatter}
    \title{Neural posterior inference with state-space models for calibrating ice sheet simulators}
    \runtitle{Neural posterior inference with state-space models}

    \begin{aug}
        \author[A]{\fnms{Bao Anh}~\snm{Vu} \ead[label=e1]{baoanh.vu@anu.edu.au}},
        \author[B]{\fnms{Andrew}~\snm{Zammit-Mangion} \thanks{Corresponding author.} \ead[label=e2]{a.zammit\_mangion@unsw.edu.au}},
        \author[C]{\fnms{David}~\snm{Gunawan}\ead[label=e3]{dgunawan@uow.edu.au}},
        \author[D]{\fnms{Felicity}~\snm{S. McCormack}\ead[label=e4]{felicity.mccormack@monash.edu}}
        \and
        \author[C]{\fnms{Noel}~\snm{Cressie}\ead[label=e5]{ncressie@uow.edu.au}}

        \address[A]{Research School of Finance, Actuarial Studies and Statistics, Australian National University, Canberra, Australian Capital Territory, Australia 2600, and Securing Antarctica's Environmental Future, School of Mathematics and Physics, University of Wollongong, Wollongong, New South Wales, Australia 2500\printead[presep={,\ }] {e1}}

        \address[B]{School of Mathematics and Statistics, University of New South Wales, Sydney, New South Wales, Australia 2052, and Securing Antarctica's Environmental Future, School of Mathematics and Physics, University of Wollongong, Wollongong, New South Wales, Australia 2500\printead[presep={,\ }] {e2}}
        
        \address[C]{Securing Antarctica's Environmental Future, School of Mathematics and Physics, University of Wollongong, Wollongong, New South Wales, Australia 2500\printead[presep={,\ }] {e3,e5}}     


        \address[D]{Securing Antarctica's Environmental Future, School of Earth, Atmosphere and Environment, Monash University, Clayton, Kulin Nations, Victoria, Australia 3800\printead[presep={,\ }]{e4}}

    \end{aug}

    \begin{abstract}
    Ice sheet models are routinely used to quantify and project an ice sheet's contribution to sea level rise. In order for an ice sheet model to generate realistic projections, its parameters must first be calibrated using observational data; this is challenging due to the nonlinearity of the model equations, the high dimensionality of the underlying parameters, and limited data availability for validation. This study leverages the emerging field of neural posterior inference for efficiently calibrating ice sheet model parameters and boundary conditions. We make use of a one-dimensional (flowline) Shelfy-Stream Approximation model in a state-space framework. A neural network is trained to infer the underlying parameters, namely the bedrock elevation and the basal friction coefficient along the flowline, based on observations of ice surface velocity and surface elevation. Samples from the approximate posterior distribution of the parameters are then used within an ensemble Kalman filter to infer latent model states, namely the ice thickness along the flowline. We show through a simulation study that our approach yields more accurate estimates of the parameters and states than a state-augmented ensemble Kalman filter, which is the current state-of-the-art. We apply our approach to infer the bed elevation and basal friction along a flowline in Thwaites Glacier, Antarctica.
\end{abstract}




\end{frontmatter}

\section{Introduction}
\label{sec:intro}
The Antarctic Ice Sheet (AIS) is the largest freshwater reservoir on Earth and a major contributor to sea level rise. 
The rate of ice mass loss from the AIS has steadily increased over the period 1992--2020, with mass loss expected to continue throughout this century~\citep{ipcc2021}. Ice mass loss will have significant impact on the biodiversity of Antarctica and its surrounding oceans as well as other coastal communities. 
Furthermore, Antarctica's contribution to global mean sea level rise is predicted to accelerate in the coming decades to centuries, with subsequent impacts on coastal flooding and erosion, salinisation of surface and ground waters, and degradation of coastal habitats such as wetlands \citep{ipcc2021}.
To prepare for, and adapt to, these consequences, accurate and reliable estimates of the evolution of the AIS and its contribution to the sea level budget are of paramount importance.

Ice sheet models are among the most reliable tools for simulating the AIS's future contribution to sea level rise. The Ice Sheet Model Intercomparison Project for CMIP6 (ISMIP6) Antarctica simulations~\citep{seroussi2024evolution} comprises the largest ensemble of ice sheet model simulations. ISMIP6 projections of 2100 sea level rise range from $0.07-0.12$ m under a low-emission scenario and from $0.20-0.53$ m under a high-emission scenario. The broad spread in the sea level rise projections between individual ice sheet models is largely attributable to differences in the ice flow models themselves: this is due to different numerics, physics, boundary and initial conditions, parameterisations, parameter values, and mesh resolution \citep{beckmann2025disentangling, seroussi2023insights}. Reducing uncertainties associated with these key factors is thus critically important for improved sea level projections.

A major source of uncertainty in ice sheet modelling lies in the conditions at the base of the ice sheet. Due to logistical challenges in obtaining measurements, observations of the bed topography underneath the ice sheet tend to be sparsely and unevenly sampled, and parameters in the friction component of the model are not directly observable \citep{mccormack2025synthetic}. While the latest bed topography data from BEDMAP3 has seen significantly increased coverage, there remain areas where observations are collected along flight lines that are separated across-track by kilometres to hundreds of kilometres~\citep{bedmap3}. Current approaches for inferring the bed topography and friction parameters often involve inversion methods applied to data of the ice surface, such as measurements of ice surface velocity and of surface elevation relative to sea level. These approaches can be broadly categorised into (i) deterministic approaches that provide a single estimate for the (spatially varying) parameters, often based on a best-fit criterion; and (ii) probabilistic approaches that provide parameter estimates with some accompanying measure of uncertainty. 

A widely applied class of deterministic approaches for making inference on unknown quantities in ice sheet models is the class of ``control methods''. The aim of these methods is to find a single estimate for the parameters that minimises a cost function quantifying the mismatch between the ice sheet model output and the observations. These methods can be either time-independent \cite[e.g.,][]{macayeal1992basal, macayeal1993, joughin2004basal, morlighem2010spatial, morlighem2013inversion} or time-dependent \cite[e.g.,][]{larour2014inferred, choi2023impact, badgeley2025increased}. The former methods aim to match modelled quantities to observations at a specific instance in time, while the latter take advantage of data that evolve over time. Despite being widely adopted among the glaciological community, control methods still suffer from a number of drawbacks. First, parameter estimates obtained using control methods tend to be non-unique, or they are sensitive to small changes in the observations and require regularisation with careful tuning~\citep[e.g.,][]{morlighem2010spatial, gillet2012greenland}. 
Second, control methods do not naturally take into account parameter uncertainties, which are important for future sea level rise projections.

Probabilistic approaches, such as the ensemble Kalman filter~\citep[EnKF,][]{evensen1994}, are better suited for uncertainty quantification. The EnKF combines information from a physical model with that from observations, taking into account measurement error and other potential sources of uncertainty such as model-reality discrepancy. The EnKF emerged as a sample-based approximation to the standard Kalman filter \citep{kalman1960new} that can also accommodate nonlinearity, making it suitable for high-dimensional problems involving partial differential equations in geophysical models. While it has been widely used for geophysical applications such as numerical weather forecasting, the EnKF has only recently been applied to problems in glaciology \citep[e.g.,][]{bonan2014etkf, gillet2020assimilation, choi2025estimation}. \cite{bonan2014etkf} uses an ensemble Kalman filter for jointly inferring the ice thickness, bed elevation, and basal friction coefficient in a one-dimensional Shallow-Ice Approximation \citep[SIA, ][]{hutter1983application}. 
\citet{gillet2020assimilation} later uses a variant of the EnKF called the error subspace ensemble transform Kalman filter~\citep[ESTKF,][]{nerger2012unification} to estimate the ice thickness, bed elevation, and basal friction coefficient in a simulation study involving a one-dimensional Shelfy-Stream Approximation~\citep[1D SSA,][]{macayeal1989large} model, which is more complex than the SIA model considered by \cite{bonan2014etkf} as it includes the transition between grounded and floating ice. Both studies use a ``state-augmentation'' approach~\citep{anderson2001ensemble}, in which the parameters (namely the spatially varying basal friction coefficient and bed elevation) are concatenated to, and updated along with, the state vector. They assume a persistence model~\citep[e.g.,][]{baek2006local, zupanski2006model} for these parameters, which in practice may lead to underestimation of the uncertainties around parameter estimates. The approach of \cite{gillet2020assimilation} has recently been extended to a 2D SSA model by \citet{choi2025estimation} in an observing system simulation experiment; however, to our knowledge, no application of the approach to real data has been made.

The EnKF can also be viewed as a Bayesian method, since it computes a probability distribution over the parameters and other quantities of interest given observed data \citep{wikle2007bayesian}. Here, we briefly review a number of studies that implement Bayesian inference in glaciology, and direct readers to \citet{gopalan2023review} for a more thorough discussion.
Earlier studies that follow a Bayesian framework \cite[e.g.,][]{berliner2008modeling, raymond2009estimating, pralong2011bayesian, brinkerhoff2016bayesian} often assume ``Gaussian-Gaussian'' models, in which the (prior) probability distribution over the parameters and the probability distribution of the data given the parameters (also known as the likelihood function) are both Gaussian, leading to some computational simplifications.
Gaussian-Gaussian models for larger (catchment-sized or continental) scale Bayesian inference were later developed by~\cite{petra2014stochastic}, \cite{isaac2015scalable}, and \cite{brinkerhoff2022variational}. Bayesian inference provides a rigorous framework for uncertainty quantification and may be made scalable using various approximating techniques; however, estimating the basal conditions from time evolving observations within a Bayesian framework is challenging because of the computations involved when using classical Bayesian machinery.
In this work, we propose a two-stage Bayesian inference approach for state-space models that can take advantage of time-evolving observations and yield estimates with superior accuracy and uncertainty quantification compared to the one-stage EnKF-based approach of~\cite{gillet2020assimilation}. In the first stage, we infer the posterior distribution of the unknown parameters, namely the bed elevation and basal friction coefficient in a 1D SSA model, using neural inference~\citep{zammit2025neural}. This stage involves training a convolutional neural network (CNN) to take observations of the surface elevation and velocity as input, and output an approximate (minimum-divergence) posterior distribution of the bed elevation and basal friction coefficient, thereby implicitly marginalising over the hidden states. Once inference has been made on the bed and friction fields, in the second stage we infer the ice geometry conditional on the estimated bed and friction fields. This is done by using samples of the bed and friction from the approximate posterior distribution as input to an EnKF to infer ice thickness.

\Copy{neural_inference}{Neural inference is a class of simulation-based inference methods~\citep{cranmer2020} that make use of neural networks to learn a complex mapping used for inference, such as the likelihood function~\citep[e.g.,][]{papamakarios2019sequential, walchessen2024neural}, the likelihood-to-evidence ratio~\citep[e.g.,][]{thomas2022likelihood, leonte2025simulation}, and the posterior distribution~\citep[e.g.,][]{papamakarios2016fast, lueckmann2017, chan2018likelihoodfree, radev2022bayesflow}; see~\cite{zammit2025neural} for a recent review. 
Neural inference is being increasingly used for inferring parameters associated with spatial processes~\citep[e.g.,][]{zammit2020deep, gerber2021fast, lenzi2021NeuralNF, richards2024neural}, spatio-temporal processes~\citep[e.g.,][]{dell2025flexible}, and point processes~\citep[e.g.,][]{lambe2025neural}. In this study, we focus on the use of neural networks to approximate the posterior distribution of basal parameters in a 1D SSA model, which we refer to as neural posterior inference (NPI).} 

There are other works in the glaciology literature that infer basal conditions using neural networks, specifically physics-informed neural networks (PINNs); see \cite{riel2021data} and \cite{cheng2024forward}. PINNs incorporate physics-based constraints into the loss function (such as constraints on the spatial smoothness of basal drag), thereby encouraging the neural network to produce estimates that satisfy these constraints.
Our approach differs from that of \cite{riel2021data} in several ways: first, we directly target a minimum-divergence approximation of the posterior distribution of both the basal friction coefficient and the bed elevation, whereas \cite{riel2021data} focus on estimating basal drag (a function of the basal friction coefficient). They do so by training a PINN to output estimates of the surface velocity and ice thickness for a given spatio-temporal point, from which estimates of the basal drag (and in turn the basal friction coefficient) are then computed using the basal sliding law in the ice sheet model. Further, we assume that the approximate posterior distribution on a suitable transformation of the ice sheet model parameters is multivariate Gaussian with a sparse precision matrix, which takes into account spatial correlations that are not considered by \cite{riel2021data}. 

The remaining sections of this article are structured as follows. Section~\ref{sec:modelling} provides details of the 1D SSA model and how it can be represented in a state-space modelling framework. Section~\ref{sec:ssa_inference} presents our proposed approach for estimating model parameters and states with neural posterior inference and the (non-augmented) EnKF. We compare the performance of our approach to that of the state-augmented EnKF in a simulation study in Section~\ref{sec:ssa_sim_study}. In Section~\ref{sec:real_data}, we apply our approach to infer the bed elevation and basal friction coefficient along a 1D flowline in Thwaites Glacier. A summary of the results and outline of directions for future research are given in Section~\ref{sec:nbe_conclusion}. Reproducible code for both the simulation study and the real data example are available at \url{https://github.com/bao-anh-vu/npi_icesheet}.

\section{Ice sheet model and its state-space representation}
\label{sec:modelling}
The Shelfy-Stream Approximation (SSA) model~\citep{macayeal1989large} is a simplification of Stokes' equations for ice flow under the assumption that the thickness of the ice is much smaller than its horizontal extent. This model also assumes that vertical shear stress is negligible and that ice flow is primarily driven by basal sliding, making it suitable for fast-flowing ice streams and floating ice shelves. The SSA model is depth-averaged: the ice velocity is taken to be uniform vertically. These simplifications reduce the computational cost of simulating ice dynamics by avoiding solving the full 3D stress tensor. 
In this study, we make use of a 1D SSA model, which we describe in more detail in Section~\ref{sec:ssa_model}. We then show in Section~\ref{sec:ssa_state_space} how this model can be expressed in a state-space framework in order to facilitate statistical calibration of its parameters.

\subsection{One-dimensional (1D) SSA model}
\label{sec:ssa_model}
\begin{figure}
    \centering
    \includegraphics[width=0.9\linewidth]{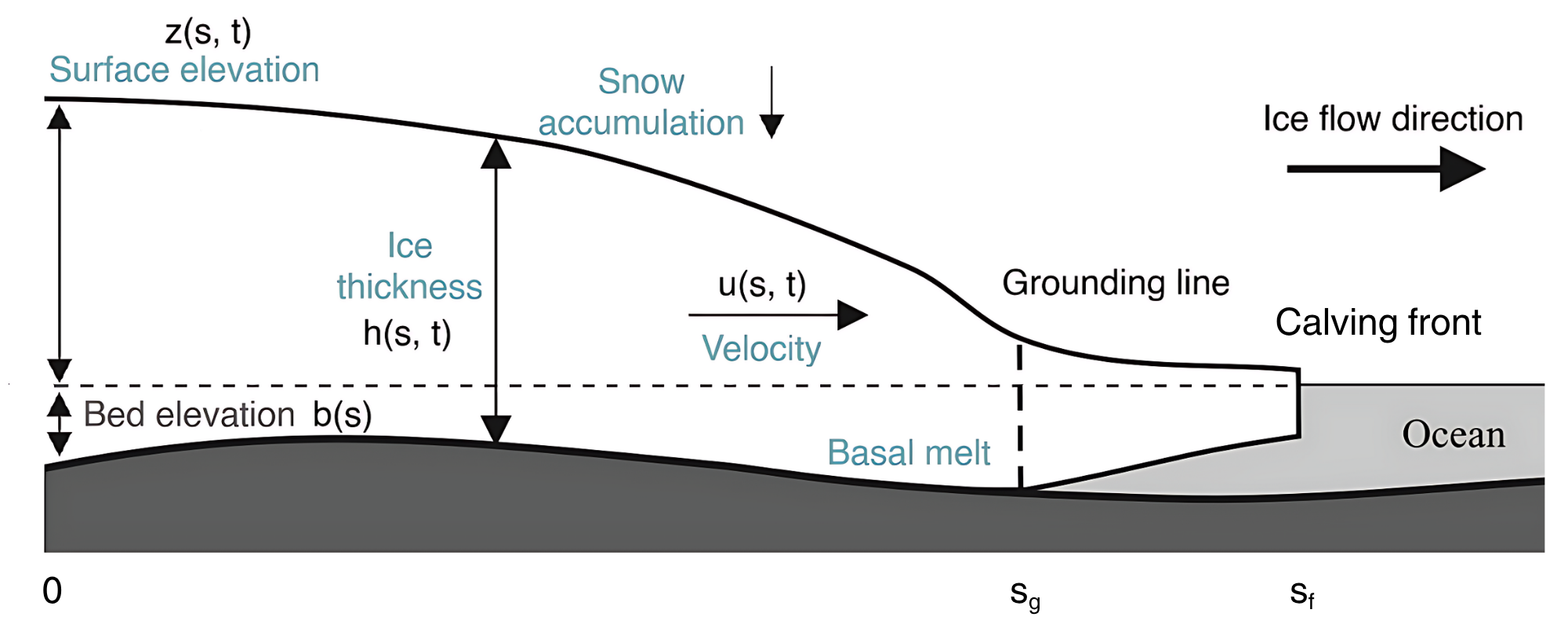}
    \caption{Schematic of a 1D ice sheet, adapted from~\cite{schoof2007marine}.}
    \label{fig:ice_sheet_schematic}
\end{figure}

Consider a glacier transect conceptualised in Figure~\ref{fig:ice_sheet_schematic}. Here, the spatial domain $\mathcal{S} = [0, s_F]$ begins at a drainage divide within the interior of the ice sheet and extends to the calving front $s_F$ (where the ice terminates in the ocean). We describe the ice dynamics on this domain using a 1D SSA model, which consists of two main partial differential equations (PDEs). The first PDE is a stress-balance equation:
\begin{equation}
    \label{eq:ssa_force_bal}
    \frac{\partial }{\partial s} \left( 2 B h(s, t) \left| \frac{\partial u(s, t)}{\partial s} \right|^{(1-n_c)/n_c}  \frac{\partial u(s, t)}{\partial s} \right) - \tau_{b}(s, t) = \rho_i g h(s, t) \frac{\partial z(s, t)}{\partial s},
\end{equation}
for $s \in (0, s_F), t \geq 0$, where, as functions of space and time respectively, 
$h(\cdot, \cdot)$ is the ice thickness, $u(\cdot, \cdot)$ is the depth-averaged horizontal velocity (hereafter referred to as `surface velocity'), and $z(\cdot, \cdot)$ is the surface elevation. The parameters $B$, $\rho_i$, and $g$ are scalars that represent the ice stiffness (a parameter in the material constitutive relation that determines the ice viscosity), ice density, and the gravitational acceleration, respectively. We specify the values we use for these parameters in Table S1 in the Supplementary Material. The parameter $n_c$ is the creep exponent in Glen's flow law~\cite[][Chapter 5]{PATERSON199478}, usually set to $n_c = 3$. The leftmost term in~\eqref{eq:ssa_force_bal} represents the longitudinal stress (caused by the ice stretching and compressing), and the term on the right-hand side represents the driving stress. The basal shear stress $\tau_b(\cdot, \cdot)$ in~\eqref{eq:ssa_force_bal} can be described by a nonlinear Weertman friction law~\citep{weertman1957sliding} for grounded ice:
\begin{equation}
    \tau_b (s, t) = c(s) |u(s, t)|^{m-1} u(s, t), \quad s \in (0, s_g), \quad t \geq 0, \label{eq:weertman_friction}
\end{equation}
where $s_g \leq s_F$ denotes the position of the grounding line (the point at which ice detaches from the underlying bed and begins to float over the ocean), and $c(\cdot)$ and $m$ are the basal friction coefficient and exponent, respectively. The basal shear stress is assumed to be zero for floating ice. The exponent $m$ is often related to the creep exponent $n_c$ by the relation $m = 1/n_c$ and is therefore usually set to $m = 1/3$~\citep{weertman1974stability, schoof2007marine, gudmundsson2012stability, pattyn2012results}. Thus, \eqref{eq:ssa_force_bal} describes a balance between the longitudinal stress, basal shear stress and driving stress.

The second PDE is a mass continuity equation describing the evolution of the ice thickness $h(s, \cdot)$, for $s \in (0, s_F)$, in time:
\begin{equation}
    \label{eq:mass_cty}
    \frac{\partial h (s, t)}{\partial t} + \frac{\partial (u(s, t) h(s, t))}{\partial s} = m_a, \quad t \geq 0,
\end{equation}
where $m_a = a_s - a_b$, with $a_s$ being the surface mass balance (mostly driven by precipitation) and $a_b$ the basal melt rate, all assumed here to be constant in space and time. 

The surface elevation $z(\cdot, \cdot)$ is defined as
\begin{equation}
    \label{eq:ssa_top_surface}
    z(s, t) =
    \begin{cases}
        b(s) + h(s, t),                                  & h(s, t) \geq (z_0 - b(s)) \frac{\rho_w}{\rho_i}, \quad \text{(grounded ice)} \\ 
        \left(1 - \frac{\rho_i}{\rho_w} \right) h(s, t), & h(s, t) < (z_0 - b(s)) \frac{\rho_w}{\rho_i}, \quad \text{(floating ice)}    
    \end{cases}
\end{equation}
for $s \in (0, s_F)$ and $t \geq 0$, where $b(\cdot)$ is the bed elevation (assumed constant in time), $z_0$ is a constant reference sea level (set to zero for this study), and $\rho_w$ is the density of water. The inequalities on the right-hand side of~\eqref{eq:ssa_top_surface} constitute the flotation condition, which determines the position of the grounding line $s_g$. As the ice geometry evolves in time, the grounding line position also evolves accordingly. At the grounding line, we assume the horizontal velocity $u(\cdot, \cdot)$ and ice thickness $h(\cdot, \cdot)$ are continuous.

\subsection{SSA model as a state-space model}
\label{sec:ssa_state_space}

The SSA model of Section~\ref{sec:ssa_model} is deterministic. However, in reality, there are many sources of uncertainty, including inherent randomness in the physical system, mismatch between model output and observations, and potential inaccuracies in the observed data due to measurement error. To account for some of these sources of uncertainty, we treat the SSA model in a statistical context and re-express it as a state-space model \citep[SSM; e.g.,][]{durbin2012time}. SSMs are useful for modelling a temporally-evolving system in which the latent model states are indirectly observed, either through noisy observations or through other quantities that are related to the latent states via a mathematical relation. 
An SSM contains two sub-models: a process model that describes the (temporal) evolution of the latent model states, and a data model that describes the relationship between the latent states and the observations. To express the SSA ice sheet model in a state-space framework, we treat the time-evolving ice thickness $h(\cdot, \cdot)$ as latent states, and assume that observations of the ice surface elevation $z(\cdot, \cdot)$ and horizontal velocity $u(\cdot, \cdot)$ are available. The dynamics of the model depend on the bed elevation $b(\cdot)$ and basal friction coefficient $c(\cdot)$, which we treat as static parameters (that are functions of space) to be estimated.

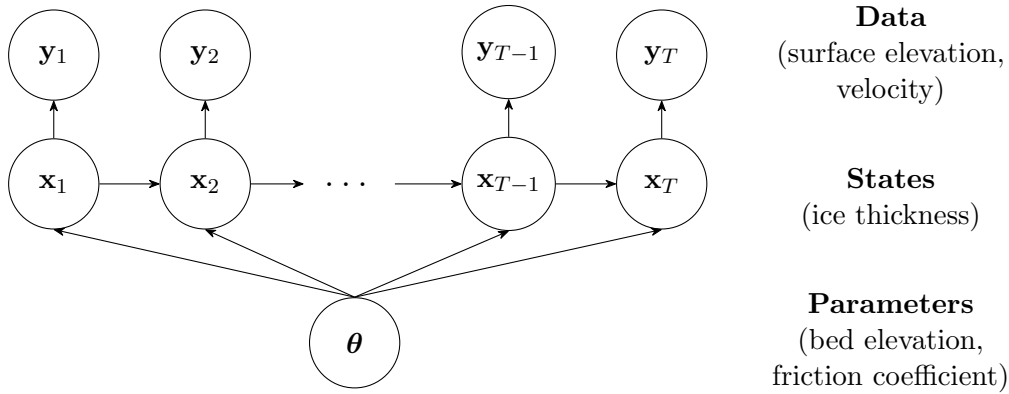
\begin{figure}
    \centering
    \begin{tikzpicture}[
    >={Stealth[round]}, 
    every node/.style={draw, minimum size=1.2cm, font=\small}, 
    node distance=0.5cm and 0.8cm, 
    ]

    \node (theta) [circle] at (3,1) { $\btheta$ };
    \node (thetaLabel) [right=4.8cm of theta, draw=none, font=\small, align=center] {\textbf{Parameters} \\ (bed elevation, \\ basal friction coefficient)};

    \node (x1) [circle, above=of theta, xshift=-4cm, yshift=0.4cm] { $\x_1$ };
    \node (x2) [circle, right=of x1] { $\x_2$ };
    \node (xT-1) [circle, right=of x2, xshift=2cm] { $\x_{T-1}$ };
    \node (xT) [circle, right=of xT-1] { $\x_T$ };
    \node (xLabel) [above = of thetaLabel, draw=none, font=\small, align=center] {\textbf{States} \\ (ice thickness)};

    \node (dots) [above=0.05cm of theta, right=0.7cm of x2, draw=none, font=\large] { $\dots$ };

    \node (y1) [circle, above=of x1] { $\y_1$ };
    \node (y2) [circle, above=of x2] { $\y_2$ };
    \node (yT-1) [circle, above=of xT-1] { $\y_{T-1}$ };
    \node (yT) [circle, above=of xT] { $\y_T$ };
    \node (yLabel) [above = of xLabel, draw=none, font=\small, align=center] {\textbf{Data} \\ (surface elevation, \\ surface velocity)};

    \draw[->] (theta.north) -- (x1.south);
    \draw[->] (theta.north) -- (x2.south);
    \draw[->] (theta.north) -- (xT-1.south);
    \draw[->] (theta.north) -- (xT.south);

    \draw[->] (x1) -- (x2);
    \draw[->] (x2) -- (dots);
    \draw[->] (dots) -- (xT-1);
    \draw[->] (xT-1) -- (xT);

    \draw[->] (x1) -- (y1);
    \draw[->] (x2) -- (y2);
    \draw[->] (xT-1) -- (yT-1);
    \draw[->] (xT) -- (yT);

\end{tikzpicture}
    \caption{The three layers of the state-space model described in Section~\ref{sec:ssa_state_space}. The latent states $\{\x_t\}$ evolve in time according to the process model, which depends on static parameters $\btheta$. Each $\y_t$ is conditionally independent of data at other time points when conditioned on the corresponding latent state $\x_t$.}
    \label{fig:SSM_diagram}
\end{figure}

We discretise the 1D spatial domain $\mathcal{S}$ into a grid with $J + 1$ equally-spaced nodes, denoted by $s_0, \dots, s_J$. We use finite differences to discretise the partial differential equations~\eqref{eq:ssa_force_bal} and~\eqref{eq:mass_cty} in time (see Section S1 of the Supplementary Material for details), and we consider the evolution of the ice sheet only at discrete time points $t \in \mathcal{T} \equiv \{1, \dots, T\}$, where each point corresponds to a simulation year. Then, the latent state (ice thickness) at time $t$, $h(\cdot, t)$, can be represented as a vector ${\h_t \equiv (h_{0, t}, h_{1, t}, \dots, h_{J, t})^\top}$, where $h_{j, t} \equiv h(s_j, t)$ is the ice thickness at spatial node $s_j$ at time $t$, for $t \in \mathcal{T}$. We similarly denote the surface velocity as $\u_t \equiv (u_{0, t}, u_{1, t}, \dots, u_{J, t})^\top$, where $u_{j, t} \equiv u(s_j, t)$, for $t \in \mathcal{T}$.
The SSA model can then be written in state-space form as
\begin{align}
    \text{Data model:} \quad \y_t    & = \mathcal{H}_t (\x_t, \btheta) + \w_t,  \quad \w_t \sim \Gau(\0, \R_t), \quad t \in \mathcal{T}, \label{eq:ssm_obs}                       \\
    \text{Process model:} \quad \x_t & = \mathcal{M}_t(\x_{t-1}, \btheta) + \v_t, \quad \v_t \sim \Gau(\0, \V_t), \quad t \in \mathcal{T}\backslash\{1\},  \label{eq:ssm_process}
\end{align}
where $\y_t \equiv (\z_t^\top, \u_t^\top)^\top$ is a vector containing (noisy) observations of the surface elevation and velocity at time $t$,
$\x_t \equiv \h_t$ is the vector of latent states (ice thickness) at time $t$,
and $\btheta \equiv (\btheta_b^\top, \btheta_c^\top)^\top$ is a vector of parameters used to construct the bed elevation $b(\cdot)$ (with $\btheta_b$) and basal friction coefficient $c(\cdot)$ (with $\btheta_c$); see Section~\ref{sec:sim_param} and Section S5 of the Supplementary Material for details. Here, $\mathcal{H}_t(\cdot)$ is an ``observation operator'' that maps the latent states to the observations at time $t$ (based on equations~\eqref{eq:ssa_force_bal} and~\eqref{eq:ssa_top_surface}), and $\mathcal{M}_t(\cdot)$ maps the latent states at time $t-1$ to those at time $t$
(by numerically solving the coupled system of PDEs~\eqref{eq:ssa_force_bal} and \eqref{eq:mass_cty} at a fine temporal resolution; see Sections S1 and S2 of the Supplementary Material). The initial ice thickness $\x_1$ is modelled as $\x_1 \sim \Gau(\m_1, \K_1)$, where $\m_1$ and $\K_1$ are the prior mean and covariance matrix, respectively. The vectors $\w_t$ and $\v_t$ are Gaussian noise terms that represent observation error and process error, respectively, with assumed known covariance matrices $\R_t$ and $\V_t$; we give details on how we construct these matrices in the simulation study in Section~\ref{sec:ssa_sim_study}. The observation error accounts for inaccuracies in the observed data due to measurement error, while the process error accounts for uncertainties in the model itself (e.g., due to simplifications in the SSA model). We assume that $\{\w_t\}$ and $\{\v_t\}$ are mutually independent and independent over time. A diagram depicting the three layers of the SSM is shown in Figure~\ref{fig:SSM_diagram}.

\section{Neural posterior inference and the ensemble Kalman filter}
\label{sec:ssa_inference}

In this study, we are primarily interested in estimating two spatially varying quantities (parameters) in the SSA model: the bed elevation and the basal friction coefficient. Since observations of the bed elevation are sparse, and one cannot directly observe the basal friction coefficient, we must inversely determine these quantities from more readily available data measured at the ice surface, such as observations of the ice velocity and surface elevation relative to sea level. Since we have some prior information on these unknown parameters, we pose the inversion problem within a Bayesian framework. 


Let $\Y \equiv (\y_1^\top, \dots, \y_T^\top)^\top$ denote the observations of the ice surface (namely the horizontal velocity and the surface elevation), and $\X \equiv (\x_1^\top, \dots, \x_T^\top)^\top$ denote the (unobserved) ice thickness for times $1$ to $T$, where each time point corresponds to a year. Our goal is to obtain the posterior distribution of the parameters, 
$p(\btheta \mid \Y) \propto p(\Y \mid \btheta) p(\btheta)$,
where $p(\Y \mid \btheta)$ is the likelihood function
and $p(\btheta)$ is the prior distribution of the parameters.
Evaluation of the likelihood function $p(\Y \mid \btheta)$ requires integrating out the high-dimensional latent states:
\begin{equation}
    p(\Y \mid \btheta) = \int p(\Y \mid \X, \btheta) p(\X \mid \btheta) \d \X, \label{eq:marginalisation}
\end{equation}
which is intractable in this case due to the nonlinearity of the SSA model. 

While the likelihood function could be estimated with a particle filter~\citep{gordon1993novel} and used within a pseudo-marginal Metropolis-Hastings \citep[PMMH,][]{andrieu2009pseudo} or a variational Bayes for intractable likelihood (VBIL) algorithm \citep{tran2017variational}, such a task is challenging in our case due to the high dimensionality of the latent states. We found that a particle filter suffers from particle degeneracy when applied to our model, leading to very high variance in the estimated likelihood, which in turn makes it difficult for the PMMH algorithm to efficiently explore the posterior distribution~\citep{deligiannidis2018correlated}. In addition, both PMMH and VBIL required many repeated evaluations of the likelihood via a particle filter, which are computationally intensive and cannot be parallelised. We have also tried a state-augmented approach, in which the states and parameters are concatenated and jointly updated using a localised particle filter~\citep{poterjoy2016localpf}, which is better optimised for models with high-dimensional states, but the issue with particle degeneracy persisted.

Recently, neural posterior inference (NPI) has emerged as a new computational statistical tool for making inference with models with intractable likelihood functions, which can circumvent several of the limitations encountered with the methods discussed above \citep[see][for a review]{zammit2025neural}. NPI only requires simulations from the likelihood (the data generating process), and it has been shown to be very effective at constructing posterior distributions in complex settings such as the problem at hand. In this work, we propose a likelihood-free approach based on NPI to obtain a minimum-divergence approximation to the posterior distribution. Our approach involves two stages:


\begin{enumerate}
    \item (Parameter inference) Train a convolutional neural network (CNN) to find an approximate posterior distribution of the parameters (bed elevation and basal friction coefficient), $p(\btheta \mid \Y)$.
    \item (State inference) Draw $L$ samples of the parameters from the approximate posterior distribution, and use each of the draws in an EnKF algorithm to make inference on the latent states (ice thickness) $\X$. 
\end{enumerate}
We give details on each of the two stages in the following sections.

\subsection{Parameter inference with neural posterior approximation}
\label{sec:stageone_cnn}

In the first stage of our approach, we seek an approximate posterior distribution $q(\btheta; \blambda)$ that is as close as possible, according to some chosen divergence metric, to the true posterior distribution, ${p(\btheta \mid \Y)}$. The form of the approximate posterior distribution $q(\btheta; \blambda)$ is arbitrary~\citep[e.g.,][]{papamakarios2016fast, radev2022bayesflow, maceda2024variational}; in this study, we define it to be a multivariate Gaussian distribution. 
In NPI, we seek
a mapping $\blambda^*(\cdot)$ from the data to the parameters of the approximate posterior distribution such that
\begin{equation}
    \blambda^*(\cdot) = \argmin_{\blambda(\cdot)} \Exp_{p(\Y)} [ \KL{p(\btheta \mid \Y)}{q(\btheta; \blambda(\Y)} ], \label{eq:argmin_kl}
\end{equation}
where $\KL{\cdot}{\cdot}$ is the (forward) Kullback--Leibler divergence~\citep{kullback1951information} given by
\begin{equation}
    \KL{p(\btheta \mid \Y)}{q(\btheta; \blambda)} = \int_{\Theta} p(\btheta \mid \Y) \log \frac{p(\btheta \mid \Y)}{q(\btheta; \blambda)} \d \btheta,
\end{equation}
and where $\Theta$ denotes the parameter space. 

We model $\blambda(\cdot)$ using a neural network $\blambda_{\bgamma}(\cdot)$, where the subscript $\bgamma$ denotes the weights and biases of the network. The problem of estimating the posterior distribution $p(\btheta \mid \Y)$ then becomes the optimisation problem of finding the weights and biases $\bgamma^*$ such that
\begin{equation}
    \bgamma^* = \argmin_{\bgamma} \Exp_{p(\Y)} [ \KL{p(\btheta \mid \Y)}{q(\btheta; \blambda_{\bgamma}(\Y))} ].
\end{equation}
This is equivalent to solving
\begin{equation}
    \bgamma^* = \argmin_{\bgamma} \int_{\Theta} \int_{\mathcal{Y}}  - \log q (\btheta; \blambda_{\bgamma}(\Y)) {p(\Y \mid \btheta)} {p(\btheta)} \d \Y \d \btheta, \label{eq:KL_int}
\end{equation}
where $\mathcal{Y}$ denotes the sample space.
We approximate~\eqref{eq:KL_int} with a Monte Carlo approximation using $N$ sets of simulated parameters and corresponding datasets, $\{ \{\btheta^{(n)}, \Y^{(n)}\}: n = 1, \dots, N\}$. Specifically, we simulate $N$ realisations of the bed elevation and basal friction coefficient and, for each realisation, we run a forward simulation of an ice sheet and use the surface elevation and velocity from the forward run as observations. The optimisation problem then becomes
\begin{equation}
    \label{eq:loss_fun}
    \bgamma^* = \argmin_{\bgamma} \frac{1}{N} \sum_{n = 1}^N - \log q \left(\btheta^{(n)}; \blambda_{\bgamma} (\Y^{(n)}) \right),
\end{equation}
where $\btheta^{(n)} \sim p(\btheta)$ and $\Y^{(n)} \sim p(\Y \mid \btheta^{(n)})$. The right hand side of~\eqref{eq:loss_fun} is the loss function that needs to be minimised when training the neural network.

For this study, we assume a Gaussian approximate posterior distribution of $\btheta$ of the form
\begin{equation}
    q(\btheta; \blambda_{\bgamma}(\Y)) \equiv \Gau \left(\bmu_{\bgamma} (\Y), \Q_{\bgamma}^{-1} (\Y) \right),
\end{equation}
where the mean $\bmu_{\bgamma} (\cdot)$ and precision matrix $\Q_{\bgamma}(\cdot)$ are functions of the data $\Y \in \mathcal{Y}$.
We parameterise the precision matrix in terms of its lower Cholesky factor, $\Q_{\bgamma}(\Y) = \L_{\bgamma}(\Y) \L_{\bgamma}(\Y)^\top$, and define
$\blambda_{\gamma}(\cdot) \equiv (\bmu_{\gamma}(\cdot)^\top, \vech(\L_{\gamma}(\cdot))^\top)^\top$, where $\vech(\cdot)$ results in lower-triangular vectorisation. 

In the simulation study and the real data example in Sections~\ref{sec:ssa_sim_study} and~\ref{sec:real_data}, \Copy{sparsity}{we use basis functions with compact support for parameter dimensionality reduction (see Section S5 of the Supplementary Material for more details). We further assume a sparse structure for the lower Cholesky factor, where we model $\L_{\bgamma} (\cdot)$ as a block diagonal matrix of the form $\L_{\bgamma} (\cdot) = \textrm{diag}(\L_{\bgamma, b}(\cdot), \L_{\bgamma, c}(\cdot))$, where
\begin{equation*}
    \L_{\bgamma, b}(\cdot) = \begin{bmatrix}
        \exp(l_{11, b} (\cdot)) &                         &                          \\
        l_{21, b} (\cdot)       & \exp(l_{22, b} (\cdot)) &                          \\
        \ddots                  & \ddots                                             \\
                                & l_{p(p-1), b} (\cdot)   & \exp(l_{p p, b} (\cdot))
    \end{bmatrix},
\end{equation*}
and $p$ is the number of basis functions used for constructing the bed elevation. We use the same structure for $\L_{\bgamma, c} (\cdot)$, the lower Cholesky factor for the basal friction coefficient.} \Copy{compact_supp}{Since the basis functions have local scope, their coefficients are only loosely correlated a posteriori, and therefore their joint posterior distribution can be effectively approximated using a sparse covariance structure.} In a toy example in Section S8 of the Supplementary Materials, we further show that NPI with a Gaussian approximate posterior with sparse covariance structure achieves similar inference accuracy to NPI with a more flexible approximate posterior based on normalising flows, but the former leads to a neural network that is much less computationally expensive to train. 

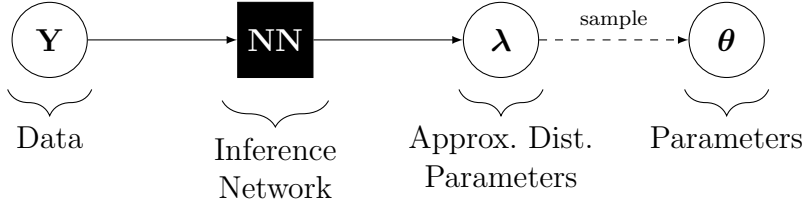
\begin{figure}
    \centering
        \begin{tikzpicture}[>=latex]
            \centering

            \node[circle, draw, minimum size=1cm, inner sep=0pt] (Z) at (2,0) {$\Y$};
            \node[rectangle, draw, minimum size=1cm, inner sep=0pt, fill=black, text=white] (NN) at (5,0) {\textbf{NN}};
            \node[circle, draw, minimum size=1cm, inner sep=0pt] (kappab) at (8,0)
            {$\blambda$};
            \node[circle, draw, minimum size=1cm, inner sep=0pt] (theta) at (11,0)
            {$\btheta$};

            \draw[->] (Z) to (NN);
            \draw[->] (NN) to (kappab);
            \draw[dashed, ->] (kappab) to node[above] {\scriptsize
                sample} (theta);

            \draw[decorate,decoration={brace,amplitude=10pt,mirror}]
            ([shift={(-0.2,-0.3)}]Z.south west) -- ([shift={(0.2,-0.3)}]Z.south east)
            node[midway,below=10pt] {Data};

            \draw[decorate,decoration={brace,amplitude=10pt,mirror}]
            ([shift={(-0.2,-0.3)}]NN.south west) -- ([shift={(0.2,-0.3)}]NN.south east)
            node[midway,below=10pt,align=center] {Inference \\ Network};

            \draw[decorate,decoration={brace,amplitude=10pt,mirror}]
            ([shift={(-0.2,-0.3)}]kappab.south west) -- ([shift={(0.2,-0.3)}]kappab.south east)
            node[midway,below=10pt, align=center] {Approx.~Dist. \\ Parameters};

            \draw[decorate,decoration={brace,amplitude=10pt,mirror}]
            ([shift={(-0.2,-0.3)}]theta.south west) -- ([shift={(0.2,-0.3)}]theta.south east)
            node[midway,below=10pt, align=center] {Parameters};


        \end{tikzpicture}

    \caption{Neural posterior inference framework for inferring ice sheet simulator parameters $\btheta$ from surface data $\Y$.}
    \label{fig:inference_neural_bayes}
\end{figure}

Once the neural network is trained, when given a new dataset it can rapidly output the corresponding mean and precision matrix of the posterior distribution. This posterior distribution allows us to generate samples of the parameters (namely the bed elevation and basal friction coefficient fields), and also construct credible intervals for these parameters. We summarise our NPI framework in Figure~\ref{fig:inference_neural_bayes}.

\subsection{State inference with the EnKF}
\label{sec:stagetwo_enkf}
Once the parameter inference stage is complete, we can use the approximate posterior distribution of the parameters for the next stage: inference on the latent states (the ice thickness). To infer the time-varying ice thickness ${\x_t \equiv (x_{0,t}, \dots, x_{J,t})^\top}$ for $t \in \mathcal{T}$, we use an EnKF algorithm~\citep{evensen1994}, which approximates the distribution of $\x_t$ given all observations up to time $t$, denoted by $\y_{1:t} \equiv (\y_1^\top, \dots, \y_t^\top)^\top$. This algorithm is widely used in geophysical applications~\citep[e.g.,][]{evensen1997advanced, allen2003ensemble}, and it has previously been applied to an SSA model by~\cite{gillet2020assimilation}. However, our work differs from that of \cite{gillet2020assimilation} in that they use the EnKF to jointly infer the parameters and latent states (which we refer to as a \textit{state-augmented} approach), whereas we use the EnKF to infer the states only, with the parameters inferred in a separate first stage using neural inference as described in Section~\ref{sec:stageone_cnn}. While our approach initially appears more complicated, we later show in Section~\ref{sec:sim_study_results} that by targeting the (marginal) posterior of $\btheta$ directly, we obtain more accurate parameter inference and better uncertainty quantification than the augmented EnKF approach. 


We define the \textit{posterior filtering distribution} of the states at time $t$  as
\begin{align}
    p^o(\x_t) & \equiv \Exp_{p(\btheta \mid \Y)} \left( p(\x_t \mid \btheta, \y_{1:t}) \right) \nonumber         \\
              & = \int p(\x_t \mid \btheta, \y_{1:t}) \, p(\btheta \mid \Y) \d \btheta, \quad t \in \mathcal{T}.
    \label{eq:post_filter_dist}
\end{align}
This posterior filtering distribution can be interpreted as the filtering distribution of the latent states (ice thickness), averaged over the posterior distribution of the parameters (bed elevation and basal friction coefficient). We approximate the posterior filtering distribution by first replacing the true posterior distribution of the parameters in~\eqref{eq:post_filter_dist} with the approximate posterior distribution obtained from the first stage of inference, namely $q(\btheta; \blambda_{\bgamma^*}(\Y))$:
\begin{align}
    \tilde{p}^o(\x_t) & \equiv \int p(\x_t \mid \btheta, \y_{1:t}) \, q(\btheta; \blambda_{\bgamma^*}(\Y)) \d \btheta, \quad t \in \mathcal{T}.
    \label{eq:post_filter_approx}
\end{align}
and then approximating $\tilde{p}^o(\x_t)$ using Monte Carlo samples,
\begin{equation}
    \tilde{p}^o(\x_t) \approx \frac{1}{L} \sum_{l=1}^L p(\x_t \mid \tilde{\btheta}^{(l)}, \y_{1:t}), \quad t \in \mathcal{T},
    \label{eq:post_filter_approx_mc}
\end{equation}
where $\tilde{\btheta}^{(l)} \sim q(\btheta; \blambda_{\bgamma^*}(\Y)), \, l = 1, \dots, L$. (Here, we use the tilde notation to distinguish the parameters sampled from the neural network output, namely $\{\tilde{\btheta}^{(l)}: l = 1, \dots, L\}$, from the simulated parameters used for training the neural network, namely ${\{\btheta}^{(n)}: n = 1, \dots, N\}$).
In practice, each summand in \eqref{eq:post_filter_approx_mc} is obtained by plugging each parameter sample $\tilde{\btheta}^{(l)}$ into a standard EnKF algorithm. This yields $L$ ensembles of the states, each with $N_e$ ensemble members say, which are then concatenated to form one large ensemble with $L N_e$ members that approximates the posterior filtering distribution $p^o(\x_t)$ given by~\eqref{eq:post_filter_dist}. We give details of the EnKF algorithm that we implement in Section S3 of the Supplementary Material. There, we also give details of the state-augmented EnKF approach of \cite{gillet2020assimilation}.

\section{Observing system simulation experiment}
\label{sec:ssa_sim_study}
We demonstrate our approach through an observing system simulation experiment (OSSE) with synthetic data generated from the 1D SSA model. We first train a neural network to learn the response of an ice sheet's geometry and dynamics to different parameter settings (bed elevation and basal friction coefficient). We then assess whether the trained neural network can accurately recover the bed elevation and basal friction coefficient based on observations of the ice sheet's surface elevation and velocity. 

To generate synthetic data, we first perform a spin-up procedure to generate an ice sheet in steady-state; we describe this procedure in Section S4 of the Supplementary Material. The steady-state ice sheet serves as a ``starting point'' from which we simulate training data for the neural network. The procedure for simulating training data is described in Section~\ref{sec:sim_train_data}. We apply our approach for parameter and state inference in Sections~\ref{sec:sim_study_stageone} and~\ref{sec:sim_study_stagetwo}, respectively, and compare the results from our approach against the state-augmented EnKF of~\cite{gillet2020assimilation} in Section~\ref{sec:sim_study_results}. 

\subsection{Simulated data}
\label{sec:sim_train_data}

\begin{figure}[t]
    \centering
    \includegraphics[width=0.6\linewidth]{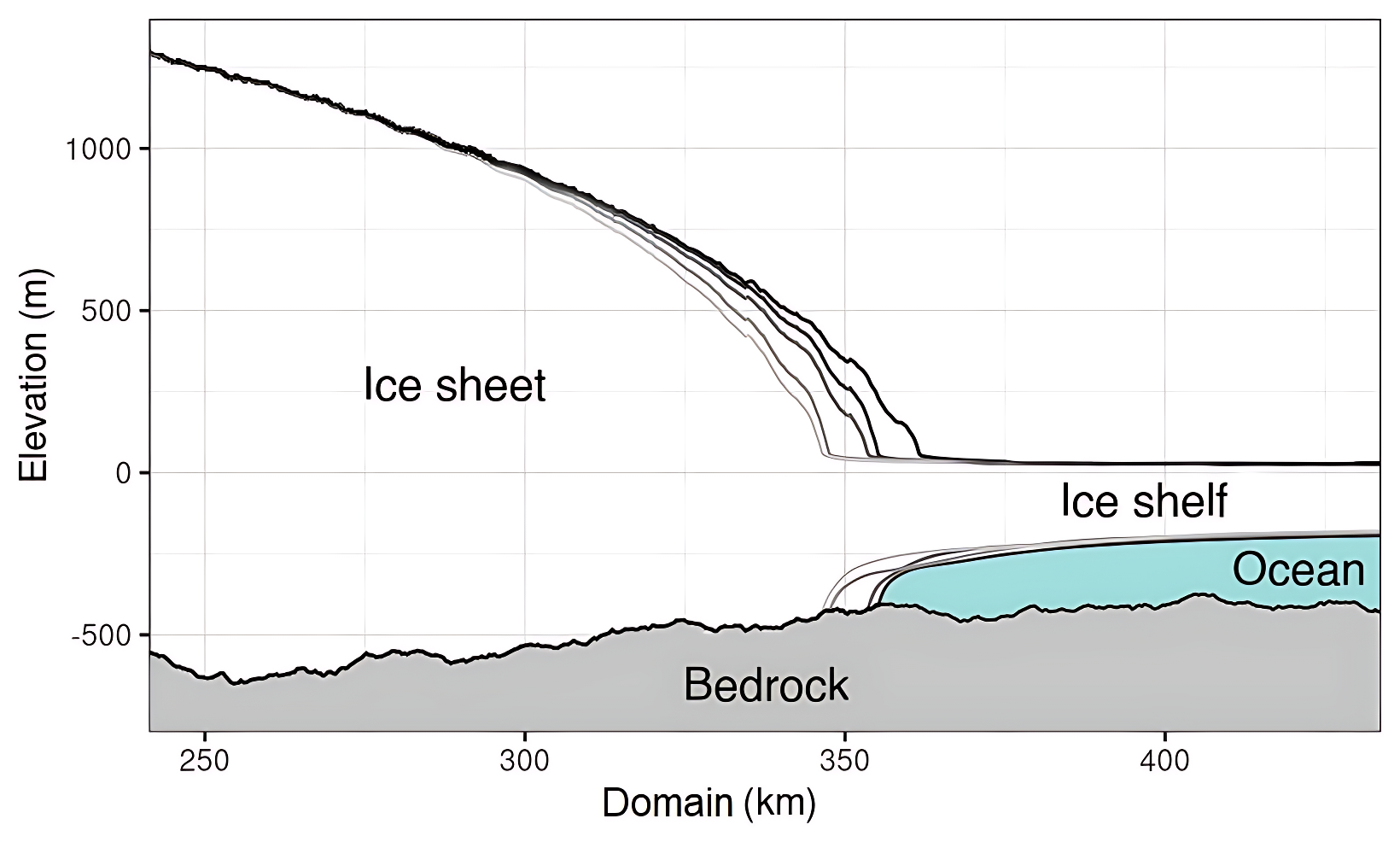}
    \caption{An evolving ice sheet with lighter colours corresponding to increasing time (one ice sheet profile per simulation year). The steady-state profile (at $t = 1$) shown in black is the initial profile from which the ice stiffness is reduced (from $B = 0.4 \stiffnessunit$ to $B = 0.3 \stiffnessunit$) to induce ice retreat (dynamics) over time.}
    \label{fig:evolving_ice_sheet}
\end{figure}

For this simulation study, we consider a flowline $D$ that starts from the ice divide $s = 0$ to a fixed calving front at $s_F = \num{800000}$~m. We discretise this domain $D$ into equal segments of length $400$~m, with the first grid point set to $s_0 = 0$~m, and the final grid point set to the ice calving front, $s_J = s_F = \num{800000}$~m, where $J = 2000$. We simulate a steady-state ice sheet on this flowline following the procedure in Section S4 of the Supplementary Material. An example of a steady-state ice sheet on a subset of $D$ is shown as a black line in Figure~\ref{fig:evolving_ice_sheet}.

Using the steady-state ice sheet geometry as the initial condition, we then simulate different ways in which the ice sheet can evolve over a 20-year period by varying the bed elevation and basal friction coefficient. In particular, for each given profile of bed elevation and basal friction coefficient, we run the SSA model forward for 20 years using a 1-year timestep, with the steady-state profile as the initial time point ($t = 1$), and with a slightly reduced ice stiffness to induce dynamics over time.
The pairs of parameters (bed elevation and basal friction coefficient) and observations (surface elevation and surface velocity) from these simulations are then used for training, validating, and testing the neural network. We describe the simulation procedure in more detail in the next section. 



\subsubsection{Simulating parameters}
\label{sec:sim_param}
Following \citet{gillet2020assimilation}, we unconditionally simulate $\num{50000}$ realisations of the basal friction coefficient from a Gaussian process with a constant mean of $\mu_c = 0.02 \fricunit$ and a squared exponential covariance function $K_c(\cdot)$ given by
\begin{equation}
    K_c(d) = \varsigma_c^2 \exp \left(-3 \left(\frac{d}{\ell_c}\right)^2 \right), \quad d = \abs{s - r}, \quad s, r \in D,
    \label{eq:fric_cov}
\end{equation}
where the range $\ell_c = \num{2500}$~m, and the variance $\varsigma_c^2 = 8.10^{-5} \fricunitsq$. 

To simulate bed elevation profiles, we first generate 50 synthetic observations of the bed elevation along the flowline that would mimic actual radar measurements and then use conditional simulation to generate bed samples. Specifically, we generate the observations by randomly selecting 50 locations along the domain and adding uncorrelated Gaussian measurement error with a standard deviation of $\sigma_b = 20$~m to the bed elevation used in the simulation of the steady-state ice sheet. Given these 50 observations, we then conditionally simulate bed profiles using Gaussian process regression~\citep{rasmussen2006gaussian} with a prescribed mean and covariance function.
The mean function is constructed by fitting a smooth curve to the bed elevation observations using local polynomial regression. To generate samples, we use the covariance function
\begin{equation}
    K_b(d) =
    \begin{cases}
        \varsigma_b^2 + \nu_b^2,                              & d = 0, \\
        \varsigma_b^2 \exp \left( \frac{-3d}{\ell_b} \right), & d > 0,
    \end{cases}
    \label{eq:bed_cov}
\end{equation}
where $\varsigma_b^2 = 4000$~m$^2$, $\ell_b = \num{50000}$~m, and $\nu_b^2 = 200$~m$^2$ is a nugget term that represents the bed elevation's fine-scale variation.

\subsubsection*{Dimensionality reduction using basis functions}
\begin{figure}
    \centering
    \includegraphics[width=0.8\linewidth]{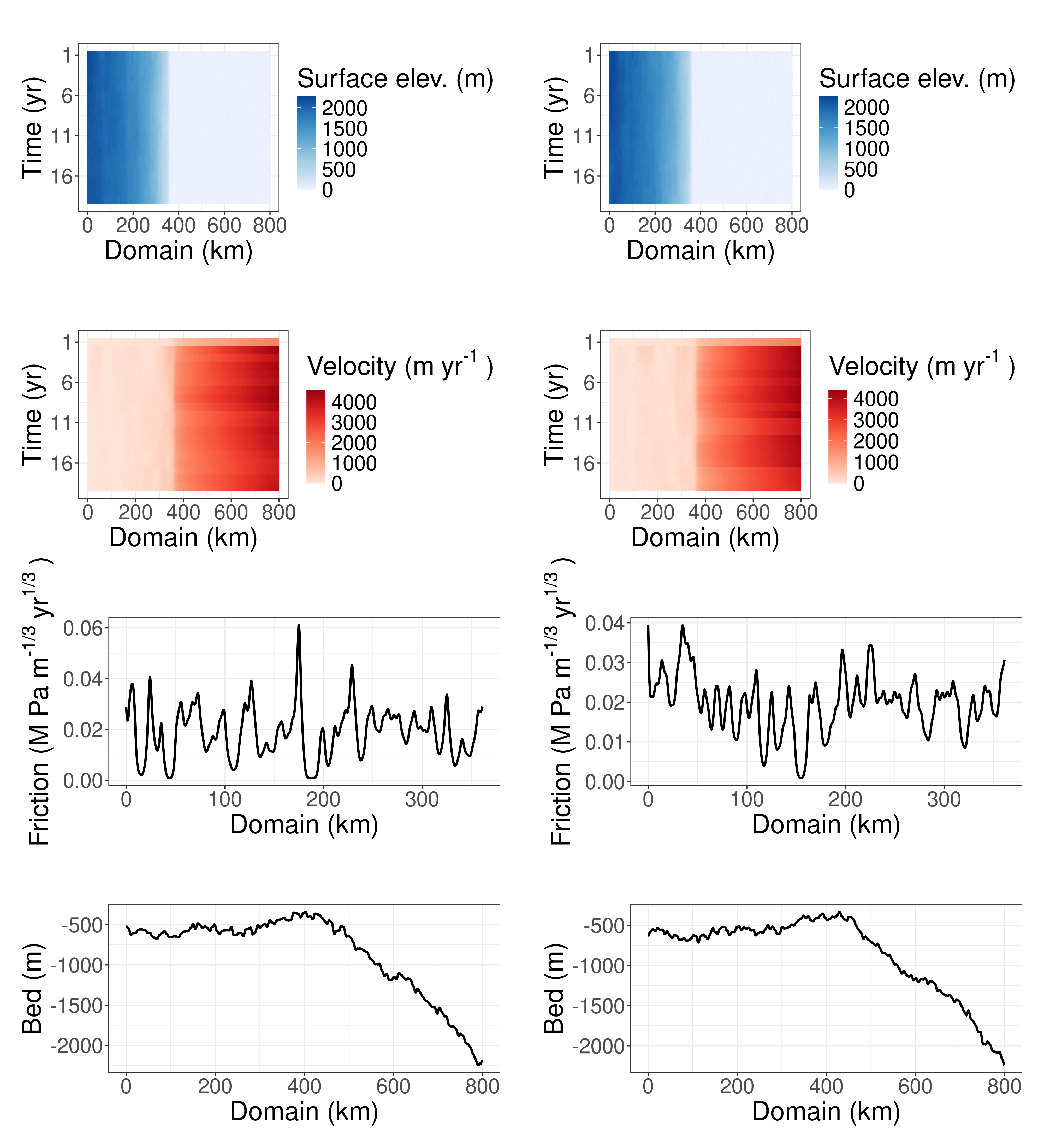}
    \caption{Plots of simulated parameters (basal friction coefficient and bed elevation, shown in the third and fourth rows) and corresponding data (surface elevation and surface velocity, shown in the first two rows) for two samples from the training data set. The plots in the first row show two examples of the surface elevation at each grid point along the flowline over 20 years, and similarly for the surface velocity in the second row. In the third row, the basal friction coefficient is plotted up to the grounding line position at the final time point (at approximately $360$~km). The fourth row shows bed elevation profiles.}
    \label{fig:simulations_01_20240320_01}
\end{figure}

To control the dimensionality of the unknown parameters, we represent the bed elevation and basal friction coefficient as linear combinations of local bisquare basis functions. To ensure positivity of the basal friction coefficient, we perform a log-transformation and fit basis functions on the log-transformed space. We fit basis functions to de-trended bed elevations (so that we are effectively inferring small-scale variations around a mean); see Section S5 of the Supplementary Material for details. For both the bed elevation and basal friction coefficient, we use $p = 150$ basis functions evenly spaced along the spatial domain, with the centres of the first and the last basis functions placed at $s_0 = 0$~m and $s_J = \num{800000}$~m, respectively.
Thus, the parameters that we make inference on are the basis function coefficients $\btheta \equiv \left({\btheta_{b}}^\top, {\btheta_{c}}^\top\right)^\top$, where $\btheta_{b} \in \mathbb{R}^{150}$ are the basis function coefficients for the (de-trended) bed elevation, and $\btheta_{c} \in \mathbb{R}^{150}$ are those for the (log-transformed) basal friction coefficient.
Two examples of simulated bed elevation and basal friction coefficient are shown in Figure~\ref{fig:simulations_01_20240320_01}.

\subsubsection{Simulating observations}
\label{sec:sim_obs}
Each set of parameters is used to construct bed elevation and friction coefficient profiles for the 20-year ice sheet model simulations. The model is run with a reduced ice stiffness compared to that used to reach steady-state (from $B = 0.4 \stiffnessunit$ to $B = 0.3 \stiffnessunit$) to simulate retreat/dynamics.
An example of a single realisation of the ice geometry evolution over the 20-year period is shown in Figure~\ref{fig:evolving_ice_sheet}. 
Annual noisy observations are then generated by adding Gaussian measurement error to the simulated surface elevations and velocities at the end of each year. For the surface elevation, we follow~\cite{gillet2020assimilation} and let the measurement error be uncorrelated with mean zero and standard deviation $\sigma_z = 10$~m at all spatial grid points and time points. For the surface velocity, we also let the measurement error be Gaussian with mean zero. However, since the surface velocity can vary between zero and several thousands of metres per year, and since larger velocities are measured with larger error, we let the standard deviation of the error at each grid point depend on the magnitude of the surface velocity. In particular, we model the measurement error standard deviation for the surface velocity at grid point $j$ at time $t$ as (in units of metres per year):
\begin{equation}
    \sigma_{u_{j, t}}  = \min(0.25 u_{j, t}, 20), \quad j = 0, \dots, J, \quad t \in \mathcal{T},
\end{equation}
where $u_{j, t}$ is the (noise-free) surface velocity at grid point $j$ at time $t$ obtained from the SSA model. \Copy{vel_measurement_err}{This model allows the measurement error to be dependent on the magnitude of the surface velocity in slow-flowing regions in the interior of the ice sheet (where the velocity can be close to zero). In other regions of the ice sheet, we follow the simulation setting of \citet{gillet2020assimilation} and set the measurement error to $20$ m yr$^{-1}$.} 

\begin{figure}[ht]
    \centering
    \begin{subfigure}[b]{0.48\textwidth}
        \centering
        \includegraphics[width=\textwidth]{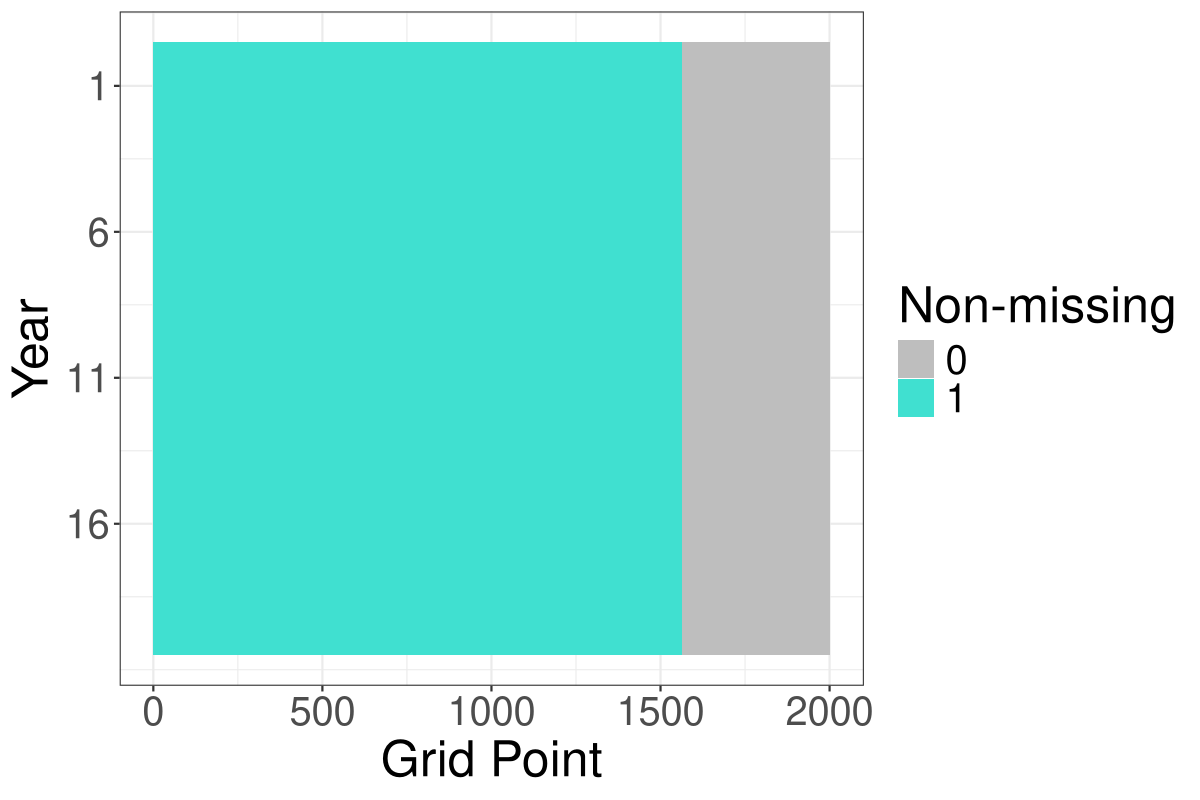}
        \caption{Surface elevation}
        \label{fig:surf_elev_missing_pattern}
    \end{subfigure}
    \hfill
    \begin{subfigure}[b]{0.48\textwidth}
        \centering
        \includegraphics[width=\textwidth]{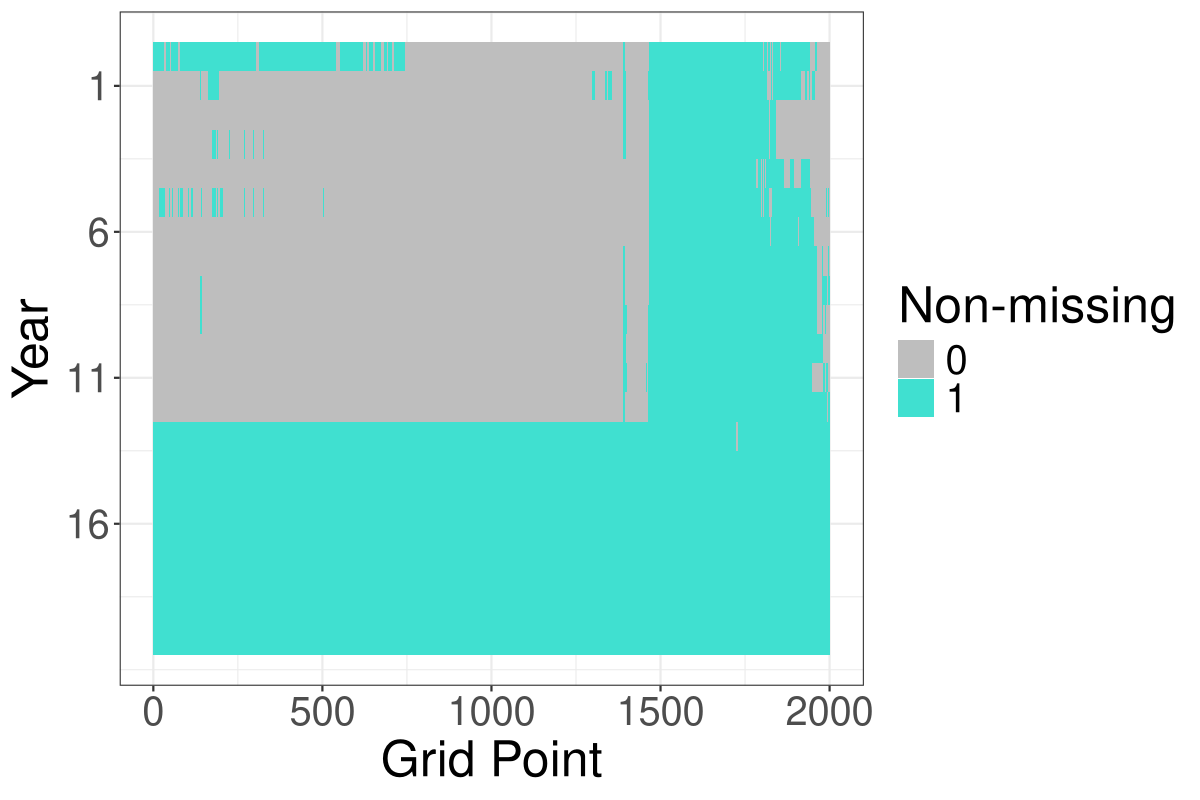}
        \caption{Surface velocity}
        \label{fig:vel_missing_pattern}
    \end{subfigure}
    \caption{Missingness pattern used for the surface elevation and velocity data (0 indicates missing).}
    \label{fig:missing_patterns}
\end{figure}

In practice, Antarctic datasets contain missing observations. For example, the data we analyse in Section~\ref{sec:real_data} does not contain measurements of elevation where the ice is floating, and observations of ice velocity in the period 2000--2012 are much sparser than those from 2013 onwards. For the neural network to be able to handle these missing observations, we follow a ``masking'' approach similar to that of \citet{wang2024missing}, where we mask the generated training data with a ``missingness pattern" that mimics the pattern of missing values in the real data. Specifically, we set the value of the surface elevation/velocity arbitrarily to zero at any spatial grid point and time point where an observation (surface elevation or velocity) is missing. (Note that any value can be used to denote missingness, as long as it is kept constant across training datasets; see, for example, \cite{wang2024missing}). The missingness patterns we use are shown in Figure~\ref{fig:missing_patterns}.

\Copy{missingness}{Our approach is not specifically reliant on a missing completely at random (MCAR) assumption, it just requires one to be able to simulate from the missingness mechanism during training (which could be MCAR, missing at random (MAR), or missing not at random (MNAR) depending on the specific application). Here, we have a known, fixed, missingness pattern and apply it across all $\num{50000}$ simulated datasets, which enables us to avoid making distributional assumptions on the occurrence of the missing values. If we wished to make the (trained) neural network generalisable to other flowlines/glaciers, then we would need to consider other possible missingness patterns, generated using some specified missingness mechanism. This problem was tackled by \citet{sainsbury2025neural}, by simulating observation locations for the training data using a Mat{\'e}rn cluster point process. There are ways to deal with missingness that do not require knowledge of the missingness mechanism \citep[e.g.,][]{sainsbury2025incomplete}, but these are typically designed under an MCAR or MAR assumption.}

\subsection{Neural network training}
\label{sec:sim_study_stageone}
For training data we use $N = \num{44500}$ sets of surface elevation and surface velocity data (as input), and corresponding basis function coefficients for the bed elevation and basal friction coefficient (as output). We use an additional $\num{5000}$ sets for validation and $500$ sets for testing.
The training and validation sets are used for training the neural network, and for monitoring training progress, respectively. The test set is reserved for subsequent benchmarking.

Since the neural network is sensitive to the scaling of the data, we first perform a standardisation step as detailed in Section S6.1 of the Supplementary Material.
With the (standardised) simulated observations, we train a CNN according to~\eqref{eq:loss_fun} to output the parameters of the approximate posterior distribution, namely $\blambda_{\bgamma^*}(\cdot) \equiv (\bmu_{\bgamma^*}(\cdot)^\top, \vech(\L_{\bgamma^*}(\cdot))^\top)^\top$. 

\begin{figure}
    \begin{center}
    \begin{tikzpicture}[>=latex, center coordinate=(kappab)]
        \centering

        \node[circle, draw, minimum size=1cm, inner sep=0pt] (Z) at (0,0) {$\Y$};
        \node[rectangle, draw, minimum size=1cm, inner sep=0pt, fill=black, text=white] (NN) at (2.5,0) {\textbf{NN}};
        \node[circle, draw, minimum size=1cm, inner sep=0pt] (kappab) at (5,0)
        {$\bkappa$};
        \node[circle, draw, minimum size=1cm, inner sep=0pt] (theta) at (7.5,0)
        {$\btheta$};
        \node[circle, draw, minimum size=1cm, inner sep=0pt] (bc) at (10,0)
        {$b(\cdot)$, $c(\cdot)$};

        \draw[->] (Z) to (NN);
        \draw[->] (NN) to (kappab);
        \draw[dashed, ->] (kappab) to node[above] {sample} (theta);
        \draw[->] (theta) to (bc);

        \draw[decorate,decoration={brace,amplitude=10pt,mirror}]
        ([shift={(-0.2,-0.3)}]Z.south west) -- ([shift={(0.2,-0.3)}]Z.south east)
        node[midway,below=10pt] {Data};

        \draw[decorate,decoration={brace,amplitude=10pt,mirror}]
        ([shift={(-0.2,-0.3)}]NN.south west) -- ([shift={(0.2,-0.3)}]NN.south east)
        node[midway,below=10pt,align=center] {Inference \\ Network};

        \draw[decorate,decoration={brace,amplitude=10pt,mirror}]
        ([shift={(-0.2,-0.3)}]kappab.south west) -- ([shift={(0.2,-0.3)}]kappab.south east)
        node[midway,below=10pt, align=center] {Approx.~Dist. \\ Parameters};

        \draw[decorate,decoration={brace,amplitude=10pt,mirror}]
        ([shift={(-0.2,-0.3)}]theta.south west) -- ([shift={(0.2,-0.3)}]theta.south east)
        node[midway,below=10pt, align=center] {Basis Fun. \\ Coefs.};

        \draw[decorate,decoration={brace,amplitude=10pt,mirror}]
        ([shift={(-0.2,-0.3)}]bc.south west) -- ([shift={(0.2,-0.3)}]bc.south east)
        node[midway,below=10pt, align=center] {Ice sheet \\Parameters };

    \end{tikzpicture}

\end{center}
    \caption{Neural posterior inference (NPI) framework for the SSA model.}
    \label{fig:npi_ssa}
\end{figure}
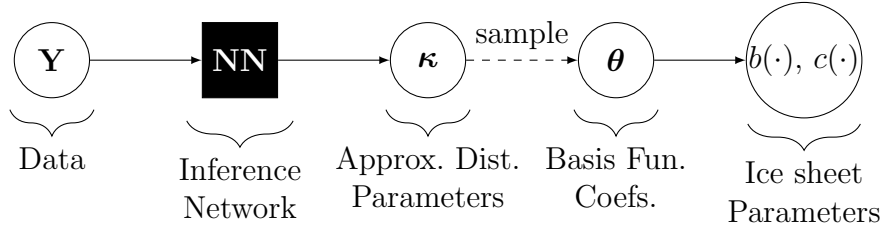

We train the neural network for 30 epochs (where an epoch corresponds to one complete pass through the entire training dataset during neural network training) and choose the model weights and biases at the epoch with the lowest validation loss as our trained model. The validation loss is the loss function~\eqref{eq:loss_fun} evaluated using the validation data. Further details on the neural network architecture are given in Section 6.2 of the Supplementary Material. We summarise our NPI framework in Figure~\ref{fig:npi_ssa}.

\subsection{State inference conditional on neural network output}
\label{sec:sim_study_stagetwo}
After obtaining an approximation for the parameter posterior distribution $q(\btheta; \blambda_{\bgamma^*}(\Y))$ from the neural network for some $\Y$, we make inference on the state variable (ice thickness).
We do this by sampling the parameters $L = 10$ times from the approximate posterior distribution, and then running an EnKF algorithm conditional on each posterior sample of the bed elevation and basal friction coefficient. Since these samples are independent, the EnKF runs can be done in parallel.

For each EnKF run, we begin with an initial ensemble of ice thickness realisations $\{\x_{1 \mid 1}^{(i)}: i = 1, \dots, N_e\}$, $N_e = 500$, where the $i$th member is given by
\begin{equation*}
    \x_{1 \mid 1}^{(i)} \equiv (x_{0, 1 \mid 1}^{(i)}, \dots, x_{J, 1 \mid 1}^{(i)})^\top,
\end{equation*}
for $i = 1, \dots, N_e$, and $J = 2000$. Here, the notation $x_{j, t \mid t}$ denotes the estimate of the ice thickness at the $j$th grid node at time $t$, given all observations up to time $t$. This ensemble at time $t = 1$ is obtained by first calculating an ice thickness ``trend'' based on the observed surface elevation at the initial time and the bed elevation from the neural network output. Specifically, we calculate the ice thickness from the surface elevation and bed elevation using a re-arrangement of~\eqref{eq:ssa_top_surface}:
\begin{align}
    h(s, t) = \begin{cases}
                  z(s, t) - b(s), \quad                         & z(s, t) \geq (1 - \frac{\rho_w}{\rho_i}) b(s), \quad s \in (0, s_g), \\ 
                  \frac{\rho_w}{\rho_w - \rho_i} z(s, t), \quad & z(s, t) < (1 - \frac{\rho_w}{\rho_i}) b(s), \quad s \in (s_g, s_F),  
              \end{cases}
\end{align}
and then we let $x_{j, 1 \mid 1} \equiv h(s_j, 1)$ for $j = 0, \dots, J$.

Since the surface elevation measurements are noisy, this mapping yields an unrealistically noisy ice thickness trend. Thus, we first smooth the trend using polynomial regression, and then add correlated Gaussian error to it to yield different realisations of the initial ice thickness. We model this added error as $\v_1 \sim \Gau(\0, \V_1)$, where $\V_1 = \D_{V, 1} \K_V \D_{V, 1} $, where $\D_{V, 1}$ is a diagonal matrix, and where the $(i, j)$th element of the matrix $\K_V$ comes from an exponential correlation function:
\begin{equation}
    K_V(s_i, s_j) =
    \exp \left(- \frac{3\abs{s_i - s_j}}{\ell_V} \right), \quad i = 0, \dots, J, \quad j = 0, \dots, J, \label{eq:process_noise_corr1}
\end{equation}
with $\ell_V = \num{50000}$~m. Here, the elements of the diagonal matrix $\D_{V, 1}$ are set to $50$~m for grounded ice and $20$~m for floating ice. These values were chosen to reproduce realistic ice geometries.

Once an ensemble of initial ice thicknesses are obtained, we run this ensemble forward for 20 years. As the ice thickness can vary from several hundred to several thousand meters, we expect the process error variance at a given spatial location to be related to the ice thickness. Therefore, we model the state-dependent process error covariance matrix as $\V_t (\x_t) = \D_{V} (\x_t) \K_V \D_{V} (\x_t)$, where $\D_{V}  (\x_t) = \diag(\sigma(x_{0, t}), \dots, \sigma(x_{J, t}))$, in which $\sigma(x_{j, t}) = \min(0.02 \, x_{j, t}, 20)$, $j = 0, \dots, J$, ${t \in \mathcal{T}}$. The resulting ensembles are then concatenated to form one large ensemble that approximates the filtering distribution of the states.


\subsection{Results from the OSSE}
\label{sec:sim_study_results}

\begin{figure}[t]
    \centering
    \begin{subfigure}[b]{0.48\textwidth}
        \centering
        \includegraphics[width=\textwidth]{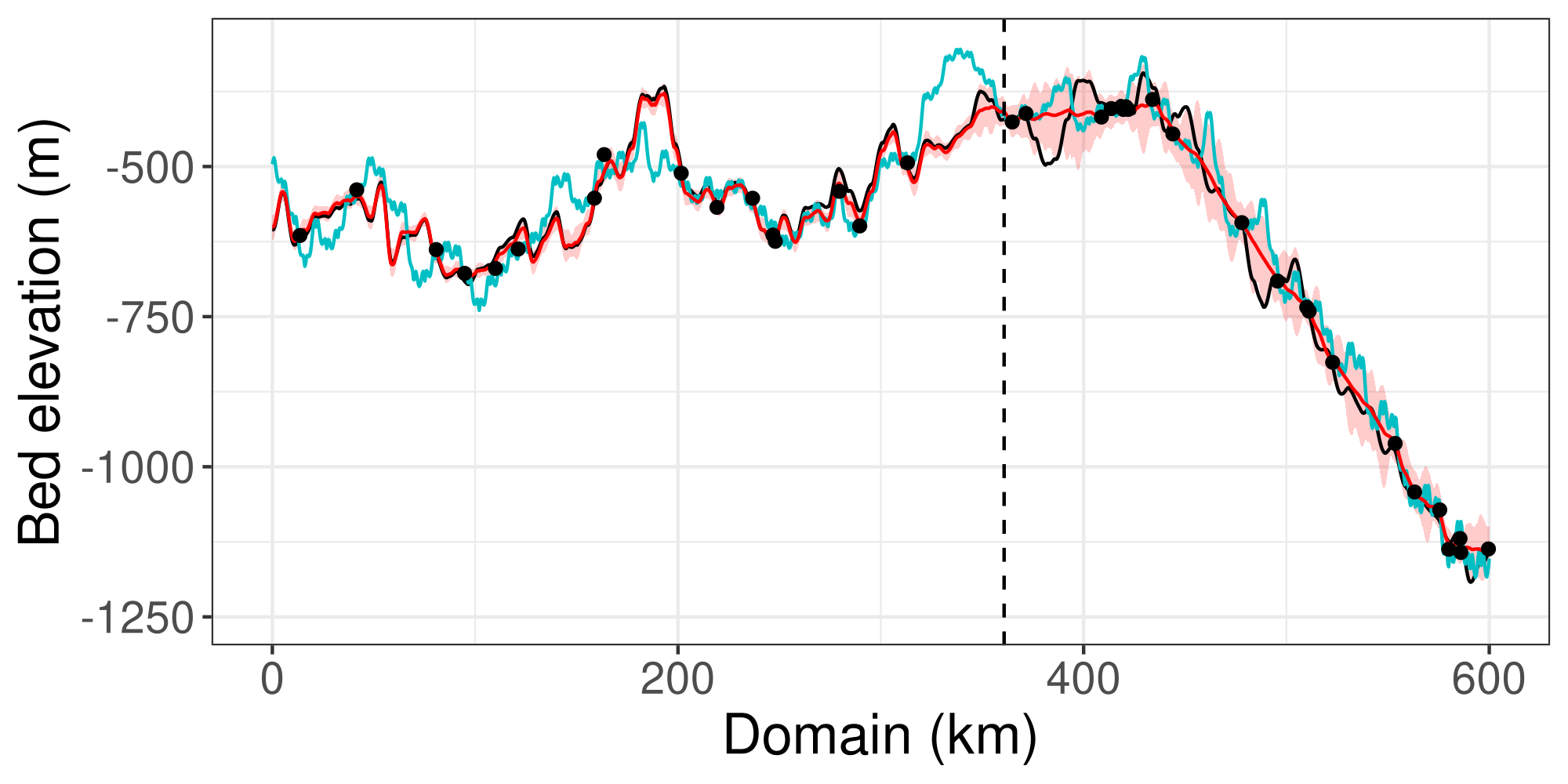}
        \caption{Sample 1.}
        \label{fig:bed_enkf_s1}
    \end{subfigure}
    \hfill
    \begin{subfigure}[b]{0.48\textwidth}
        \centering
        \includegraphics[width=\textwidth]{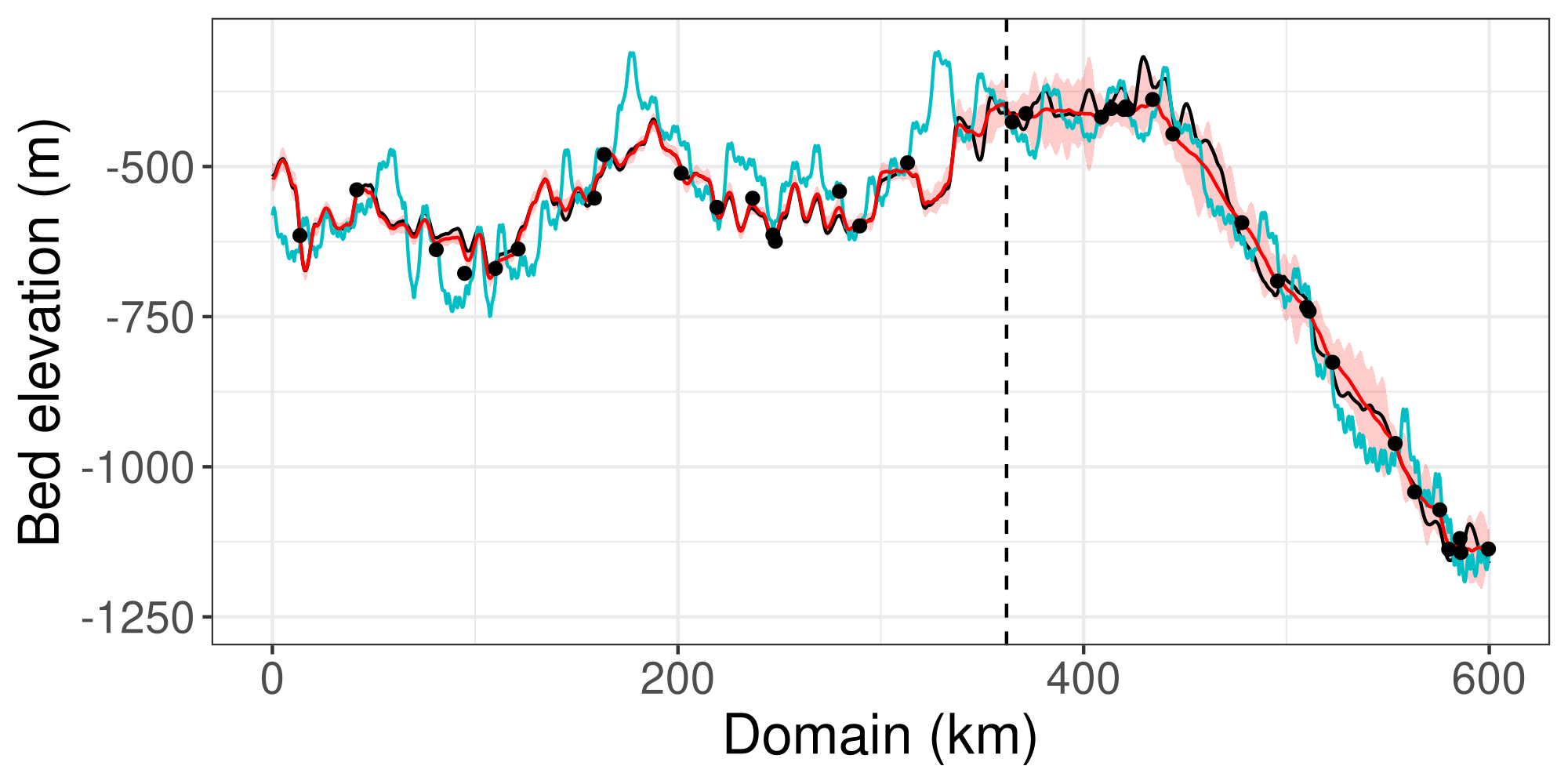}
        \caption{Sample 2.}
        \label{fig:bed_enkf_s2}
    \end{subfigure}
    \caption{Comparison of bed elevation estimates from our NPI approach (red) and augmented EnKF (blue) for two samples from the test dataset. The true bed elevation is plotted as a solid black line. Black points denote observations of bed elevation used to construct a prior distribution for $\btheta_b$. The dotted vertical line shows the grounding line position at the final time point. The pale red shading in the background shows the 95\% credible intervals from NPI. Intervals are also plotted for the augmented EnKF, but they are difficult to see due to low ensemble spread.}
    \label{fig:bed_inference}
\end{figure}

Estimates of the bed elevation from our approach and from the state-augmented EnKF~\citep{anderson2001ensemble} for two randomly chosen samples from the test dataset are shown in Figure~\ref{fig:bed_inference}. Upstream of the grounding line, where the ice is grounded, estimates of the bed elevation are very close to the true bed elevation; this is most likely because the dynamics of the ice are highly dependent on the bed elevation in this region, and therefore the surface elevation and surface velocity are highly informative. Downstream of the grounding line, where the ice is no longer attached to the bed (and thus the dynamics of ice flow are no longer affected by the bed elevation), posterior estimates of the bed elevation are more variable, and they revert to the prior mean.

\begin{figure}[t]
    \centering
    \begin{subfigure}[b]{0.48\textwidth}
        \centering
        \includegraphics[width=\textwidth]{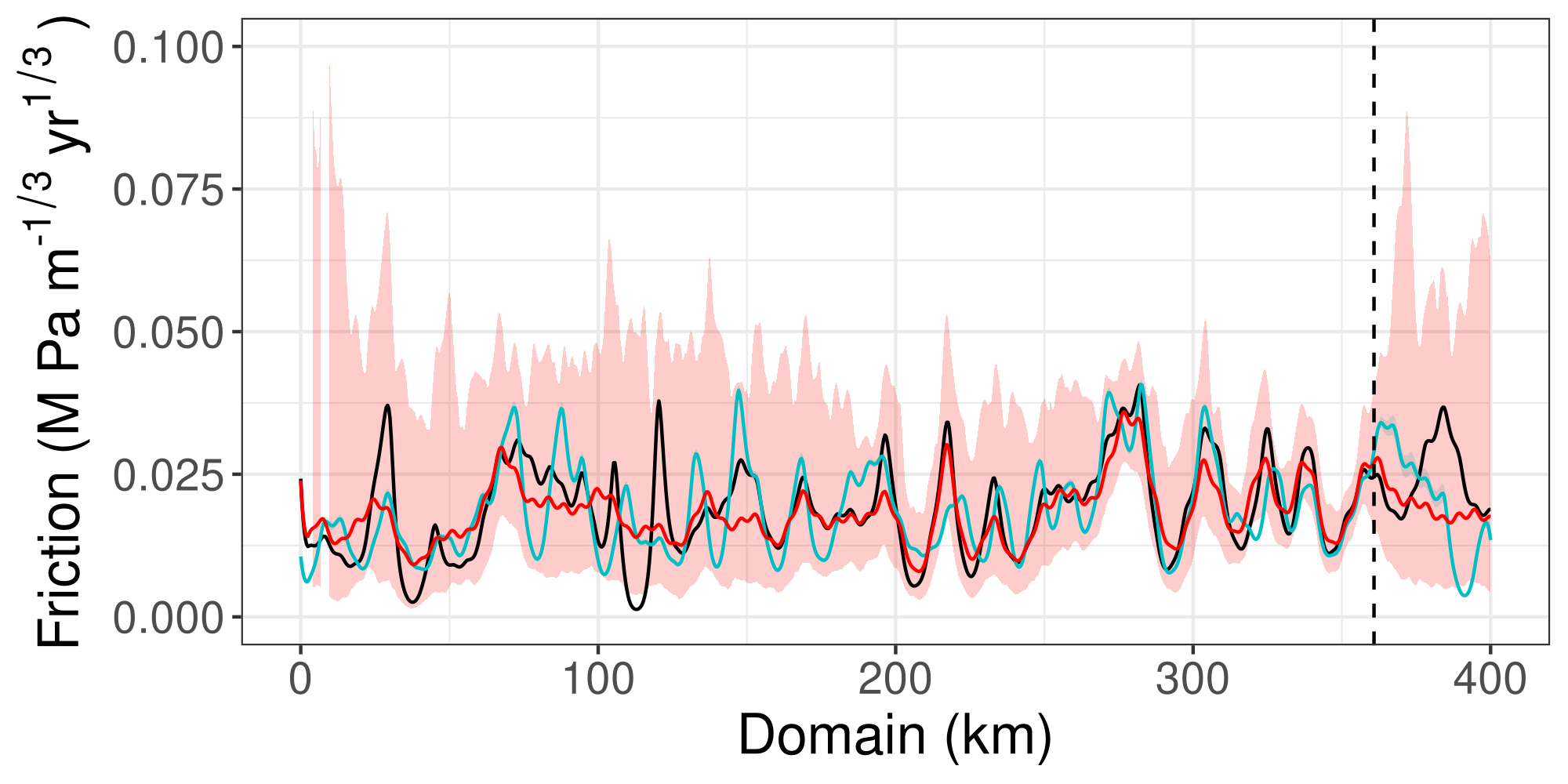}
        \caption{Sample 1.}
        \label{fig:friction_enkf_s1}
    \end{subfigure}
    \hfill
    \begin{subfigure}[b]{0.48\textwidth}
        \centering
        \includegraphics[width=\textwidth]{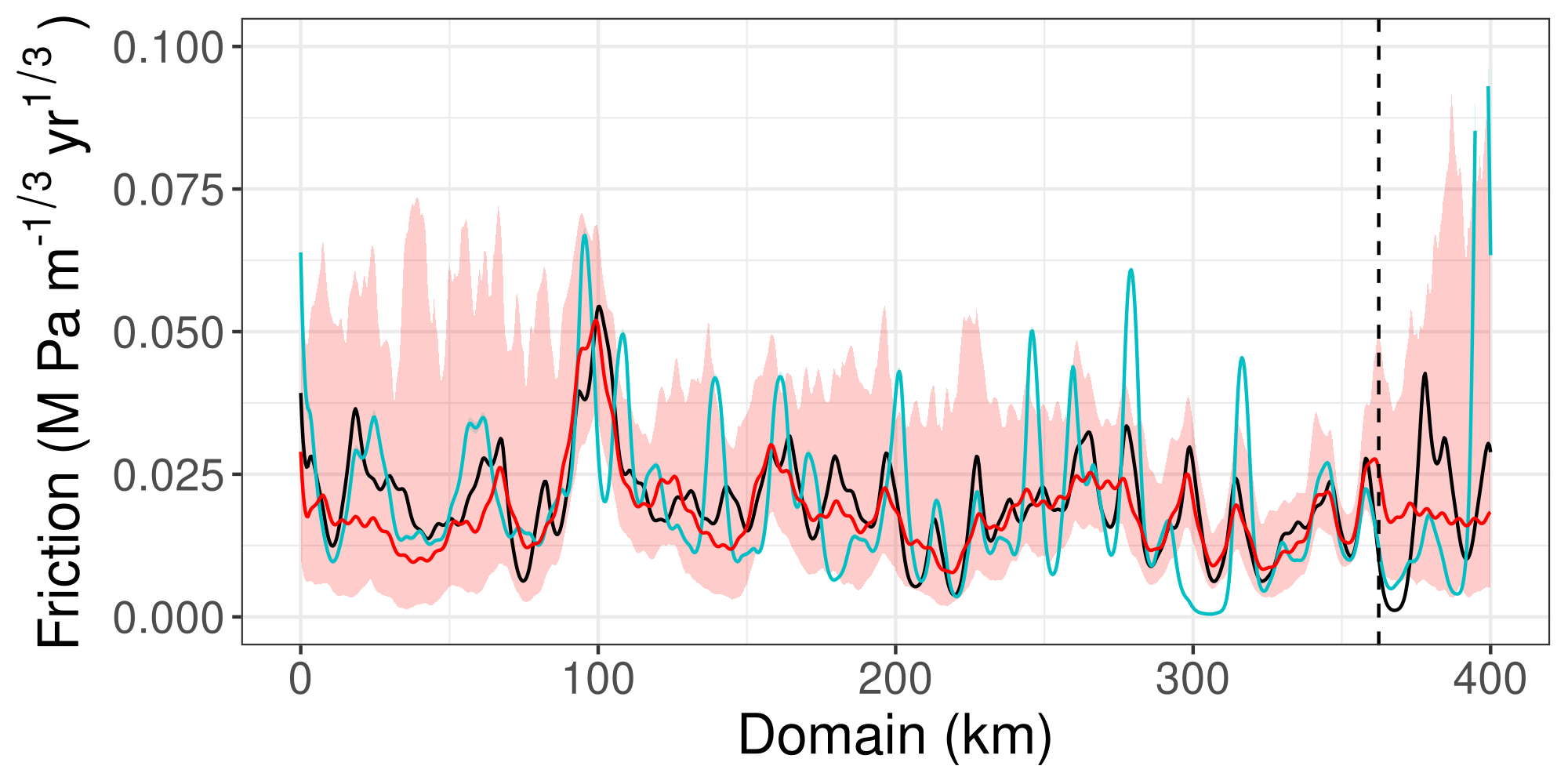}
        \caption{Sample 2.}
        \label{fig:friction_enkf_s2}
    \end{subfigure}
    \caption{Comparison of basal friction coefficient estimates from our NPI approach (red) and augmented EnKF (blue) for two samples from the test dataset. The true basal friction coefficient is plotted as a solid black line. The dotted vertical line shows the grounding line position at the final time point. The pale red shading in the background shows the 95\% credible intervals from NPI. Intervals are also plotted for the augmented EnKF, but they are difficult to see due to low ensemble spread.}
    \label{fig:friction_inference}
\end{figure}

A plot of the basal friction coefficient estimates is shown in Figure~\ref{fig:friction_inference}.
In the case of the augmented EnKF, the ensemble members are very similar, and uncertainty intervals are difficult to detect visually. For our NPI approach, the 95\% credible intervals are wide around the ice divide ($s_0 = 0)$, where ice flows very slowly, but then they narrow close to the grounding line as the ice velocity increases, suggesting that it is easier to estimate the basal friction coefficient in areas where ice is flowing more rapidly. This finding is consistent with the study of~\cite{cheng2024forward}, who used a PINN to find (point) estimates of the basal friction coefficient based on surface velocity, surface elevation, and ice thickness data, and who found that prediction errors for the basal friction coefficient are larger in regions of slow-moving ice.

\begin{figure}[ht]
    \centering
    \begin{subfigure}[b]{0.49\textwidth}
        \centering
        \includegraphics[width=\textwidth]{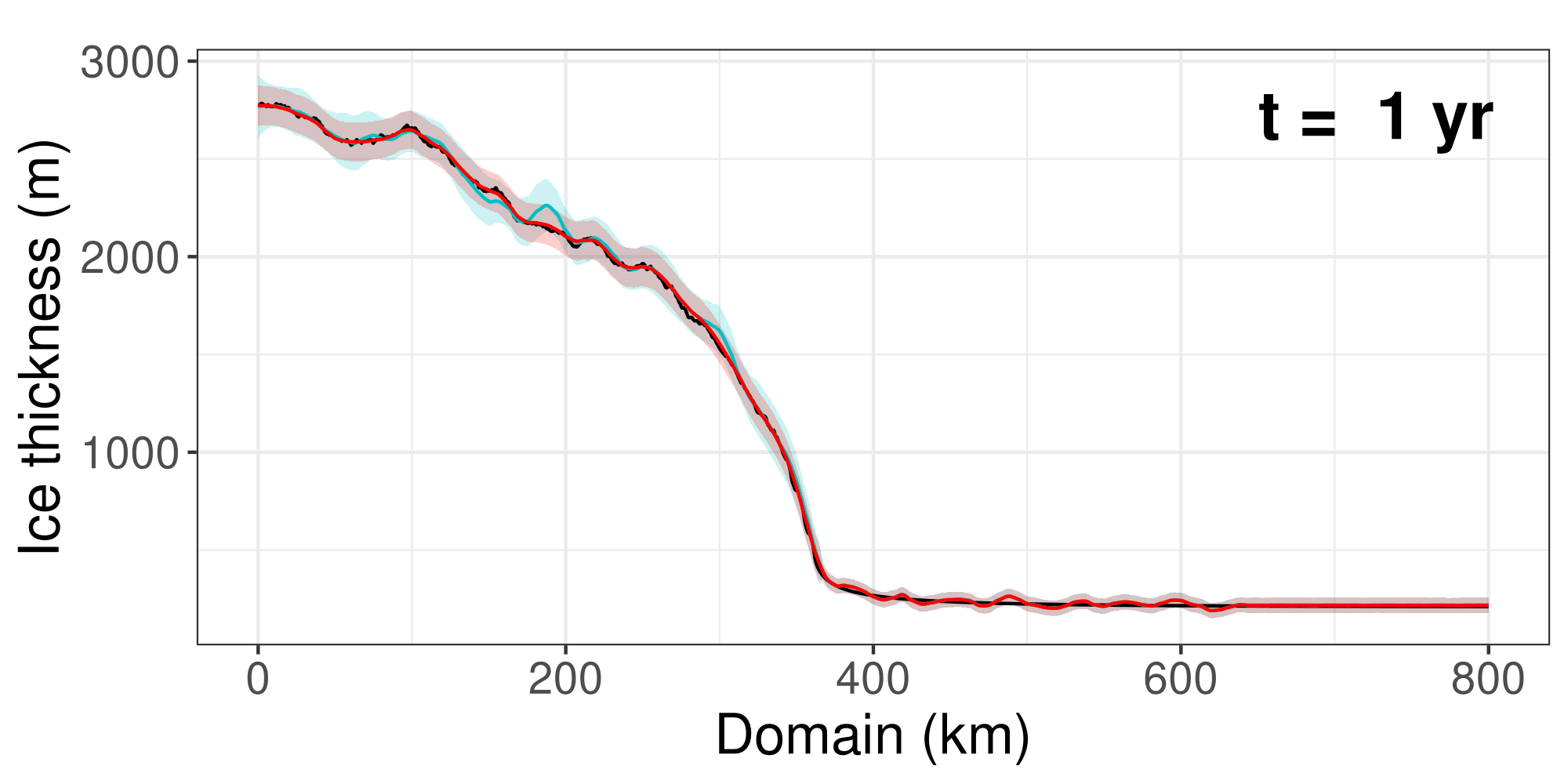}
    \end{subfigure}
    \hfill
    \centering
    \begin{subfigure}[b]{0.49\textwidth}
        \centering
        \includegraphics[width=\textwidth]{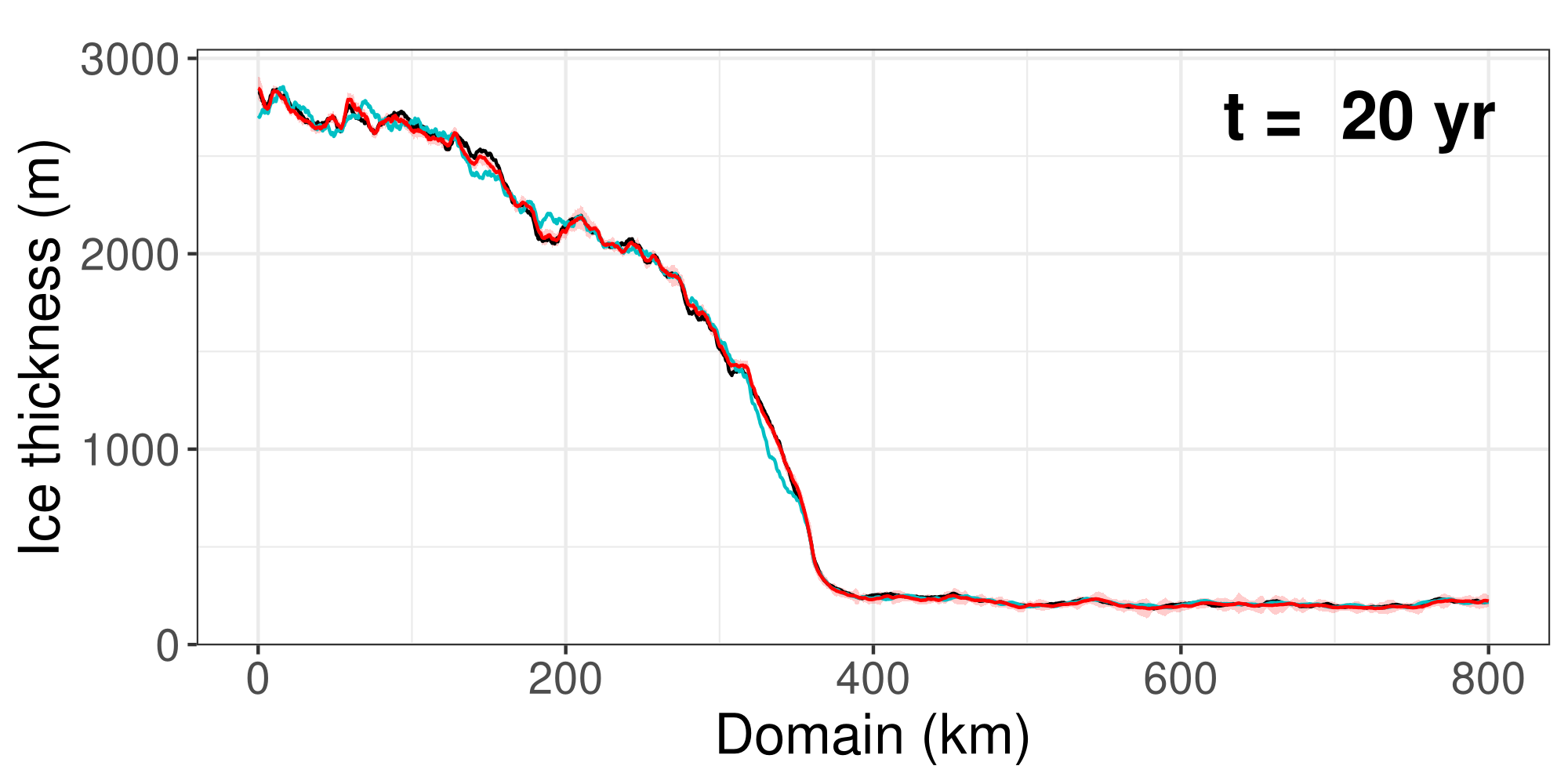}
    \end{subfigure}
    \caption{Comparison of ice thickness estimates between our NPI approach (red) and augmented EnKF (blue) at the initial time (left) and after 20 years (right), for a randomly selected sample from the test dataset. The true ice thickness is plotted in black. The pale shadings in the background show 95\% prediction intervals for the two approaches.}
    \label{fig:thickness_inference}
\end{figure}

Next, we compare the results of state inference between our NPI approach and the augmented EnKF in Figure~\ref{fig:thickness_inference}. Estimates of the ice thickness from NPI are very close to the true ice thickness. In the plot of the initial ensemble, uncertainties are represented by the pale shading. This uncertainty quickly reduces over time as the ensemble members become less dispersed. Note that we are able to infer ice thickness everywhere in the domain, despite having missing data (recall Figure~\ref{fig:missing_patterns}).

We compare the root mean squared error (RMSE) and the continuous ranked probability score (CRPS) for the two methods in Table~\ref{tab:rmse}; for both metrics, smaller values indicate better performance. The RMSE and CRPS are averaged over all spatial grid points and over 10 randomly chosen samples from the test dataset. The table shows that our two-stage inference approach yields more accurate estimates than the augmented EnKF for both the latent states and the parameters. In Table~\ref{tab:rmse}, we also compare the empirical coverage of the 95\% credible intervals from both methods. NPI yields coverage values that are much closer to the nominal $0.95$ level compared to the augmented EnKF, which exhibits substantial under-coverage for both the parameters (bed elevation, basal friction coefficient) and the state (ice thickness). This suggests that our approach produces more reliable uncertainty quantification than the augmented EnKF.

The neural network took approximately 11 minutes to train on an NVIDIA H100 GPU but, once trained, it produced posterior estimates in as little as 1 second. In comparison, parameter estimation with the augmented EnKF took around 12 minutes. While the second stage (state inference) in our approach requires multiple runs of the EnKF, these can be done in parallel, and take approximately the same amount of time as a single run of the augmented EnKF. Therefore, our approach, while offering superior performance, requires slightly more computing effort than the augmented EnKF algorithm.

\begin{table}[t!]
    \centering
    \caption{Comparison of the RMSE, CRPS, and coverage between our approach (NPI) and the state-augmented EnKF (Aug. EnKF) for inference on the parameters (bed elevation, basal friction coefficient) and states (ice thickness) at the final time point (20 years). The RMSE, CRPS and coverage values are averaged over 10 samples from our test dataset.}
    \begin{tabular}{|l|rr|rr|rr|}
        \hline
        \multirow{2}{*}{} & \multicolumn{2}{c|}{RMSE}      & \multicolumn{2}{c|}{CRPS} & \multicolumn{2}{c|}{Coverage}                                                                                         \\ \cline{2-7}
                          & \multicolumn{1}{c|}{Aug. EnKF} & \multicolumn{1}{c|}{NPI}  & \multicolumn{1}{c|}{Aug. EnKF} & \multicolumn{1}{c|}{NPI} & \multicolumn{1}{c|}{Aug. EnKF} & \multicolumn{1}{c|}{NPI} \\ \hline
        Bed elevation     & \multicolumn{1}{r|}{56.9008}   & \textbf{25.9586}          & \multicolumn{1}{r|}{2579.2892} & \textbf{879.9988}        & \multicolumn{1}{r|}{0.0590}    & \textbf{0.8902}          \\
        Friction coef.    & \multicolumn{1}{r|}{0.0080}    & \textbf{0.0051}           & \multicolumn{1}{r|}{0.2664}    & \textbf{0.1341}          & \multicolumn{1}{r|}{0.0598}    & \textbf{0.9615}          \\
        Ice thickness     & \multicolumn{1}{r|}{40.8692}   & \textbf{11.1891}          & \multicolumn{1}{r|}{1762.2959} & \textbf{679.7939}        & \multicolumn{1}{r|}{0.4336}    & \textbf{0.9895}          \\ \hline
    \end{tabular}
    \label{tab:rmse}
\end{table}



\section{Application to Thwaites Glacier, Antarctica}
\label{sec:real_data}
\begin{figure}[t]
    \centering
    \includegraphics[width=0.8\textwidth]{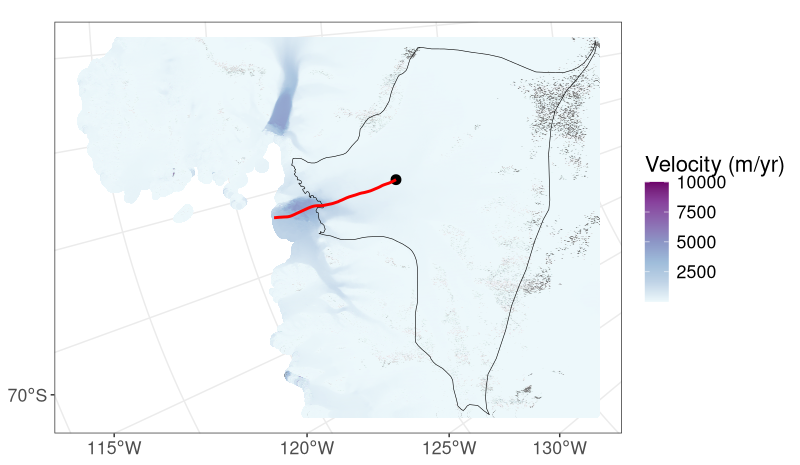}
    \caption{Map of Thwaites Glacier with the chosen flowline indicated by the red line, and the flowline start point indicated by a black dot. The background image shows surface velocity observations from the MEaSUREs dataset.}
    \label{fig:thwaites_flowline}
\end{figure}

In this section, we apply our approach to infer the bed elevation and friction coefficient under a flowline along Thwaites Glacier in Antarctica. Thwaites Glacier is of particular interest to glaciologists due to its rapid retreat and potential contribution to sea level rise \citep{ehrenfeucht2026impacts, rignot2026thirty}.
The chosen flowline is approximately $250$ km long and extends from the interior of the ice sheet into the ocean, as visualised in Figure~\ref{fig:thwaites_flowline}.

\subsection{Datasets}

In this study, we use the NASA MEaSUREs gridded surface elevation change and surface velocity datasets \citep{its_live, nilsson2022elevation} over the period 2010-2020.
Surface velocity data are available at an annual resolution. For the surface elevation data, which is available at a monthly resolution, we aggregate the data to an annual resolution by taking the average of the surface elevation over all the months in each year. For conditional simulations of bed elevation profiles (conditional on sparse observations), which are necessary for generating training data for the neural network, we make use of (un-gridded) bed elevation data along flight tracks from the BedMap3 dataset \citep{bedmap3}. For climate forcings used in the ice sheet model, we use a constant snow accumulation rate computed by taking the average over the annual values from 1979--2016 in the Regional Atmospheric Climate Model version 2~\citep[RACMO2,][]{racmo2} dataset. For the floating ice shelf, we also use a constant basal melt rate ($a_b$) obtained by taking the average of annual melt rates from 1992--2017 from the MEaSUREs basal melt dataset~\citep{measures_basal_melt}.

As the spatial resolutions of these datasets differ, we pre-process the datasets to a common grid along the flowline. We discretise the flowline into $J = 2000$ evenly spaced segments, with a grid spacing of approximately $123.40$ m. For the surface elevation, surface velocity, and snow accumulation rate, we re-grid the original data onto the flowline by taking the average of the nearest 4 observations to each grid point (within a certain radius) as the value of that grid point. This radius is equal to the grid spacing in the original datasets: $1920$ m for the surface elevation, $120$ m for the velocity, and $27.5$ km for the snow accumulation rate.
For the bed elevation observations, we take the average of the nearest 4 observations within a $200$ m radius around each grid point. 
As across-track observations can be sparse, there are cases where no observations are made within this radius, and we treat the bed elevation at these grid points as missing. This leads to 72 bed elevation observations along the flowline, which are used in the simulation of training data.

As part of this study, we also do out-of-sample forecasting of the ice surface elevation and surface velocity for assessing the quality of our parameter inference, and we reserve surface elevation and surface velocity data from the year 2020 for this purpose. The training data are thus constructed for a 10-year period (2010--2019).


\subsection{Generation of training data}
\label{sec:train_data_gen_real}
Training data for the neural network are generated by running the ice sheet model thousands of times while varying the parameters (bed elevation and basal friction coefficient). The procedure for simulating different realisations of bed elevation and basal friction coefficient is similar to that in Section~\ref{sec:sim_train_data}, with minor adjustments to the value of the ice stiffness and basal melt rate, and assumptions are made for the ice thickness beyond the grounding line, since no observations of surface elevation or thickness are available in this region of floating ice. Details for the ice sheet model initialisation and generation of the training data are given in Section S7 of the Supplementary Material.

\subsubsection*{Discrepancy between simulated and observed data}
One challenge that arose during the generation of the training data is the presence of a systematic discrepancy between the simulated and observed ice sheet surface elevation and velocity from the actual glacier. Simulated surface elevations from our model reflected an ice sheet that is retreating more quickly than observed in reality, and with surface velocity that is lower than the observed velocity, particularly near the grounding line region where ice dynamics become more complex. This may be due to the inability of the one-dimensional model to capture lateral stresses from valley walls that buttress the glacier, and hence decrease the mass flux.
To account for the mismatch between simulated and actual observations, we introduce a ``model discrepancy'' term as in \citet{simmach}.
Specifically, we alter the data model~\eqref{eq:ssm_obs} to
\begin{align}
    \y_t         & = \tilde{\y}_t + \bdelta, \label{eq:model_discrepancy}                                                      \\
    \tilde{\y}_t & = \mathcal{H}_t (\x_t, \btheta) + \w_t,  \quad \w_t \sim \Gau(\0, \R_t), \quad t \in \mathcal{T}, \nonumber
\end{align}
where $\bdelta$ is the $2(J+1)$-dimensional model discrepancy term that accounts for differences between observations and the ice sheet model output that cannot be explained by measurement noise alone.
We assume that the model discrepancy $\bdelta$ varies spatially but not temporally. 

While the discrepancy term is typically modelled as a Gaussian process whose hyperparameters are learned during inference~\citep{simmach}, here we take a simpler approach and assume that the discrepancy is deterministic. This is a reasonable assumption because the discrepancies across different simulations have similar large-scale structure; see Figure S3
in Section S7.3
of the Supplementary Material for some representative examples. We estimate the discrepancy term $\bdelta$ by computing the temporally averaged difference between the observed surface elevation and the model output over many simulations drawn from our prior distribution. That is, we rearrange~\eqref{eq:model_discrepancy} to
\begin{equation}
    \bdelta = \y_t - [\mathcal{H}_t (\x_t, \btheta) + \w_t],  \quad \w_t \sim \Gau(\0, \R_t), \quad t \in \mathcal{T},
    \label{eq:model_discrepancy_rearranged}
\end{equation}
and estimate it using Monte Carlo samples,
\begin{align}
    \tilde{\bdelta} & \equiv \frac{1}{N_t}\sum_{t = 1}^{N_t}
    [\y_t - \left(\mathcal{H}_t (\x_t, \btheta)  + \w_t \right)], \nonumber                                                                                                                                                      \\
                    & \approx \frac{1}{N_t N_{sim}} \sum_{t=1}^{N_t} \sum_{i=1}^{N_{sim}} \bigl\{\y_t - \mathcal{H}_t (\x_t, \btheta^{(i)}) \bigr\}, \quad \btheta^{(i)} \sim p(\btheta), \label{eq:estimated_model_discrepancy}
\end{align}
where $N_t = 10$ is the number of years for which observations are available, $N_{sim} = 1000$ is the number of simulations (from the prior) used to compute the average discrepancy, and where we have omitted $\w_t$ from~\eqref{eq:estimated_model_discrepancy} since it has mean $\0$. The resulting averaged discrepancy is a $2(J+1)$-dimensional spatial vector. We then fit a spline to $\tilde{\bdelta}$ to obtain a ``smoothed'' discrepancy term, which we denote as $\hat{\bdelta}$. Using this smoothed discrepancy enables adjustment of systematic biases in the model output without affecting the fine-scale variability. 

\begin{figure}[ht]
    \centering
    \includegraphics[width=0.85\textwidth]{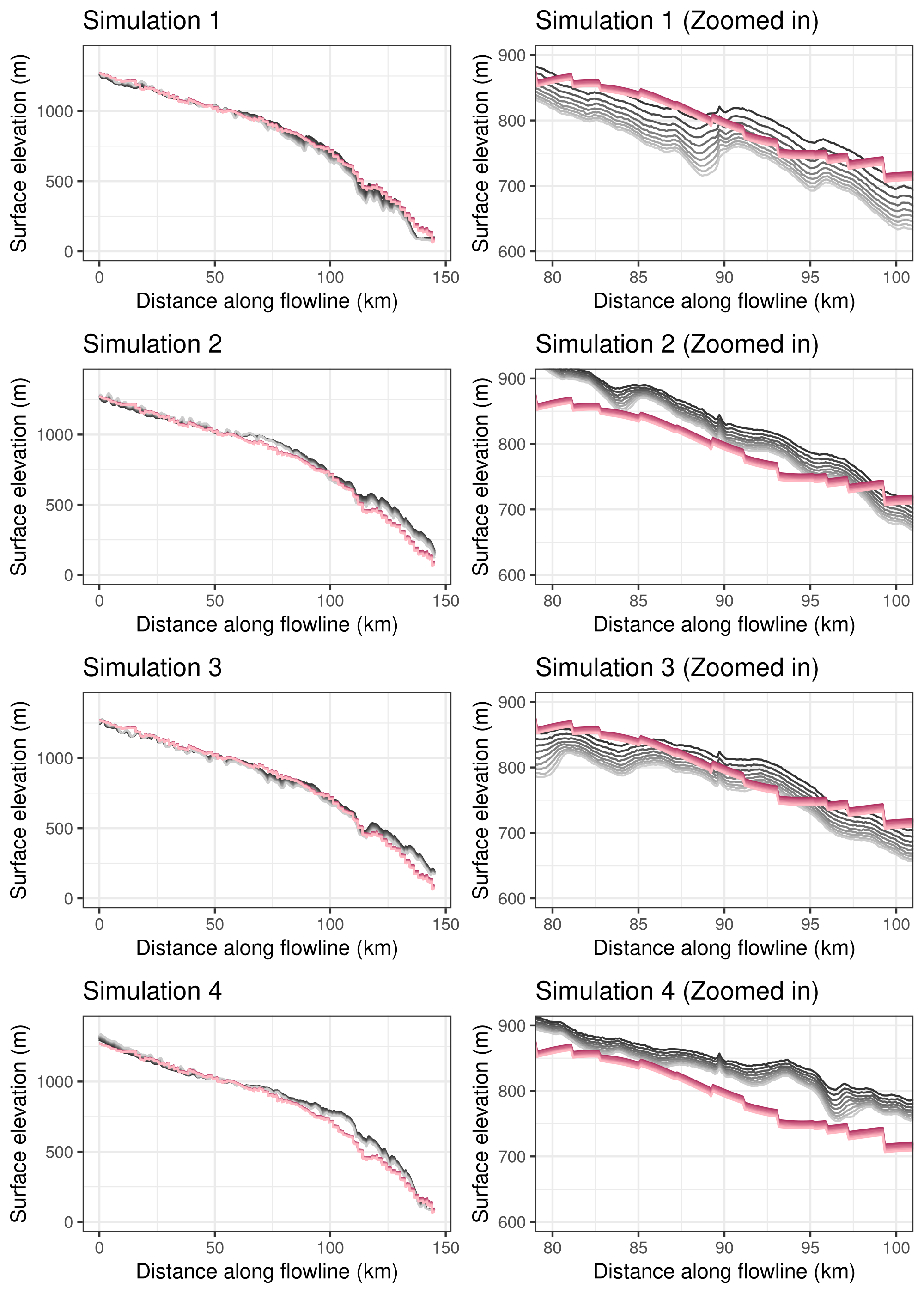}
    \caption{Adjusted observations of surface elevation (left column) along the Thwaites Glacier flowline for the years 2010-2019, with zoomed-in plots (right column). The adjustment is done by subtracting the estimated model discrepancy from the observations. Red lines represent adjusted observations, and grey lines represent simulated observations from our prior distribution. In both cases, lighter lines indicate increasing time.}
    \label{fig:se_obs_adjusted_20241111}
\end{figure}
\begin{figure}[t]
    \centering
    \includegraphics[width=0.85\textwidth]{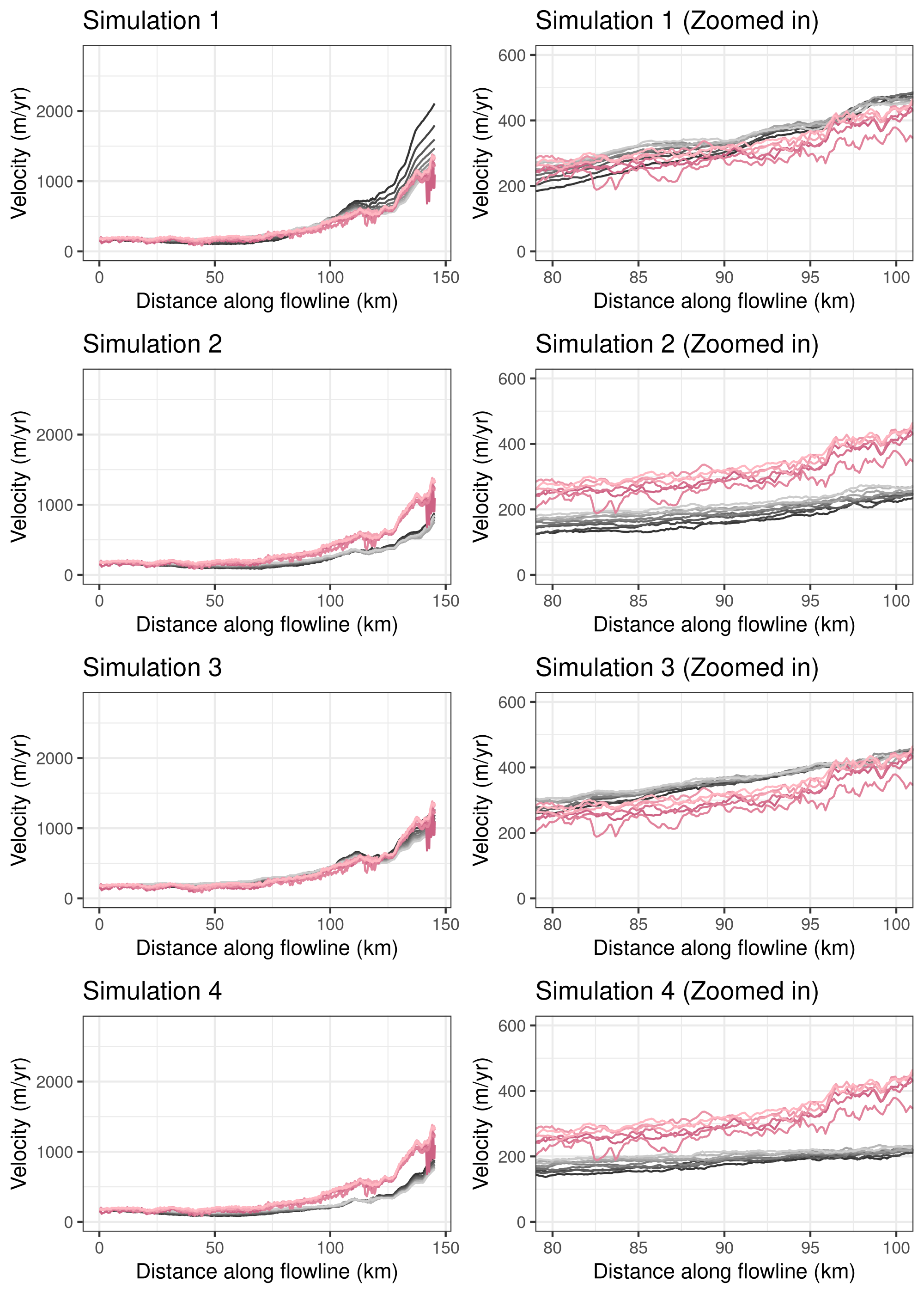}
    \caption{Adjusted observations of the surface velocity (left column), along the Thwaites Glacier flowline for the years 2010-2019, with zoomed-in plots (right column). The adjustment is done by subtracting the estimated model discrepancy from the observations. Red lines represent adjusted observations, and grey lines represent simulated observations from our prior distribution. In both cases, lighter lines indicate increasing time.}
    \label{fig:vel_obs_adjusted_20241111}
\end{figure}

Equation~\eqref{eq:model_discrepancy} implies that, when generating training data for the neural network, the estimated discrepancy $\hat{\bdelta}$ is added to simulations of the surface elevation in order to better align model output with actual observations; this ensures that the neural network is trained on ``adjusted'' model simulations. However, in practice, rather than adding this discrepancy term to each set of simulated observations, we instead train the neural network on simulations from the model in the same way as done in the OSSE, and only make use of the discrepancy after training by subtracting it from the observations. These ``adjusted'' observations are then used as input for the trained neural network to make inference on the underlying bed elevation and basal friction coefficient of Thwaites Glacier. 
Comparisons of the adjusted observations to simulations from the ice sheet model under $p(\btheta)$ are shown in Figures~\ref{fig:se_obs_adjusted_20241111} and~\ref{fig:vel_obs_adjusted_20241111}.

\subsection{Results}

\begin{figure}[t]
    \centering
    \includegraphics[width=\linewidth]{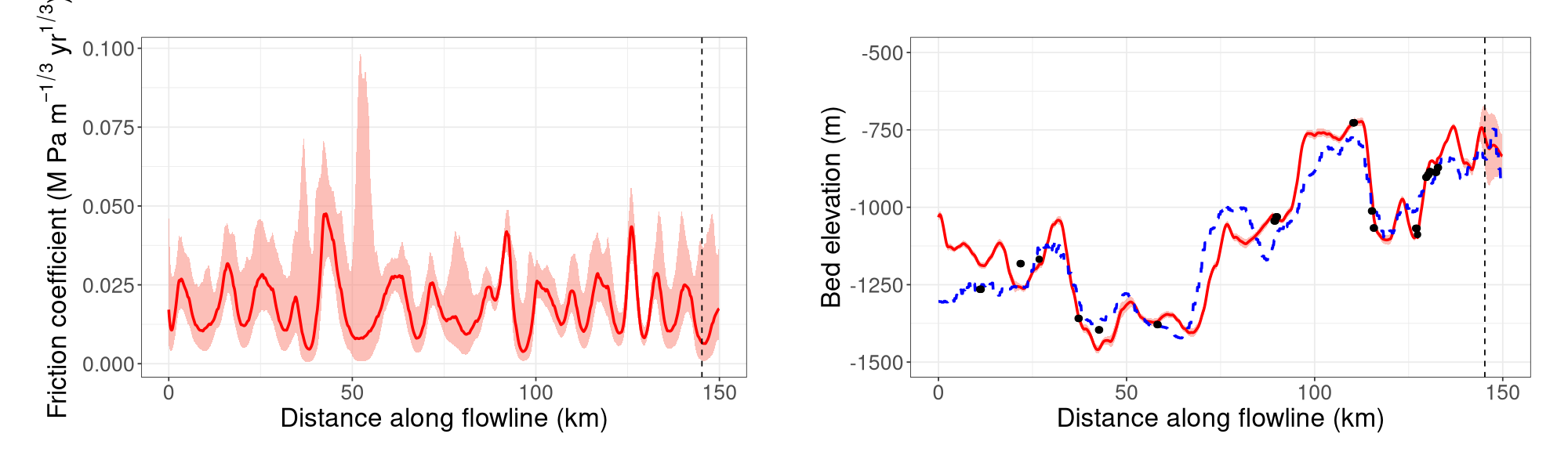}
    \caption{Posterior inferences for the basal friction coefficient (left) and bed elevation (right) along the Thwaites Glacier flowline. For both plots, the solid red line represents the posterior mean, while the red shaded region indicates the 95\% credible interval. Observations of the bed elevation from BedMap3 are overlaid as black dots. The blue dashed line shows bed elevation data from BedMachine v3. The observed grounding line is indicated with a dashed vertical line.}
    \label{fig:bed_fric_inference}
\end{figure}

Inferences for the bed elevation and basal friction coefficient along the Thwaites Glacier flowline are shown in Figure~\ref{fig:bed_fric_inference}. For the bed elevation, we compare the estimates from our NPI approach with the bed elevation from BedMachine version 3, which is another data product that reconstructs the Antartic bed topography using (deterministic) mass conservation methods~\citep{morlighem2020deep, bedmachinev3}. The BedMachine estimates are two-dimensional, and are ``mapped'' to the 1D grid of the flowline by taking the average of the nearest 4 estimates around each grid point on the flowline. In general, we find that the bed elevation estimate from our NPI captures the large-scale features (peaks and troughs) seen in the BedMachine estimates. However, there are some differences in the elevation of these peaks and troughs between the NPI estimates and the BedMachine estimates. These differences could be due to the different ice sheet models used for reconstruction, or differences in the climate forcings used in these models. Overall, the NPI estimates appear to be reasonable and capture key features of the bed elevation along the flowline.

Posterior uncertainty on the basal friction coefficient is higher in the ice sheet interior than close to the grounding line. There are neither observations nor ground truth for the basal friction coefficient, so instead we perform posterior predictive checks in the next section to assess the quality of our inferences. 

\subsubsection*{Posterior predictive checks}

For the year 2020, which was withheld for testing purposes, we compare the observed surface elevation and surface velocity data with the predictive distribution of the surface elevation and surface velocity. These predictive distributions are generated by running the ice sheet model (from the year 2010) 1000 times, each with a posterior sample of the parameters (bed elevation and basal friction coefficient) as input, extracting the surface elevation and surface velocity from each simulation, and then adjusting the ensemble with the discrepancy term described in Section S7.3.
We also compare the posterior predictive distribution to the prior predictive distribution, which is generated by running the ice sheet model with 1000 parameter samples drawn from our prior distribution.

\begin{figure}[ht]
    \centering
    \begin{subfigure}[b]{0.48\textwidth}
        \centering
        \includegraphics[width=\textwidth]{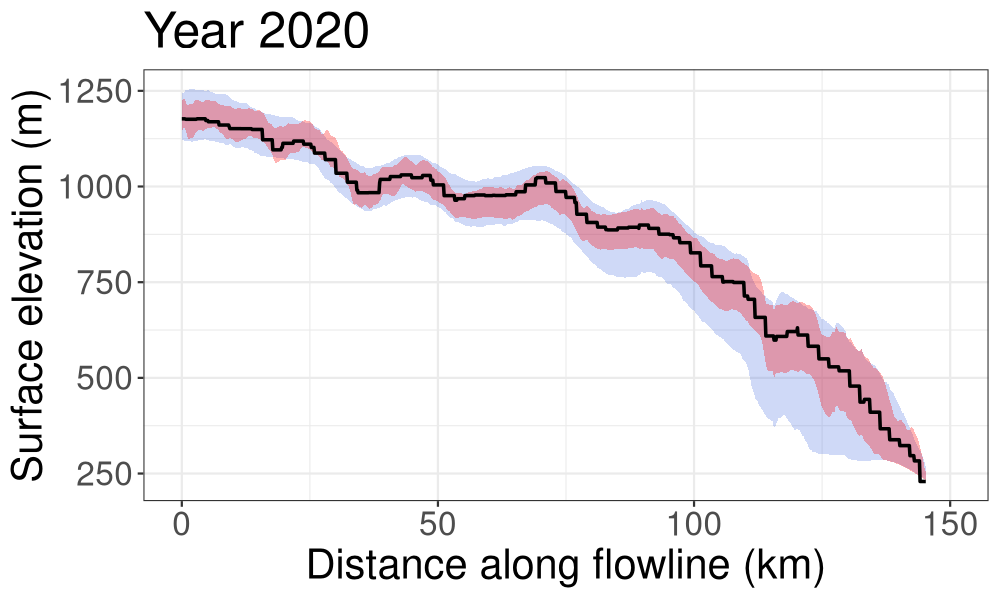}
        \caption{Surface elevation}
        \label{fig:post_pred_se_year2020_20241111}
    \end{subfigure}
    \hfill
    \begin{subfigure}[b]{0.48\textwidth}
        \centering
        \includegraphics[width=\textwidth]{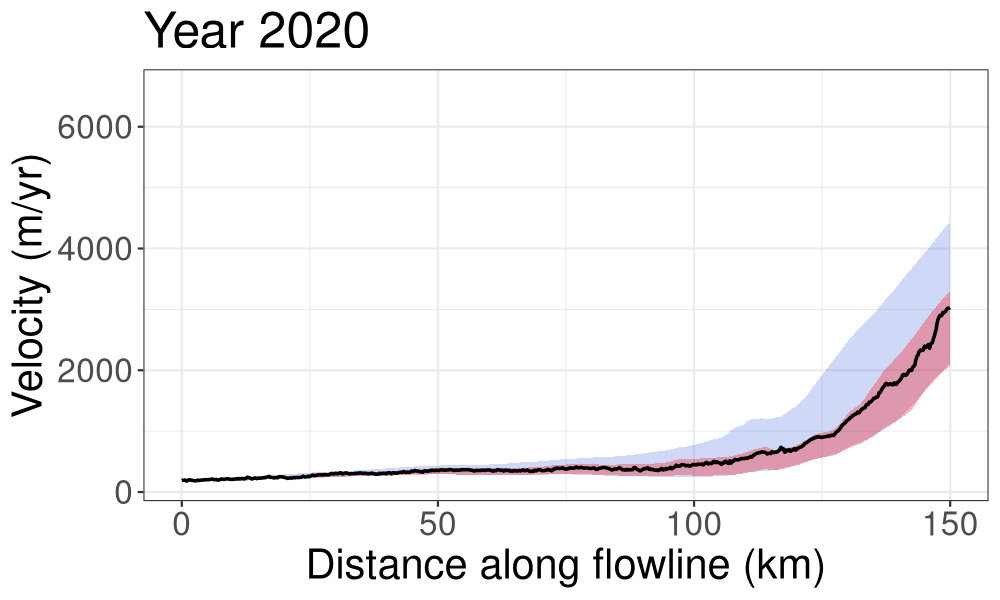}
        \caption{Surface velocity}
        \label{fig:post_pred_vel_year2020_20241111}
    \end{subfigure}
    \caption{Out-of-sample posterior predictive check for the surface elevation and surface velocity along the Thwaites Glacier flowline for the year 2020. Blue and red shaded regions indicate the 95\% prediction interval for the prior and the posterior predictive distributions, respectively. Black lines represent the observations from the MEaSUREs dataset.}
    \label{fig:post_pred_year2020}
\end{figure}

A plot of the pointwise $95\%$ prediction intervals for the prior and posterior predictive distribution of the surface elevation for the year 2020 is shown in Figure~\ref{fig:post_pred_year2020}. The observed surface elevation from the MEaSUREs dataset is also plotted for comparison.
These figures show that the posterior predictive distributions tighten around the observed surface elevation compared to the prior predictive distributions. As expected, once calibrated with estimates of the bed elevation and basal friction coefficient from the neural network, the ice sheet simulator yields outputs that match observations more closely than those obtained when using uncalibrated (prior) parameter values.

\subsection{Change in ice sheet area along the flowline}

Inferred bed elevation and basal friction coefficient are typically used to project ice sheet behaviour and estimate potential contributions to sea level rise. Here, we do a calculation to demonstrate this idea, bearing in mind that the model we employ is one-dimensional and therefore only able to yield estimates of the area, rather than the volume, of the slice of ice along the flowline.

As floating ice does not contribute to sea level, the contribution from a glacier or an ice sheet is calculated based on the so-called `volume above flotation', which is the volume of land ice that has potential to contribute to sea level rise if melted. The area above flotation (AAF) can be calculated using the following formula~\citep{goelzer2020calculating}:
\begin{equation}
    \text{AAF} = \sum_{j=0}^{J} \max \left[\left(h(s_j, t) + \min(0, b(s_j)) \frac{\rho_w}{\rho_i} \right), 0 \right] \Delta_s, \label{eq:vaf}
\end{equation}
where
$\Delta_s$ is the grid spacing.

We calculate the posterior predictive AAF along the flowline of Thwaites Glacier over the years 2010-2020 by applying~\eqref{eq:vaf} to posterior predictive samples of ice thickness, which are obtained by running the ice sheet simulator forward from the year 2010 using samples from our parameter posterior distribution.
We note that the estimated AAF for 2020 is an ``out-of-sample forecast'', as we do not make use of observations from 2020 in our inference procedure. For comparison, we also calculate the AAF based on the observed surface elevation from MEaSUREs, using our posterior mean as the estimate of the bed elevation in the calculation.

\begin{figure}[t]
    \centering
    \includegraphics[width=0.8\textwidth]{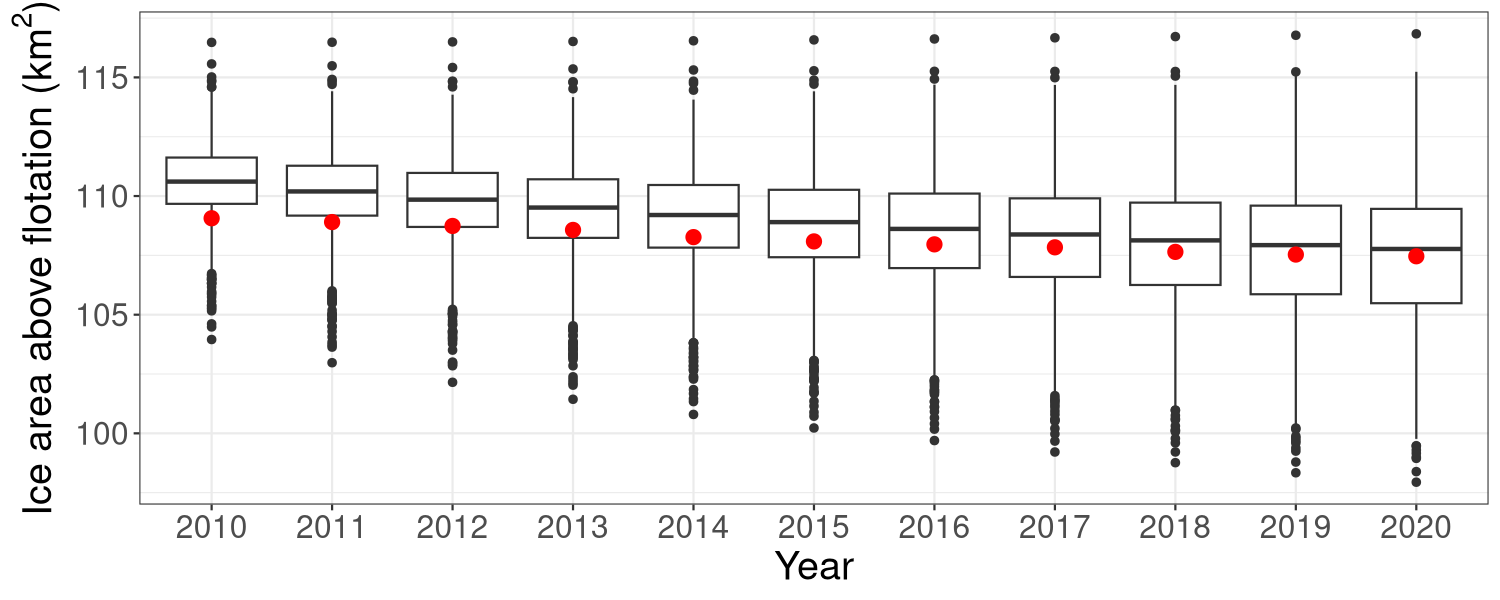}
    \caption{Box plots of posterior predictive area above flotation (AAF) along the Thwaites Glacier flowline, calculated using posterior samples of bed elevation and friction coefficient as input to the ice sheet model. Red dots indicate observed AAF, calculated using observed surface elevation from MEaSUREs and the posterior mean bed elevation.}
    \label{fig:ice_vaf_boxplot_20241111}
\end{figure}

The box plots of the posterior predictive samples of AAF for each year from 2010 to 2020 are shown in Figure~\ref{fig:ice_vaf_boxplot_20241111}. 
The decreasing trend in the predictive AAF is consistent with the decrease in the observed AAF, although the predicted rate of thinning over time seems to be slightly higher than what is observed. As the years pass, the predicted AAFs appear to align more closely to the observed AAFs; however, uncertainty around our AAF estimates also grows. This growing uncertainty is expected, and is likely a consequence of diverging model trajectories under the different parameter inputs (corresponding to samples from our posterior distribution).

\section{Conclusion}
\label{sec:nbe_conclusion}

In this article, we developed a two-stage Bayesian approach that combines NPI and an EnKF for making inference on parameters and latent states of complex, nonlinear SSMs, particularly the 1D SSA ice sheet model. In the first stage, we infer the posterior distribution over these spatially varying fields by training a convolutional neural network to take in observations at the ice surface (namely surface velocity and surface elevation), and output the parameters of the posterior distribution. In the second stage, we use posterior samples in an EnKF to make inference on the latent model states (thickness of the ice sheet), which are spatial processes that are temporally evolving. We show through a simulation study that our approach provides estimates of the parameters and latent states that are more accurate than the current state-of-the-art approach (the augmented EnKF). We then apply our approach to real data from a flowline in Thwaites Glacier, West Antarctica, and we obtain physically plausible estimates of the bed elevation and basal friction coefficient fields.

The work presented in this article can be extended in several directions. \Copy{posterior_investigate}{First, more flexible choices for the posterior approximation, such as Gaussian mixtures~\citep{papamakarios2016fast} or normalising flows~\citep{ardizzone2019analyzing, radev2022bayesflow} could be explored.} \Copy{amortisation}{Second, it may be possible to use the neural network in an \textit{amortised} manner: given new data from another flowline where the ice geometry and dynamics are similar to the one upon which the training data is based, the trained neural network will yield estimates of the bed elevation and basal friction coefficient in as little as one second.
The neural network could be made more robust to different glacier geometries and dynamics by training it on data that are representative of more than one glacier.} Third, the chosen 1D SSA model could be replaced by a higher-dimensional ice sheet model that can more closely simulate the geometry and dynamics of Antarctic (and other) ice sheets; this is a requirement for long-term sea-level rise projections. In principle, our approach can accommodate more complex ice sheet models, but multiple simulations from higher-dimensional models are more computationally intensive, although parallelisable. In addition, emulators of model output based on neural networks~\citep[e.g.,][]{jouvet2023IceflowModelEmulator} could be built to replace expensive simulators.
Having obtained encouraging results on a 1D model along a flowline, we are hopeful that our NPI framework can be extended to higher-dimensional ice sheet models in future work.

\section*{Significance Statement}
Mathematical models of ice flow are routinely used to project an ice sheet's contribution to sea level rise. The reliability of these projections depends on accurately calibrating the model parameters using observational data, which is challenging due to the complexity of the model equations and limited data availability for validation. This study develops a neural-network-based approach for efficiently calibrating ice sheet model parameters and quantifying their uncertainty. Through simulation experiments, we show that our approach yields more accurate parameter and uncertainty estimates than a current state-of-the-art approach. We then apply our approach to infer the parameters along a flowline in Thwaites Glacier, Antarctica, and discuss connections between inference on the parameters and sea level rise estimation.

\begin{acks}[Acknowledgments]
    This work was supported by Australian Research Council (ARC) SRIEAS Grant SR200100005 Securing Antarctica’s Environmental Future. 
    FSM was also supported under an ARC Discovery Early Career Research Award (DE210101433). This work contributes to delivering the Australian Antarctic Science Decadal Strategy. Further, this material is based upon work supported by the Air Force Office of Scientific Research under award number FA2386-23-1-4100 (AZ-M and NC). We are grateful to the editors and reviewers for their careful reading and helpful feedback on earlier submissions of this article. We would also like to thank Matthew Sainsbury-Dale for helping with the manuscript.

\end{acks}


\bibliographystyle{imsart-nameyear} 
\bibliography{bibliography.bib}       




\newpage
\supplementnumbering
\setcounter{page}{1}
\setcounter{section}{0}
\setcounter{equation}{0}
\setcounter{table}{0}
\setcounter{figure}{0}

\begin{supplement}
\begin{frontmatter}
\title{Neural posterior inference with state-space models for calibrating ice sheet simulator}
\runtitle{Supplementary Material}

\begin{aug}
    \author[A]{\fnms{Bao Anh}~\snm{Vu}},
    \author[B]{\fnms{Andrew}~\snm{Zammit-Mangion} \thanks{Corresponding author.}},
    \author[C]{\fnms{David}~\snm{Gunawan}},
    \author[D]{\fnms{Felicity}~\snm{S. McCormack}}
    \and
    \author[C]{\fnms{Noel}~\snm{Cressie}}

    \address[A]{Research School of Finance, Actuarial Studies and Statistics, Australian National University, Canberra, Australian Capital Territory, Australia 2600, and Securing Antarctica's Environmental Future, School of Mathematics and Physics, University of Wollongong, Wollongong, New South Wales, Australia 2500\printead[presep={,\ }] {e1}}

    \address[B]{School of Mathematics and Statistics, University of New South Wales, Sydney, New South Wales, Australia 2052, and Securing Antarctica's Environmental Future, School of Mathematics and Physics, University of Wollongong, Wollongong, New South Wales, Australia 2500\printead[presep={,\ }] {e2}}
    
    \address[C]{Securing Antarctica's Environmental Future, School of Mathematics and Physics, University of Wollongong, Wollongong, New South Wales, Australia 2500\printead[presep={,\ }] {e3,e5}}     


    \address[D]{Securing Antarctica's Environmental Future, School of Earth, Atmosphere and Environment, Monash University, Clayton, Kulin Nations, Victoria, Australia 3800\printead[presep={,\ }]{e4}}

\end{aug}

\end{frontmatter}
    To distinguish between the numbering of the sections, equations, tables and figures in the main paper and in the supplement, we use the following rule: sections, equations, tables and figures in the supplement are numbered with an `S' prefix, while those in the main paper are not. For example, the first section in the supplement is Section~S1, the first equation in the supplement is equation~(S1), and so on.
\end{supplement}


\section[SSA model: numerical solution using finite differences]{Numerical solution of the SSA model using finite differences}
\label{sec:finite_diff}
This section describes how to numerically solve the one-dimensional (1D) SSA model of Section 2.1,
which has no known analytical solution. Numerical solutions for this model can be obtained via finite differences~\citep[e.g.][]{van2013fundamentals} or finite elements~\citep[e.g.][]{braess2007finite}. Here, in the 1D case, we use finite differences.

The stress balance equation (1)
and the mass continuity equation (3) in the main paper
form a coupled system that can be iteratively solved for the surface velocity and ice thickness. One begins with some initial conditions for the velocity and ice thickness, and then the ice thickness is propagated forward in time using (3)
to obtain an updated thickness. This new thickness is then substituted into (1) 
and solved for an updated velocity. The process of alternately solving for the ice thickness and surface velocity can be repeated until the change in ice thickness is smaller than a set threshold, at which point the ice sheet is said to be in steady state.

We specify the following boundary conditions. At the ice divide, a Dirichlet boundary condition and a Neumann boundary condition are applied to the surface velocity and ice thickness, respectively:
\begin{align}
    u(0, t)                             & = 0, \quad t \geq 0,
    \label{eq:velocity_bc}                                                             \\
    \frac{\partial h(0, t)}{\partial s} & = 0, \quad t \geq 0. \label{eq:thickness_bc}
\end{align}
At the calving front $s = s_F$, we also apply a boundary condition based on hydrostatic equilibrium~\citep{schoof2007marine}:
\begin{equation}
    \label{eq:hydrostatic_equi}
    2 B h(s_F, t) \left| \frac{\partial u(s_F, t)}{\partial s} \right|^{(1-n_c)/n_c}  \frac{\partial u(s_F, t)}{\partial s} = \frac{1}{2} \left( 1 - \frac{\rho_i}{\rho_w} \right) \rho_i g h(s_F, t)^2, \quad t \geq 0,
\end{equation}
where the constants $B, \rho_i, \rho_w$, and $n_c$ are defined in Table~\ref{tab:ssa_parameters}.

Now consider an equally-spaced grid having $J+1$ grid intervals on the domain $D = [-\Delta_s/2, s_F + \Delta_s/2]$, with spacing $\Delta_s = s_F/J$. Define the centres of the grid intervals as the spatial nodes, and let $s_j = j \Delta_s, j = 0, 1, \dots, J$, so that $s_0 = 0$ corresponds to the inflow boundary and $s_J = s_F$ to the calving front. At time $t$, the ice thickness $h(\cdot, \cdot)$ and horizontal velocity $u(\cdot, \cdot)$ are approximated by vectors
\begin{equation}
    \h(t) \equiv (h_{0}(t), h_{1}(t), \dots, h_{J}(t))^\top, \quad \u(t) \equiv (u_{0}(t), u_{1}(t), \dots, u_{J}(t))^\top,
\end{equation}
where $h_{j}(t) \equiv h(s_j, t)$ and $u_{j}(t) \equiv u(s_j, t)$, for $j = 0,\dots, J$ and $t \geq 0$. We also approximate the surface elevation $z(\cdot, \cdot)$ by $\z(t) \equiv (z_{0}(t), z_{1}(t), \dots, z_{J}(t))^\top$, to be used later in solving (1) in the main paper.  


Using a forward finite difference scheme, the mass continuity equation (3) can be approximated by
\begin{equation}
    \label{eq:mass_cty_fd}
    \frac{h_{j}(t) - h_{j}(t-\Delta_t)}{\Delta_t}
    + \frac{u_{j}(t-\Delta_t)\, h_{j}(t-\Delta_t)
        - u_{j-1}(t-\Delta_t)\, h_{j-1}(t-\Delta_t)}{\Delta_s}
    = m_a,
\end{equation}
for spatial nodes $j = 1, \dots, J$ and timesteps $t = 2 \Delta_t, 3 \Delta_t, \dots, n_t \Delta_t$, with $\Delta_t = T/n_t$, where $n_t$ is a user-specified number of time steps. Rearrange~\eqref{eq:mass_cty_fd} as
\begin{align}
    \label{eq:mass_cty_rearr}
    h_j(t) & = h_j(t-\Delta_t) \nonumber                                                \\
           & \quad - \frac{\Delta_t}{\Delta_s} \left[ u_j(t-\Delta_t)\, h_j(t-\Delta_t)
        - u_{j-1}(t-\Delta_t)\, h_{j-1}(t-\Delta_t) \right]
    + m_a \Delta_t,
\end{align}
and let
\begin{equation*}
    a^{(1)}_{j-1, t-\Delta_t} \equiv \frac{\Delta_t}{\Delta_s} u_{j-1}(t-\Delta_t),
    \qquad
    a^{(2)}_{j, t-\Delta_t} \equiv 1 - \frac{\Delta_t}{\Delta_s} u_{j}(t-\Delta_t)
\end{equation*}
Then as in~\citet[][Section 6.3]{cressie2011statistics}, ~\eqref{eq:mass_cty_rearr} can be written as a matrix equation
\begin{align*}
    \begin{bmatrix}
        h_0(t)     \\
        h_1(t)     \\
        h_2(t)     \\
        \vdots     \\
        h_{J-1}(t) \\
        h_J(t)
    \end{bmatrix}
     & =
    \begin{bmatrix}
        0~~~      & a^{(1)}_{0, t-\Delta_t} + a^{(2)}_{1, t-\Delta_t} & 0~~~                    & \hdots                    & 0                         & 0                       \\
        0~~~      & a^{(1)}_{0, t-\Delta_t} + a^{(2)}_{1, t-\Delta_t}                                                                                                             \\
        0~~~      & a^{(1)}_{1, t-\Delta_t}                           & a^{(2)}_{2, t-\Delta_t}                                                                                   \\
        \vdots~~~ &                                                   & \ddots                  & \ddots                                                                          \\
        0~~~      &                                                   &                         & a^{(1)}_{J-2, t-\Delta_t} & a^{(2)}_{J-1, t-\Delta_t}                           \\
        0~~~      &                                                   &                         &                           & a^{(1)}_{J-1, t-\Delta_t} & a^{(2)}_{J, t-\Delta_t}
    \end{bmatrix}
    \begin{bmatrix}
        h_0(t-\Delta_t)     \\
        h_1(t-\Delta_t)     \\
        h_2(t-\Delta_t)     \\
        \vdots              \\
        h_{J-1}(t-\Delta_t) \\
        h_J(t-\Delta_t)
    \end{bmatrix} \\
     & \quad +
    \begin{bmatrix}
        m_a \Delta_t \\
        m_a \Delta_t \\
        m_a \Delta_t \\
        \vdots       \\
        m_a \Delta_t \\
        m_a \Delta_t
    \end{bmatrix},
\end{align*}
where, in the first two rows, we enforce $h_{0}(t) = h_{1}(t)$ to approximate the boundary condition~\eqref{eq:thickness_bc} for all $t$. The above matrix equation can be written as
\begin{equation}
    \label{eq:thickness_evol_model}
    \h(t) = \A(\u(t-\Delta_t)) \h(t-\Delta_t) + m_a \Delta_t \1.
\end{equation}
This forward difference scheme is numerically stable under the condition~\citep[][Section 3.3]{bueler2009shallow} 
$$\frac{\max_{j}(u_{j}(t - \Delta_t)) \Delta_t}{\Delta_s} < 1,$$
where $u_{j}(t)$ is the velocity at spatial node $j$ and time $t$, $\max_{j}(\cdot)$ is an operator that takes the maximum over all spatial nodes $j$.

Note that the matrix $\A(\cdot)$ is a function of $\u(t-\Delta_t)$, so the evolution of the ice thickness $\h(t)$ depends on $\u(t-\Delta_t)$. Thus, we need $\u(t)$ for $t \geq 0$. 
We explain how to find $\u(t)$ below.

Consider now the problem of solving for the surface velocity at a single time point $t$. Assuming that the ice thickness $h(\cdot, \cdot)$ and surface elevation $z(\cdot, \cdot)$ at time $t$ are known, the velocity can be found by solving (1) in the main paper. To reduce notational clutter, we suppress the time dependence for now and rewrite this equation as
\begin{equation}
    \label{eq:ssa_linear_time_suppressed}
    \frac{\partial }{\partial s} \left( 2 B h(s) \left| \frac{\partial u(s)}{\partial s} \right|^{(1-n_c)/n_c} \frac{\partial u(s)}{\partial s} \right) - c(s) |u(s)|^{m-1} u(s) = \rho_i g h(s) \frac{\partial z(s)}{\partial s}, 
\end{equation}
where $u(s), h(s)$ and $z(s)$ implicitly refer to the surface velocity, ice thickness and surface elevation at some fixed time $t$. Equation~\eqref{eq:ssa_linear_time_suppressed} can be solved with fixed-point iteration by starting with an initial guess $u^{(0)} (\cdot)$ for the velocity, and then iteratively computing $u^{(k)} (\cdot), k = 1, \dots$, until the difference between the previous iterate $u^{(k-1)} (\cdot, \cdot)$ and the current iterate $u^{(k)} (\cdot)$ falls below a chosen threshold. Each velocity iterate $u^{(k)} (\cdot)$ is obtained by solving the following approximation to~\eqref{eq:ssa_linear_time_suppressed}:
\begin{align}
    \label{eq:ssa_linear}
    \frac{\partial }{\partial s} \left( 2 B h(s) \left| \frac{\partial u^{(k-1)}(s)}{\partial s} \right|^{(1-n_c)/n_c} \frac{\partial u^{(k)}(s)}{\partial s} \right) & - c(s) |u^{(k-1)}(s)|^{m-1} u^{(k)}(s) \nonumber              \\
                                                                                                                                                                      & \quad \quad = \rho_i g h(s) \frac{\partial z(s)}{\partial s}, 
\end{align}
which is linear in $u^{(k)} (\cdot)$, with coefficients that depend on $u^{(k-1)}(\cdot)$.

To solve this equation for $u^{(k)} (\cdot)$, first let
\begin{align*}
    W^{(k-1)}(s)       & \equiv 2 B h(s) \left| \frac{\partial u^{(k-1)}(s)}{\partial s} \right|^{(1-n_c)/n_c}, \\
    \gamma^{(k-1)} (s) & \equiv c(s) |u^{(k-1)}(s)|^{m-1},                                                      \\
    \beta(s)           & \equiv \rho_i g h(s) \frac{\partial z(s)}{\partial s},
\end{align*}
for $s \in (0, s_F)$.
Equation~\eqref{eq:ssa_linear} then becomes
\begin{equation}
    \label{eq:force_balance_short}
    \frac{\partial }{\partial s} \left( W^{(k-1)}(s) \frac{\partial u^{(k)} (s)}{\partial s} \right) - \gamma^{(k-1)}(s) u^{(k)}(s) = \beta(s), \quad k = 1, \dots
\end{equation}
Using finite differences, \eqref{eq:force_balance_short} can be approximated by
\begin{equation}
    \label{eq:force_balance_discr}
    \frac{W^{(k-1)}_{j + 1/2} (u^{(k)}_{j+1} - u^{(k)}_{j}) - W^{(k-1)}_{j - 1/2} (u^{(k)}_{j} - u^{(k)}_{j-1}) }{(\Delta_s) ^2} - \gamma_{j}^{(k-1)} u^{(k)}_{j} = \beta_{j},
\end{equation}
where the terms $\{W^{(k-1)}_{j+1/2}\}$ are defined on mid-point locations as
\begin{equation*}
    W^{(k-1)}_{j+1/2} \equiv \begin{cases}
        2B h_{j+1/2} \left| \Delta_{u^{(k-1)}_j} \right|^{\frac{1-n_c}{n_c}}, & j = 0, \dots, J-1,        \\
        W^{(k-1)}_{j-1/2},                                                    & j = J, \quad k = 1, \dots
    \end{cases}
\end{equation*}
with
\begin{equation*}
    \Delta_{u^{(k-1)}_j} \equiv
    \begin{cases}
        \dfrac{u^{(k-1)}_1 - u^{(k-1)}_0}{\Delta_s},          & j = 0,                                \\
        \dfrac{u^{(k-1)}_{j+1} - u^{(k-1)}_{j-1}}{2\Delta_s}, & j = 1, \dots, J-1, \quad k = 1, \dots 
    \end{cases}
\end{equation*}
and
\begin{equation*}
    h_{j+1/2} \equiv \frac{1}{2} (h_{j+1} + h_j), \quad j = 0, \dots, J-1, \quad k = 1, \dots
\end{equation*}
as used in the software by~\cite{buelercode}.
The terms $\{W^{(k-1)}_{j + 1/2}\}$ are computed at mid-point locations for numerical stability~\citep{bueler2005exact}. The terms $\{\gamma^{(k-1)}_j\}$ and $\{\beta_j\}$ are defined as
\begin{equation*}
    \gamma^{(k-1)}_j \equiv c_j \abs{u^{(k-1)}_j}^{m-1}, \quad j = 0, \dots, J, \quad k = 1, \dots
\end{equation*}
with $c_j = c(s_j), j = 0, \dots, J$, and
\begin{equation*}
    \beta_j \equiv \rho_i g h_j \Delta_{z_j}, \quad j = 0, \dots, J, 
\end{equation*}
where 
\begin{equation*}
    \Delta_{z_j} \equiv
    \begin{cases}
        \dfrac{z_1 - z_0}{\Delta_s},          & j = 0,             \\
        \dfrac{z_{j+1} - z_{j-1}}{2\Delta_s}, & j = 1, \dots, J-1, \\
        \dfrac{z_J - z_{J-1}}{\Delta_s},      & j = J.
    \end{cases}
\end{equation*}

Rearrange~\eqref{eq:force_balance_discr} to
\begin{equation}
    \label{eq:force_balance_rearr}
    W^{(k-1)}_{j + 1/2} u^{(k)}_{j+1} - \left[ W^{(k-1)}_{j + 1/2} + W^{(k-1)}_{j - 1/2} + (\Delta_s) ^2 \gamma^{(k-1)}_j \right] u^{(k)}_j + W^{(k-1)}_{j - 1/2} u^{(k)}_{j-1} = (\Delta_s)^2 \beta_j.
\end{equation}

On our discretised domain, the velocity boundary condition~\eqref{eq:velocity_bc} can be written as
\begin{equation}
    \label{eq:velocity_bc_discr}
    u_0^{(k)} = 0, 
\end{equation}
and the hydrostatic equilibrium condition~\eqref{eq:hydrostatic_equi} can be approximated by introducing an extra spatial node $u_{J+1}$ such that
\begin{equation}
    \label{eq:hydrostatic_equi_discr}
    \frac{u^{(k)}_{J+1} - u^{(k)}_{J-1}}{2 \Delta_s} = u_F,
\end{equation}
where $u_F = \frac{\rho_i g}{4 B} \left( 1 - \frac{\rho_i}{\rho_w} \right) h_J$. By rearranging~\eqref{eq:hydrostatic_equi_discr} to make $u^{(k)}_{J+1}$ the subject of the formula and substituting into~\eqref{eq:force_balance_rearr} with $j = J$, this node can be rewritten in terms of other nodes, and then eliminated.

Equations~\eqref{eq:force_balance_rearr}, \eqref{eq:velocity_bc_discr} and \eqref{eq:hydrostatic_equi_discr} can be combined to form a tri-diagonal system
\begin{equation}
    \label{eq:u_tridiag}
    \begin{bmatrix}
        1                                                                                                    \\
        W^{(k-1)}_{1/2} & D^{(k-1)}_{11}  & W^{(k-1)}_{3/2}                                                  \\
                        & W^{(k-1)}_{3/2} & D^{(k-1)}_{22, t} & W^{(k-1)}_{5/2}                              \\
                        &                 & \ddots            & \ddots                                       \\
                        &                 & W^{(k-1)}_{J-3/2} & D^{(k-1)}_{(J-1)(J-1)} & W^{(k-1)}_{J - 1/2} \\
                        &                 &                   & W^{(k-1)^*}_{J-1/2}    & D^{(k-1)^*}_{JJ}
    \end{bmatrix}
    \begin{bmatrix}
        u^{(k)}_0     \\
        u^{(k)}_1     \\
        u^{(k)}_2     \\
        \vdots        \\
        u^{(k)}_{J-1} \\
        u^{(k)}_J     \\
    \end{bmatrix} =
    \begin{bmatrix}
        0                        \\
        (\Delta_s)^2 \beta_1     \\
        (\Delta_s)^2 \beta_2     \\
        \vdots                   \\
        (\Delta_s)^2 \beta_{J-1} \\
        \beta^*_J                \\
    \end{bmatrix},
\end{equation}
where
\begin{equation*}
    D^{(k-1)}_{jj, t} = - [W^{(k-1)}_{j+1/2} + W^{(k-1)}_{j-1/2} + (\Delta_s)^2 \gamma^{(k-1)}_j], \quad j = 1, \dots, J-1.
\end{equation*}
For $j = J$ (the last row in the system), there are adjustments for boundary conditions:
\begin{align*}
    W^{(k-1)^*}_{J-1/2} & = 2 W^{(k-1)}_{J-1/2},                                             \\
    D^{(k-1)^*}_{JJ}    & = - [2 W^{(k-1)}_{J-1/2} + (\Delta_s)^2 \gamma^{(k-1)}_J],         \\
    \beta_J^*           & = - 2 u^{(k)}_F \Delta_s W^{(k-1)}_{J-1/2} + (\Delta_s)^2 \beta_J.
\end{align*}
Solving this tri-diagonal system yields the vector ${\u^{(k)} (t) \equiv (u^{(k)}_0 (t), u^{(k)}_1 (t), \dots, u^{(k)}_J(t))^\top}$, where we now return the time point `$t$' to the notation. 



\section{Derivations: SSA model as a state space model}
\label{sec:ssa_state_space_supp}

In order to make statistical inference, we re-express the SSA model of Section S2.1 as a state-space model \citep[SSM; e.g.,][]{durbin2012time}. In a state-space model, the system ``states'' are not directly observed, but are instead inferred from noisy observations or through other quantities that are related to the latent states via some mathematical and/or statistical relations.




Recall that in Section S1 we discretise the 1D flowline into a grid with $J + 1$ equally-spaced intervals centred on $s_0, \dots, s_J$ (hereafter referred to as spatial ``nodes''), and denote the gridded ice thickness $h(\cdot, t)$ at time $t$ as a vector ${\h(t) \equiv (h_{0}(t), h_{1}(t), \dots, h_{J}(t))^\top}$. We similarly denote the surface velocity as $\u(t) \equiv (u_{0}(t), u_{1}(t), \dots, u_{J}(t))^\top$, for $t \geq 0$.
Now, define a set of discrete time indices $\mathcal{T} = \{1, \dots, T\}$, where the first index corresponds to the initial time point and each subsequent index denotes another time point in specified time units (e.g., years). Typically the gap between two consecutive time indices is many time steps $\Delta_t$ of the numerical model; for example, in the OSSE of Section 4,
$\Delta_t = 1/52$~year; in the real data example of Section 5,
$\Delta_t = 1/100$~year; and in both cases, $\mathcal{T} = \{1, \dots, T\}$ years. Let $\h_t$ and $\u_t$ denote the ice thickness and surface velocity at times $t \in \mathcal{T}$, respectively. Setting the latent process vector $\x_t = \h_t$, and accounting for unexplained process forcing, we derive the process model:
\begin{equation}
    \label{eq:ssa_process_model}
    \x_t = \mathcal{M}_t(\x_{t-1}, \btheta) + \v_t, \quad \v_t \sim \Gau(\0, \V_t), \quad t \in \mathcal{T},
\end{equation}
where $\mathcal{M}_t(\x_{t-1}, \btheta)$ captures the process of running~\eqref{eq:thickness_evol_model} at a fine temporal resolution ($\Delta_t$) for $n_t$ steps to propagate the state vector $x_t$ from year $t-1$ to year $t$).


To derive the data model, we write the vector of observed data at time $t \in \mathcal{T}$ as ${\y_t \equiv (\z_t^\top, \u_t^\top)^\top}$, where $\z_t \equiv (z_{0, t}, \dots, z_{J, t})^\top$ and $\u_t \equiv (u_{0, t}, \dots, u_{J, t})^\top$ are vectors of (noisy) observed surface elevation and observed surface velocity at each spatial node $j = 0, \dots, J$, respectively (see Section 2.2).
The data model depends on two parameters: the bed elevation $b(\cdot)$ and the basal friction coefficient $c(\cdot)$, which are assumed static in time. 
Then the data model can be written as:
\begin{align}
    \label{eq:ssa_data_model}
    \y_t =
    \begin{bmatrix}
        \z_t \\ \u_t
    \end{bmatrix} +
    \w_t
    = \mathcal{H}_t
    (\x_t, \btheta) + \w_t,
    \quad \w_t \sim \Gau(\0, \R_t), \quad t \in \mathcal{T}.
\end{align}
The measurement error covariance $\R_t = \diag(\R_{z_t}^\top, \R_{u_t}^\top)^\top$ is a diagonal matrix with elements $\R_{z_t} = (\sigma_{z_0}^2, \dots, \sigma_{z_J}^2)^\top$ and $\R_{u_t} = (\sigma_{u_0}^2, \dots, \sigma_{u_J}^2)^\top$, which contain the measurement error variances for the surface elevation and surface velocity at each grid node, respectively.
The observation operator $\mathcal{H}_t$ is composed of two parts: the (linear) mapping from the ice thickness to the surface elevation $\z_t$, given by (4) in the main paper,
and the (nonlinear) mapping from the ice thickness to the surface velocity $\u_t$, given by the stress balance equation (1) in the main paper. 
We describe each of these components below. 

Let $\b \equiv (b(s_0), \dots, b(s_J))^\top$. Since the mapping (4) from the ice thickness to the surface elevation depends on the position of the grounding line, we split each of the vectors $\z_t$, $\x_t$ and $\b$ into $\z_t = (\z_{g, t}^\top, \z_{f, t}^\top)^\top$, $\x_t = (\x_{g, t}^\top, \x_{f, t}^\top)^\top$, and $\b = (\b_g^\top, \b_f^\top)^\top$, where the subscripts $g$ and $f$ denote spatial nodes where the ice is grounded and floating, respectively. A discretised version of (4) can then be approximated in terms of $\z_t$, $\h_t$, and $\b$ as
\begin{equation}
    \label{eq:z_data_model}
    \z_t = \F_{1, t} \x_t + \F_{2, t} \b,
\end{equation}
where
\begin{equation}
    \F_{1, t} =
    \begin{bmatrix}
        \I_{g, t} & \0_{g, t}                                         \\
        \0_{f, t} & \left(1 - \frac{\rho_i}{\rho_w} \right) \I_{f, t}
    \end{bmatrix}, \quad
    \F_{2, t} =
    \begin{bmatrix}
        \I_{g, t} & \0_{g, t} \\
        \0_{f, t} & \0_{f, t}
    \end{bmatrix},
\end{equation}
and $\I_{g, t}, \I_{f, t}$, and $\0_{g, t}$, $\0_{f, t}$ are respectively identity and zero matrices of appropriate sizes, for $t \in \mathcal{T}$.
Note that, since the grounding line evolves with time, the spatial nodes where the ice is grounded changes at every time point $t$, but the length of $\z_t$ does not change.

Following a process of discretising the partial differential equation (1) (see~\eqref{eq:u_tridiag} in Section~\ref{sec:finite_diff}), the mapping from the ice thickness to the surface velocity can be written as
\begin{equation}
    \label{eq:u_data_model}
    \u_t = \D_t^{-1} (\x_t, \cb) \bbeta_t (\x_t, \b),
\end{equation}
where $\D_t$ is a tri-diagonal matrix whose elements are functions of the ice thickness $\x_t$ and the basal friction coefficient $\cb \equiv (c(s_0),\dots,c(s_J))^\top$, and $\bbeta_t$ is a vector that depends on $\x_t$ and the bed elevation $\b$. These quantities are implicitly defined in Section~\ref{sec:finite_diff}.


In the case where observations are not available at every grid point (a case we consider in the simulation study of Section 4), the observation model can be modified by pre-multiplying the observation operator with an incidence matrix $\C_t$ of size $N_{Y,t} \times (J+1)$, where $N_{Y,t}$ is the number of available observations, and also by modifying the measurement error covariance matrix accordingly:
\begin{equation}
    \label{eq:ssa_data_model_missing}
    \y_t =
    \C_t \mathcal{H}_t \left(
    \x_t,
    \btheta \right) +
    \w_t, \quad \w_t \sim \Gau(\0, \C_t \R_t \C_t^\top), \quad t \in \mathcal{T}.
\end{equation}
Here, for each year $t \in \mathcal{T}$, $\C_t$ is an incidence matrix where each row is a $(J+1) \times 1$ vector that contains the value one at the element/grid node where an observation is available, and zero elsewhere. Note that $\C_t$ is time-dependent as, for a given grid point, observations may be available at some time points and not at others.

\section{Ensemble Kalman filtering}
\label{sec:standard_enkf}

This section outlines the standard ensemble Kalman filter (EnKF) algorithm~\citep{evensen1994}, as well as the state-augmented EnKF approach for joint state and parameter estimation~\citep{anderson2001ensemble}.

\subsection{The standard EnKF algorithm}
While the EnKF is only statistically optimal
for linear, Gaussian SSMs, it has been found in practice that the EnKF can be effective in applications involving nonlinear models~\citep[e.g.,][]{evensen1997advanced, allen2003ensemble}.
The algorithm alternates between two steps: a ``forecast'' step in which the ensemble members are evolved in time using the physical model, and an ``analysis'' step in which the model forecast is adjusted using observations. The standard EnKF assumes that the parameters are known; we suppress the dependence of the data and process model on the parameters in this section for notational brevity. 
At each time $(t-1) \in \mathcal{T}$, the EnKF approximates the filtering distribution of the states, $p(\x_{t-1} \mid \y_{1:t-1})$, with an ensemble $\X_{t-1 \mid t-1} = \left(\x_{t-1 \mid t-1}^{(1)}, \dots, \x_{t-1 \mid t-1}^{(N_e)}\right)$ of size $\N_e$, where
\begin{equation}
    \x_{t-1 \mid t-1}^{(i)} \sim \Gau(\x_{t-1 \mid t-1}, \P_{t-1 \mid t-1}), \quad i = 1, \dots, N_e, \quad t \in \mathcal{T}.
\end{equation}
Here, the notation $\x_{t-1 \mid t-1}$ denotes the expectation of the latent variables at time $t-1$ given all observations up to time $t-1$, and $\P_{t-1 \mid t-1}$ is a sample-based covariance matrix that is initialised to an initial value at time $t = 1$ and approximated based on the ensemble members in subsequent timesteps. In the forecast step, this ensemble is propagated in time by applying the evolution equation (6) in the main paper 
to each member to yield a forecast (prior) ensemble:
\begin{equation}
    \x_{t \mid t - 1}^{(i)} = \mathcal{M}_{t-1} (\x_{t-1 \mid t-1}^{(i)}) + \v_t, \quad \v_t \sim \Gau(\0, \V_t),  \quad i = 1, \dots, N_e, 
\end{equation}
where $\mathcal{M}_t(\cdot)$ is the state evolution operator at time $t$ (see Section~\ref{sec:ssa_state_space_supp}), and $\V_t$ is the process error covariance matrix. 
The forecast distribution $p(\x_t \mid \y_{1:t-1})$ is then approximated by a Gaussian distribution with the (sample-based) mean and covariance matrix estimated respectively from the forecast ensemble as
\begin{equation}
    \hat{\x}_{t \mid t-1} = \frac{1}{N_e} \sum_{i=1}^{N_e} \x_{t \mid t-1}^{(i)}, 
\end{equation}
and
\begin{equation}
    \widehat{\P}_{t \mid t-1} = \frac{1}{N_e - 1} \sum_{i=1}^{N_e}  (\x_{t \mid t-1}^{(i)} - \hat{\x}_{t \mid t-1})(\x_{t \mid t-1}^{(i)} - \hat{\x}_{t \mid t-1})^\top. 
\end{equation}
In the analysis step, the forecast ensemble is adjusted with new observations $\y_t$ by computing
\begin{equation}
    \x_{t \mid t}^{(i)} = \x_{t \mid t-1}^{(i)} + \widehat{\K}_t (\y_t - \Tilde{\y}_t^{(i)}), \quad i = 1, \dots, N_e, 
\end{equation}
where $\Tilde{\y}_t^{(i)} = \mathcal{H}_t (\x_{t \mid t-1}^{(i)}) + \w_t^{(i)}, \w_t^{(i)} \sim \Gau(\0, \R_t)$, and $\widehat{\K}_t$ is a sample-based approximation of the Kalman gain matrix given by~\citet{houtekamer2001sequential}:
\begin{equation}
    \widehat{\K}_t = \A_1 (\A_2 + \R_t)^{-1}, 
\end{equation}
where the matrices $\A_1$ and $\A_2$ are computed as
\begin{equation*}
    \A_1 = \frac{1}{N_e - 1} \sum_{i=1}^{N_e} (\x_{t \mid t-1}^{(i)} - \hat{\x}_{t \mid t-1})(\mathcal{H}_t (\x_{t \mid t-1}^{(i)}) - \overline{\mathcal{H}_t (\x_{t \mid t-1})})^\top,
\end{equation*}
and
\begin{equation*}
    \A_2 = \frac{1}{N_e - 1} \sum_{i=1}^{N_e} (\mathcal{H}_t (\x_{t \mid t-1}^{(i)}) - \overline{\mathcal{H}_t (\x_{t \mid t-1})})(\mathcal{H}_t (\x_{t \mid t-1}^{(i)}) - \overline{\mathcal{H}_t (\x_{t \mid t-1})})^\top,
\end{equation*}
with
\begin{equation*}
    \overline{\mathcal{H}(\x_{t \mid t-1})} \equiv \frac{1}{N_e} \sum_{i=1}^{N_e} \mathcal{H}(\x_{t \mid t-1}^{(i)}). 
\end{equation*}
The filtering distribution of the states at time $t$, $p(\x_t \mid \y_{1:t}, \btheta)$, is now approximated by a Gaussian distribution with mean and covariance estimated respectively from the updated ensemble:
\begin{equation}
    \hat{\x}_{t \mid t} = \frac{1}{N_e} \sum_{i=1}^{N_e} \x_{t \mid t}^{(i)}, 
\end{equation}
and
\begin{equation}
    \widehat{\P}_{t \mid t} = \frac{1}{N_e - 1} \sum_{i=1}^{N_e} (\x_{t \mid t}^{(i)} - \hat{\x}_{t \mid t})(\x_{t \mid t}^{(i)} - \hat{\x}_{t \mid t})^\top. 
\end{equation}
The algorithm alternates between the forecast and the analysis steps until the last observation has been assimilated.

\subsection{State-augmented EnKF}

The EnKF algorithm relies on known values of the parameters $\btheta$. In the case of unknown $\btheta$, the state vector in the EnKF can be \textit{augmented} with the parameters, so that model states and parameters can be updated at the same time. The parameters are treated as static, but it is common for others to assume a persistence model that allows them to be temporally indexed \citep[e.g.,][]{baek2006local, zupanski2006model}:
\begin{equation}
    \btheta_t = \btheta_{t-1}, \quad t \in \mathcal{T}\backslash \{1\}.
\end{equation}
The state vector can then be written as $\x^*_t = (\x_t^\top, \btheta_t^\top)^\top$ for $t \in \mathcal{T}$, and the forecast and update steps are performed based on the augmented state vector $\x^*_t$. This state-augmentation strategy was employed by \cite{gillet2020assimilation} to make inference on the bed elevation and basal friction coefficient in a 1D SSA model. In practice, we find that the persistence model leads to an underestimation of parameter uncertainty. We show in Section 4
of the main paper that our two-stage inference approach offers superior accuracy and uncertainty quantification to the state-augmented EnKF.

\section{Simulating a steady-state ice sheet}
\label{sec:steady_state_ice_sheet}
This section describes the process of simulating a steady-state ice sheet for use as initial conditions for the OSSE in Section 4 of the main paper. Simulating a steady-state ice sheet involves running the 1D SSA model for (typically) several thousands of simulation years until the maximum rate of change in ice thickness over the entire model domain is smaller than a chosen threshold. The model needs to be initialised with some initial ice thickness and velocity, as well as some specification of the bed elevation and basal friction coefficient.

For this simulation study, we consider a flowline $D =[0,s_F]$ that starts from the ice divide $s = 0$ to a fixed calving front at $s_F = \num{800000}$~m. We discretise this domain $D$ into equal segments each of length $\Delta_s = 400$~m, with the first grid point set to $s_0 = 0$, and the final grid point set to $s_J = s_F = \num{800000}$~m, where $J = 2000$.
We set the initial ice sheet geometry to
\begin{equation}
    h_{init}(s) = 2000 - \frac{2000}{\num{800000}}s, \quad s \in D,
\end{equation}
in units of m, and the initial velocity to
\begin{equation}
    u_{init}(s) = 0.001s, \quad s \in D,
\end{equation}
in units of m yr$^{-1}$.

For the bed elevation, we follow the approach of~\cite{durand2011impact} to generate a bed geometry $b(\cdot)$ in units of m that is the sum of a trend component $b_{trend}(\cdot)$ and a small-scale roughness component $b_r(\cdot)$:
\begin{equation*}
    b(s) = b_{trend}(s) + b_r(s), \quad s \in D,
\end{equation*}
where the trend is a piecewise linear function
\begin{equation*}
    b_{trend} (s) = \begin{cases}
        - 600 + \frac{s}{1000},          & s \leq \num{450000}, \\
        - 150 - 5(\frac{s}{1000} - 450), & s > \num{450000},
    \end{cases}
\end{equation*}
and the small-scale roughness component $b_r(\cdot)$ is computed at 200~m resolution using random midpoint displacement~\citep{fournier1982computer}, an algorithm commonly used in artificial landscape generation. In 1D, this algorithm finds the midpoint of a segment, adds a random value drawn from a $\Gau(0, \sigma_r^2)$ distribution to the bed elevation at the midpoint. The algorithm starts with the segment equal to the entire domain $D$, and the process of adding noise is then repeated for each half-segment. The procedure of splitting and adding noise to the midpoints is done for a pre-specified number of recursions. With every recursion, the standard deviation $\sigma_r$ is decreased by a factor of $2^\alpha$, where $\alpha$ is a roughness factor. In this study, we use 12 recursions, with an initial standard deviation $\sigma_r = 500$~m and a roughness factor $\alpha = 0.7$.

For the basal friction coefficient $c(\cdot)$ (in units of MPa m$^{-1/3}$ yr$^{1/3}$), we follow~\cite{gillet2020assimilation} and use the following function, which is a product of two sine waves with different wavelengths:
\begin{equation*}
    c(s) = 0.02 + 0.015 \sin \left(5 \frac{2 \pi s}{l_D} \right) \sin \left(100 \frac{2 \pi s}{l_D} \right), \quad s \in D,
\end{equation*}
with $l_D = 800$ km.

The values of other parameters needed for simulating the steady-state ice sheet are given in Table~\ref{tab:ssa_parameters}. We initialise the 1D SSA model with the specified initial ice thickness, velocity, bed elevation, and basal friction coefficient, along with the parameters in Table~\ref{tab:ssa_parameters}, and then run the model until the maximum rate of change in ice thickness across the whole domain is less than $0.05$ m per year. 
The final time point in this simulation gives our steady-state ice sheet. In our simulation, the ice sheet reached steady-state after approximately 3220 simulation years.

\begin{table}[t]
    \centering
    \caption{Parameter values for the steady-state simulation in the OSSE of Section 4, following \cite{gillet2020assimilation}.}
    \begin{tabular}{|c|r|l|}
        \hline
        Parameter                 & Parameter value & Unit             \\
        \hline
        $B$ (before steady-state) & 0.4             & $\stiffnessunit$ \\
        $B$ (after steady-state)  & 0.3             & $\stiffnessunit$ \\
        $n_c$                     & 3               & N/A              \\
        $m$                       & 1/3             & N/A              \\
        $a_s$                     & 0.5             & m yr$^{-1}$      \\
        $a_b$                     & 0.0             & m yr$^{-1}$      \\
        $\rho_i$                  & 910.0           & kg m$^{-3}$      \\
        $\rho_w$                  & 1028.0          & kg m$^{-3}$      \\
        $\Delta_t$                & 1/52            & yr               \\
        $\Delta_s$                & 400             & m                \\
        \hline
    \end{tabular}
    \label{tab:ssa_parameters}
\end{table}

\section{Training data generation: dimensionality reduction for the parameters}
\label{sec:dim_reduc}
This section describes how we reduce the dimensionality of the parameters (bed elevation and basal friction coefficient) using basis functions. We apply this dimensionality reduction approach for both the OSSE of Section~4 and the real data example of Section~5.

We generate bed elevation simulations through the conditional simulation process outlined in Section 4.1.1. 
We then de-mean the $n$th simulation by computing
\begin{equation*}
    \tilde{b}^{(n)} = b^{(n)}(s) - \frac{1}{N} \sum_{n=1}^N b^{(n)}(s), \quad n = 1, \dots, N, \quad s \in D,
\end{equation*}
to obtain the ``fluctuations'' around the mean bed elevation for all $s \in D$.
Then we represent each of these fluctuations as
\begin{align}
    \tilde{b}^{(n)}(s) & = \bphi_{b}(s) \btheta^{(n)}_{b}, \quad n = 1, \dots, N,
\end{align}
where $\bphi_{b}(s)$ is $p_1$-dimensional comprising bisquare basis functions of the form:
\begin{equation}
    \label{eq:bed_bisquare}
    \phi_{b}(s) = \left(1 - \left(\frac{\abs{s - s_i^*}}{R_b} \right)^2  \right)^2 \mathbb{I} (\abs{s - s_i^*} < R_b), \quad i = 1, \dots, p_1,
\end{equation}
where $s_i^*$ denotes the $i$th basis function centroid, $R_b = 5000$~m, and $\btheta_{b}^{(n)}$ is a $p_1$-dimensional vector of corresponding basis function weights. The value of $R_b$ was chosen to reproduce realistic bed elevations. We simulate the basal friction coefficient as detailed in Section 4.1.1, 
replacing any negative values with a small positive constant (we use $10^{-6}$). We then take the logarithm of each realisation to obtain $\tilde{c}^{(n)}(s) = \log(c^{(n)}(s))$, for $n = 1, \dots, N$, and then represent each realisation as $\tilde{c}^{(n)}(s) = \bphi_{c}(s) \btheta^{(n)}_{c}$, where $\bphi_{c}(s)$ is $p_2$-dimensional and of the form~\eqref{eq:bed_bisquare}, with $R_b$ replaced by $R_c = 8000$~m,
and $\btheta_{c}^{(n)}$ is a $p_2$-dimensional vector of corresponding basis function weights. Here, $R_c$ was chosen so that the basis function representations closely reproduce the original (simulated) basal friction coefficients.
For simulations of both the bed elevation and basal friction coefficient, we use $p_1 = p_2 = 150$ basis functions, evenly spaced along the 1D spatial domain, with the centres of the first and the last basis functions placed at $s_0 = 0$~m and $s_J = \num{800000}$~m, respectively. (Note, however, that $p_1$ and $p_2$ need not be the same in general.) 

\section{Details for training the CNN}

This section provides additional details on the standardisation of the training data and the neural network architecture used in both the OSSE of Section~4 and the real data example in Section~5.

\subsection{Data standardisation}
\label{sec:data_std}
The simulated data in this study consists of $\num{50000}$ sets of $300$ basis function coefficients for the bed elevation and basal friction coefficient, with corresponding surface elevation and surface velocity data. Since the neural network is sensitive to the scaling of the data, we first perform a standardisation step. The surface elevation at grid point $s_j$ at time $t$ from the $n$th simulation is standardised according to
\begin{equation}
    \tilde{z}^{(n)}_{j, t} = \frac{z^{(n)}_{j, t} - m_z}{s_z}, \quad n = 1, \dots, N, \quad j = 0, \dots, J, \quad t \in \mathcal{T}.
\end{equation}
where
\begin{align}
    m_z & = \frac{1}{N(J+1)T} \sum_{n=1}^N \sum_{j=0}^J \sum_{t=1}^T z^{(n)}_{j, t},                      \\
    s_z & = \sqrt{\frac{1}{N(J+1)T - 1} \sum_{n=1}^N \sum_{j=0}^J \sum_{t=1}^T (z^{(n)}_{j, t} - m_z)^2}.
\end{align}
The surface velocity data is similarly standardised. The resulting standardised data are then used as input for the neural network.

We also standardise the training output (basis function coefficients of the bed elevation fluctuations) according to
\begin{equation}
    \tilde{\theta}^{(n)}_{b, p} = \frac{\theta^{(n)}_{b, p} - m_{\theta_b}}{s_{\theta_b}}, \quad n = 1, \dots, N,
\end{equation}
where
\begin{equation}
    m_{\theta_b} = \frac{1}{N p_1} \sum_{n=1}^N \sum_{p=1}^{p_1} \theta^{(n)}_{b, p}, \quad
    s_{\theta_b} = \sqrt{\frac{1}{N p_1 - 1} \sum_{n=1}^N \sum_{p=1}^{p_1} (\theta^{(n)}_{b, p} - m_{\theta_b})^2}. \label{eq:standardise_output}
\end{equation}
The basis function coefficients for the log-transformed basal friction coefficient are similarly standardised.

\subsection{Neural network architecture and training}
\label{sec:nn_architecture}
We use a neural network with the following architecture:
\begin{itemize}
    \item The input is a multidimensional array of size $(J+1) \times T \times N_C$ where $J = 2000$, $T = 21$ time points (years, including the initial year) for the simulation study and $T = 10$ years for the real data example, and $N_C = 2$ ``channels'' (one for the surface elevation, and one for the surface velocity).
    \item The hidden layers are composed of three convolutional layers of width $32, 64$, and $128$, with convolutional kernels of size $5 \times 5$, $3 \times 3$, and $2 \times 2$, respectively. We use ReLU activation functions:
          \begin{equation}
              g(z) \equiv \max(0, z). \label{eq:relu}
          \end{equation}
    \item Each convolutional layer is followed by a max-pooling layer with a $2 \times 2$ kernel and a stride of 2. 
    \item The output from the last max-pooling layer is flattened and passed through a dropout layer with dropout rate $0.5$, and finally it is passed to an output layer of size $d_{out}$, where $d_{out} = 3(p_1 + p_2) - 2$ is the total number of elements needed to construct the mean $\bmu_{\bgamma}(\cdot)$ and Cholesky factor $\L_{\bgamma} (\cdot)$. 

\end{itemize}

The neural network is trained using the ADAM optimiser~\citep{kingma2014adam} with a learning rate of 0.00005 and a batch size of 64. The optimiser is run for 30 epochs, and and the model weights and biases at the epoch with the lowest validation loss is chosen as our trained model. The neural network implementation can be found in the R scripts \texttt{run\_cnn\_posterior.R} and \texttt{create\_model.R} on the github repository \url{https://github.com/bao-anh-vu/npi_icesheet/}.
\section{Thwaites Glacier experiment: Training data generation}
\label{sec:train_data_gen_real_supp}

\subsection{Simulating parameters}
To obtain simulations of the bed elevation conditional on BedMap3 observations, we fit to the bed elevation observations a Gaussian process with an exponential covariance function
\begin{equation}
    K(s_i, s_j) = \varsigma_b^2 \exp \left(-\frac{\abs{s_i - s_j}}{\ell_b} \right), \quad i, j = 0, \dots, J,
\end{equation}
where the parameters $\varsigma_b$ and $\ell_b$ are chosen by maximum likelihood. We then simulate bed elevation profiles from the fitted Gaussian process model. For the basal friction coefficient, we unconditionally simulate from a Gaussian process with a mean of $0.02 \fricunit$ and an exponential covariance function in the same way as shown in the OSSE; see Sections~4.1.1
and~\ref{sec:dim_reduc}.

\subsection{Simulating observations}
For each pair of parameters (bed elevation and basal friction coefficient), we run an ice sheet model simulation. We initialise the model with the surface elevation from the year 2010. For the surface velocity, since observations in the early years (2010--2014) are sparse, we instead use the average surface velocity over the years 2015--2019 as the initial surface velocity. 
We note that surface elevation data is only available on grounded ice (for floating ice, only data on the monthly change in surface elevation is available, but not on the absolute elevation itself). Therefore, for each simulation, we reconstruct the surface elevation over the floating portion of the flowline based on the ice thickness at the grounding line $h_g$, under the assumption that ice thickness decays exponentially from the grounding line towards the ocean:
\begin{equation}
    h(s_j) = h_{g} \exp(-0.002 (j - j_g)), \quad j = j_g, j_g+1, \dots, J,
\end{equation}
where $j_g$ is the index of the grid point at the grounding line location, and the value $-0.002$ is chosen to achieve a gradual decrease in ice thickness towards the calving front. For different simulations, the thickness at the grounding line varies, but typically this procedure results in an ice shelf that is approximately $900$~m thick at the grounding line and thins to approximately $500$~m at the calving front.

For the climate forcings, we take the average snow accumulation rate from \citet{racmo2} over the years 1979--2016 at each grid point along the flowline and use this temporal average as the snow accumulation rate for all years. For the grounded portion of the flowline, we assume no basal melt. For the floating ice shelf, we use a constant basal melt rate obtained by taking the average of the rates from MEaSUREs~\citep{measures_basal_melt} over the extent of the ice shelf and over time (1992--2017), which results in a rate of approximately $-33$m per year across the shelf. We set the ice stiffness to $B = 0.6 \stiffnessunit$, which is higher than the value used in the simulation study. This was needed because the 1D SSA model does not capture all the physical processes affecting ice flow in Thwaites Glacier, such as lateral drag from valley walls. The increased ice stiffness prevents excessive ice retreat. 



\begin{figure}[t]
    \centering
    \includegraphics[width=\textwidth]{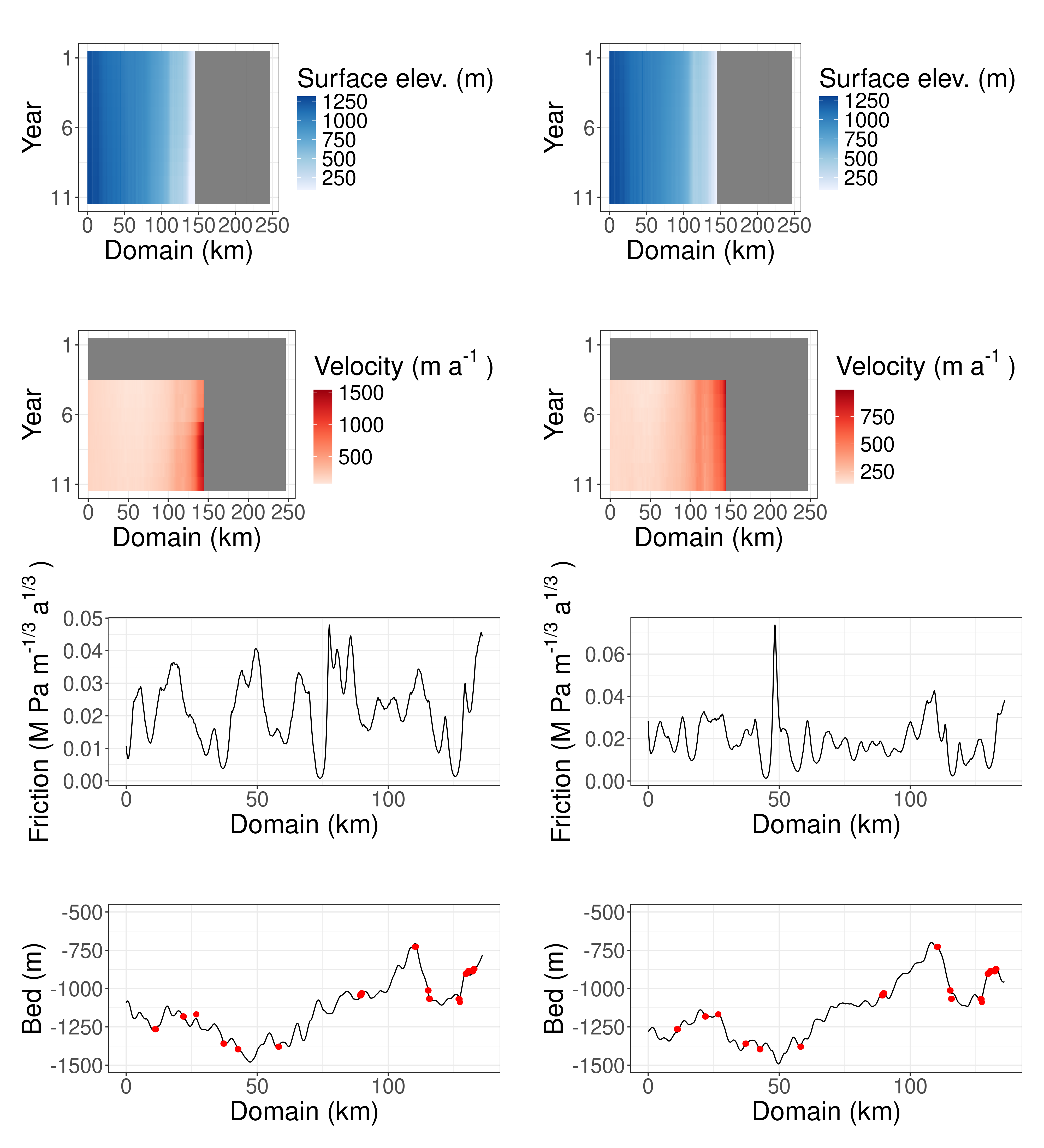}
    \caption{Plots of the simulated parameters (bed elevation and basal friction coefficient, shown in the last two rows) and corresponding data (surface elevation and surface velocity, shown in the first two rows) for two samples from the training dataset for the real-data example in Section~5. Grey areas in the surface elevation and surface velocity plots indicate locations with missing data. Red dots in the last row indicate bed elevation observations from BedMap3.}
    \label{fig:sim_params}
\end{figure}

We ignore the first 5 years of each simulation, treating them as ``warm up'' to remove any non-physical artefacts that may result from inconsistencies between data products, and then we run the model for an additional 10 years to simulate the ice evolution over the years 2010--2019. From each simulation, we extract the annual surface velocity and elevation and add Gaussian measurement error to mimic actual observations. For this experiment, we assume the measurement error has zero mean and short-range correlations. The surface elevation measurement error is generated from a Gaussian distribution with an exponential covariance function given by
\begin{equation}
    \text{Cov}(s_i, s_j) = \varsigma^2 \exp\left(-\frac{|s_i - s_j|}{\rho}\right), \label{sec:real_measurement_err}
\end{equation}
where $\rho = 5$ km, and $\varsigma = 0.5$ m. The value for the measurement error standard deviation $\varsigma$ is chosen to reflect the typical magnitude of measurement errors from the MEaSUREs dataset. For the surface velocity, the added error is assumed to come from a Gaussian distribution, $\v_u \sim \Gau(\0, \V_u)$, where $\V_u = \D_{u} \K_u \D_{u}$ and $\K_u$ is a matrix whose $(i, j)$th element also comes from exponential covariance function~\eqref{sec:real_measurement_err} with $\varsigma = 1$. The matrix $\D_{u}$ is a diagonal matrix in which the diagonal values $d_{u,j}, j = 0, \dots, J$ contain measurement error standard deviations that are chosen to reflect the magnitude of errors from MEaSUREs. Specifically, we map the recorded measurement error from the MEaSUREs dataset to the flowline by taking the average measurement error of the nearest four observations around a grid node $s_j$, then set the value of $d_{u, j}$ to one-third of the averaged measurement error (based on the property of the Gaussian distribution where $99\%$ of the probability mass lies within 3 standard deviations of the mean).
The resulting noisy surface elevation and surface velocity, along with the corresponding bed elevation and basal friction coefficient inputs, form the training data for the neural network (following standardisation; see Section~\ref{sec:data_std}). Two examples of simulated surface elevation and velocity time series and their corresponding bed elevation and basal friction coefficient inputs are shown in Figure~\ref{fig:sim_params}.

\subsection{Discrepancy between modelled and observed data}
\label{sec:model_discr}
\begin{figure}
    \centering
    \includegraphics[width=0.85\textwidth]{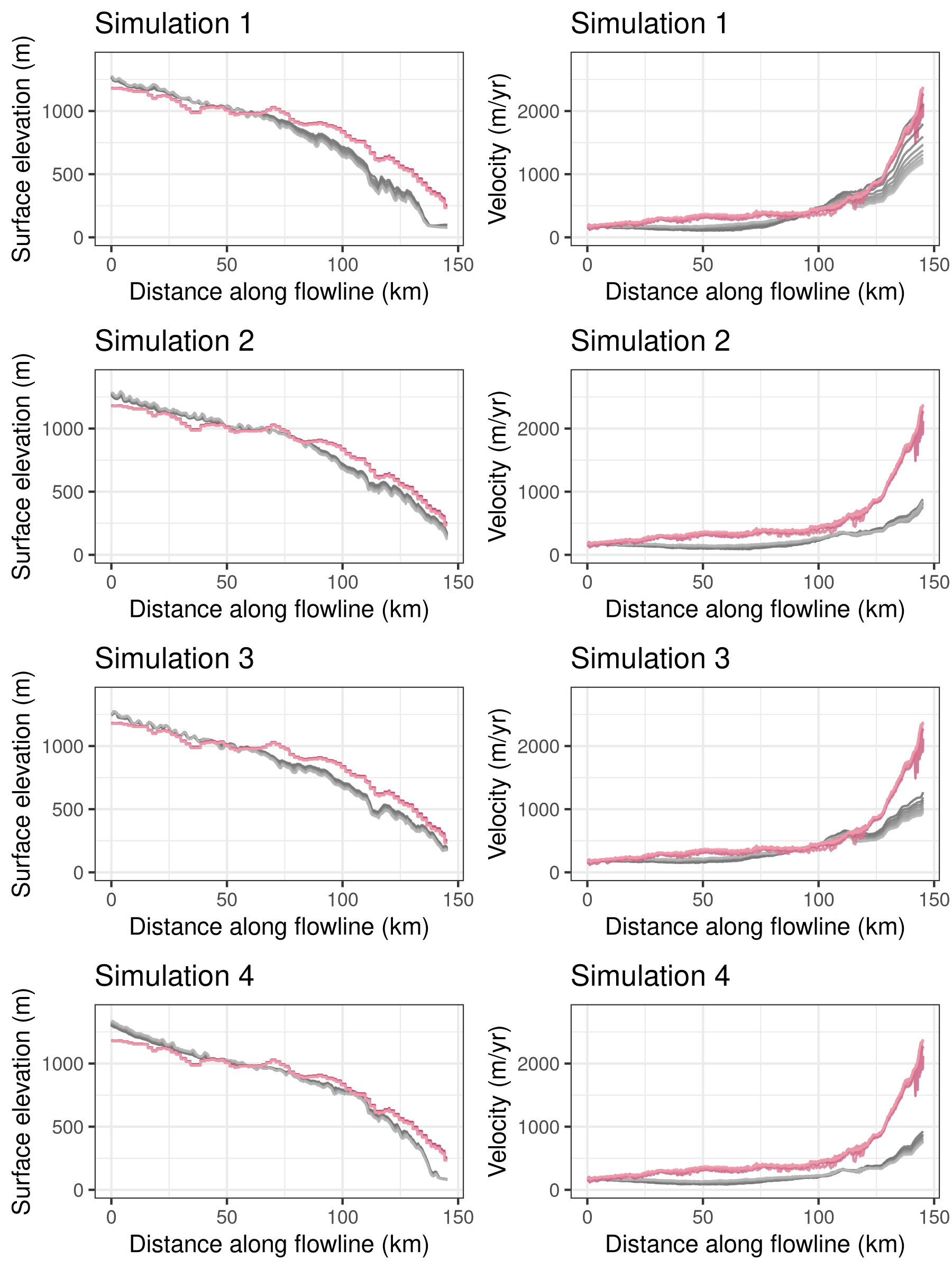}
    \caption{Examples of simulated observations of surface elevation (left column) and surface velocity (right column) along the Thwaites Glacier flowline based on several realisations of bed elevation and basal friction coefficient. Grey lines represent simulated surface elevation/velocity, with lighter lines corresponding to increasing years from 2010 to 2019. Red lines represent the observed surface elevation and surface velocity from MEaSUREs, with increasing lightness corresponding to increasing years.}
    \label{fig:se_sims_20241111}
\end{figure}

\begin{figure}[t]
    \centering
    \includegraphics[width=0.85\textwidth]{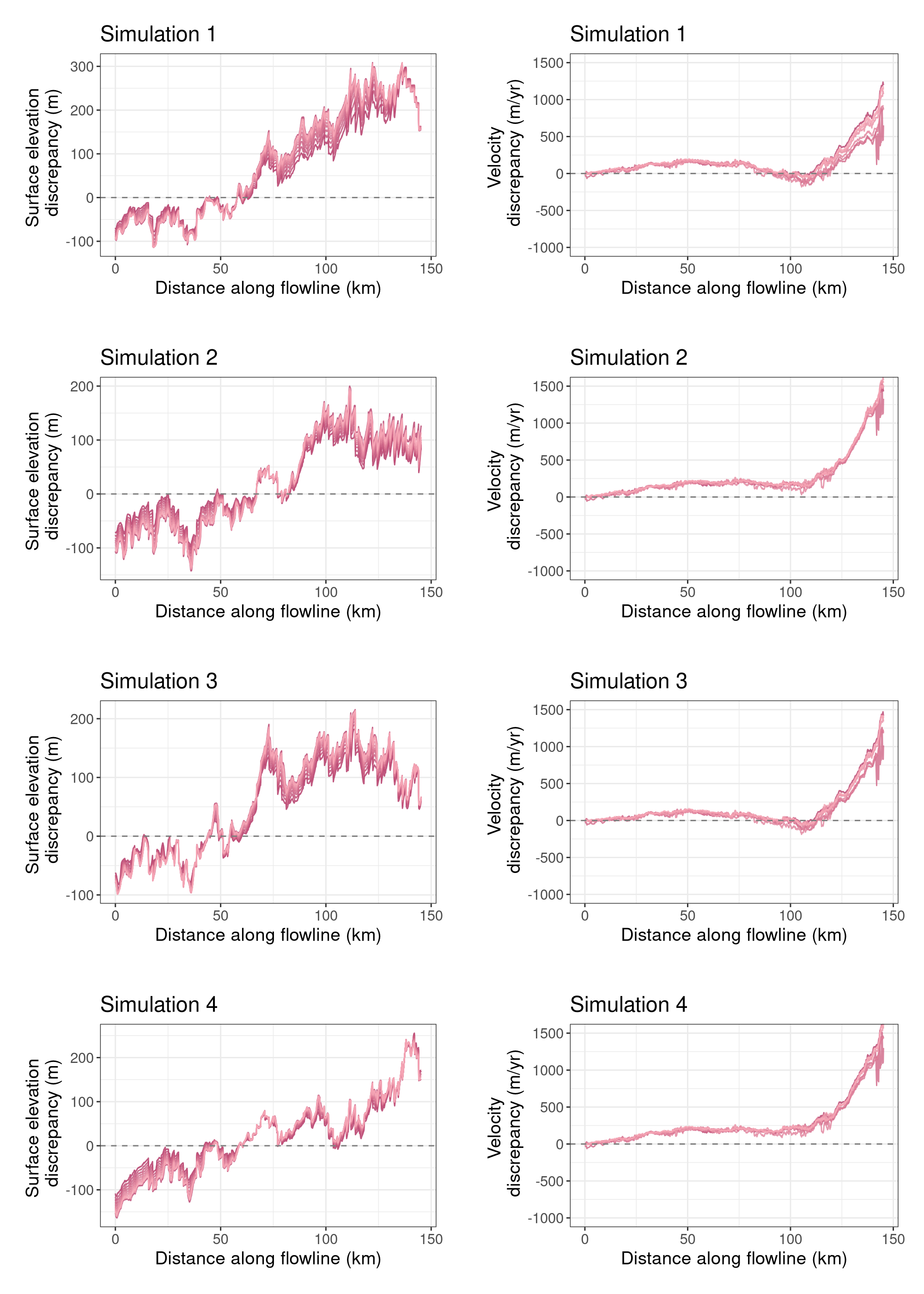}
    \caption{Examples of model discrepancy between simulated and observed surface elevation along the Thwaites Glacier flowline for several simulations (corresponding to those from Figure~\ref{fig:se_sims_20241111}). Coloured lines (in increasing lightness) represent the discrepancy with increasing years from 2010 to 2019.}
    \label{fig:se_discrepancy_20241111}
\end{figure}

Two examples of simulated bed elevation and basal friction coefficient, along with H\"{o}vmoller plots of the corresponding surface elevation and surface velocity, are shown in Figure~\ref{fig:se_sims_20241111}. 
Note the presence of a structured discrepancy between the simulations and the observations from the MEaSUREs dataset, which are shown in Figure~\ref{fig:se_discrepancy_20241111}.

\clearpage
\newpage
\section{Choice of approximate posterior}
This section contains a small simulation study with a univariate linear Gaussian state-space model, where we compare the performance of neural posterior inference under two choices of posterior approximation: (1) a Gaussian posterior with diagonal covariance matrix, and (2) posterior based on normalising flows. The model is given by
\begin{align}
    y_t &= x_t + \epsilon_t, \quad \epsilon_t \sim N(0, \sigma_{\epsilon}^2), \quad t = 1, \dots, T,\\
    x_t &= \phi x_{t-1} + \eta_t, \quad \eta_t \sim N(0, \sigma_{\eta}^2), \quad t = 2, \dots, T, \\
    x_1 &\sim N \left(0, \frac{\sigma_{\eta}^2}{1 - \phi^2} \right),
\end{align}
where $T = 100$, and we wish to make inference on the parameters $\phi, \sigma_{\eta}$ and $\sigma_{\epsilon}$. In the neural network, we work with the transformed parameters $\btheta \equiv (\theta_{\phi}, \theta_{\eta}, \theta_{\epsilon})^\top$ where $\theta_{\phi} = \tanh^{-1}(\phi), \theta_{\eta} = \log(\sigma_{\eta}^2)$, and $\theta_{\epsilon} = \log(\sigma_{\epsilon}^2)$. 
    
For both the case of a Gaussian approximate posterior and the case of posterior based on normalising flows, we use a convolutional neural network with similar structure to that used in the main paper. In particular,
\begin{itemize}
    \item The input is a time series of length $T = 100$.
    \item The hidden layers are composed of three one-dimensional convolutional layers of width $32, 64$, and $128$, with convolutional kernels of size $5$, $3$, and $2$, respectively. We use ReLU activation functions. 
    \item Each convolutional layer is followed by a max-pooling layer with a kernel of length 2 and a stride of 2. 
    \item The output from the last max-pooling layer is flattened and passed through a dropout layer with dropout rate $0.25$, then mapped to 9 ``summary statistics'' using a Dense layer.
    \item The summary statistics are mapped to the parameters of the posterior distribution (mean and variance parameters in the case of a Gaussian posterior, and flow parameters in the case of normalising flows) using a multilayer perceptron. 
\end{itemize}

For both forms of approximate posterior, neural network training is implemented using the \texttt{NeuralEstimators} package~\citep{neuralestimators}. We simulate $\num{50000}$ samples for training and another $\num{5000}$ for validation. At each training epoch, the validation loss (which is the negative log posterior) is recorded, and a ``minimum validation loss'' over all the epochs up to the current one is computed. Training terminates when the validation loss over the most recent 5 epochs has not decreased below the recorded minimum validation loss.  

\begin{figure}
    \centering
    \includegraphics[width=\linewidth]{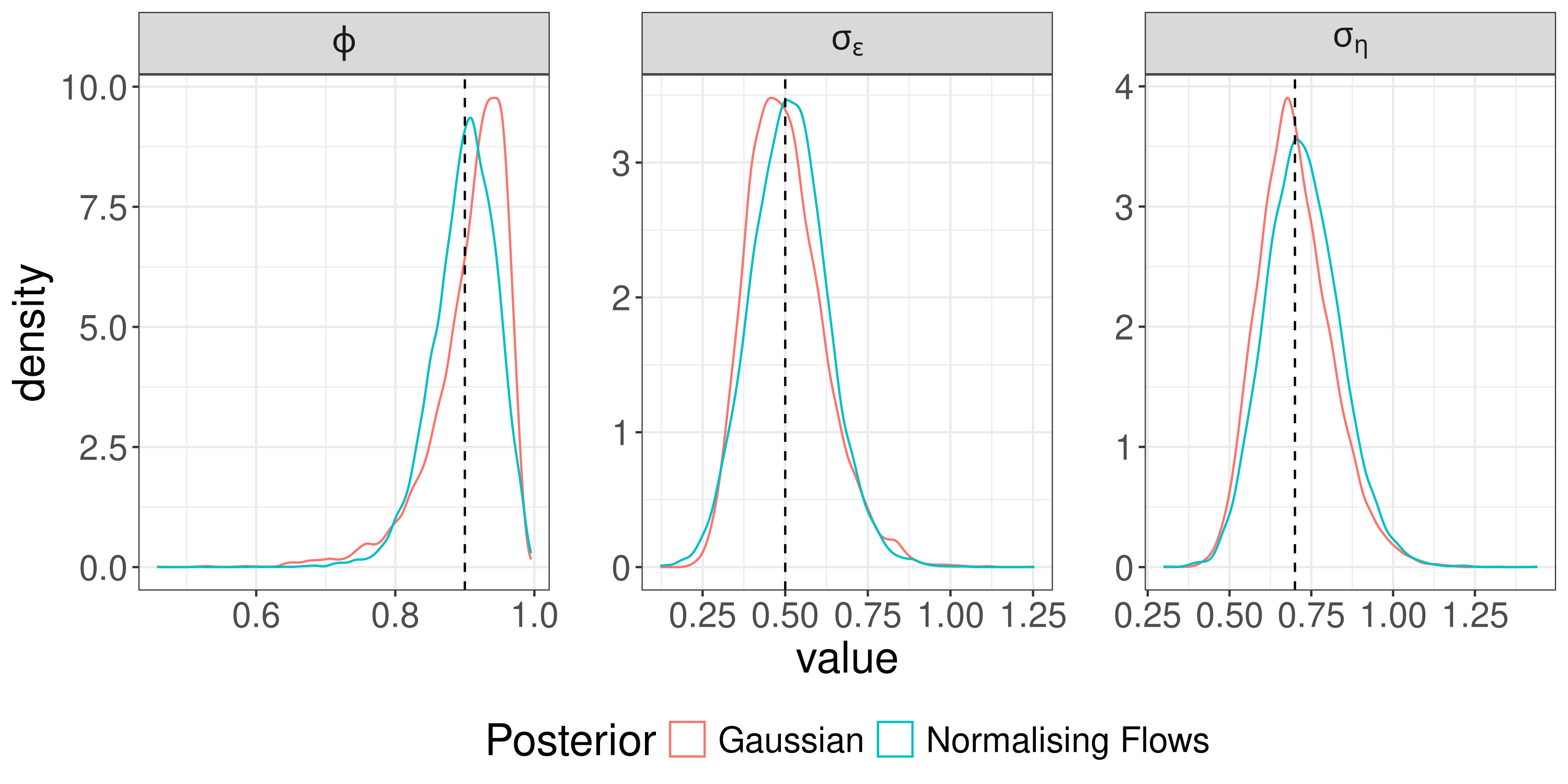}
    \caption{Comparison of approximate posterior distributions obtained via NPI, using a Gaussian approximate posterior with sparse (diagonal) covariance matrix (red) versus using normalising flows (blue) based on one ``observed'' dataset. True parameter values are indicated with vertical dotted lines.}
    \label{fig:gauss_vs_nf}
\end{figure}

We then assess the posterior distributions from both cases based on a randomly generated data sample that we treat as ``observed" data. This ``observed'' sample is generated using parameter values $\phi = 0.9, \sigma_{\eta} = 0.7, \sigma_{\epsilon} = 0.5$. A comparison of the resulting posterior distributions is shown in Figure~\ref{fig:gauss_vs_nf}. The Gaussian approximate posterior distribution is very close to that based on normalising flows. However, the former is less computationally expensive to train and takes approximately 357 seconds, while the latter takes approximately 1122 seconds on a 2x Intel Xeon Gold 6348 CPU.
    
\end{document}